\documentclass[12pt,twoside,english]{extarticle}
\usepackage{lmodern}
\usepackage{helvet}
\usepackage[T1]{fontenc}
\usepackage[latin9]{inputenc}
\usepackage{color}
\usepackage{babel}
\usepackage{float}
\usepackage{mathrsfs}
\usepackage{amsmath}
\usepackage{amsthm}
\usepackage{amssymb}
\usepackage{geometry}
\usepackage{setspace}
\usepackage[pdfusetitle,
 bookmarks=true,bookmarksnumbered=false,bookmarksopen=false,
 breaklinks=true,pdfborder={0 0 0},pdfborderstyle={},backref=false,colorlinks=true]
 {hyperref}

\makeatletter

\providecommand{\tabularnewline}{\\}

\numberwithin{equation}{section}
\numberwithin{table}{section}
\numberwithin{figure}{section}
\usepackage[natbibapa]{apacite}
\theoremstyle{definition}
\newtheorem{example}{\protect\examplename}[section]
\theoremstyle{plain}
\newtheorem{assumption}{\protect\assumptionname}
\theoremstyle{definition}
\newtheorem{defn}{\protect\definitionname}[section]
\theoremstyle{plain}
\newtheorem{thm}{\protect\theoremname}[section]
\theoremstyle{plain}
\newtheorem{cor}{\protect\corollaryname}[section]
\theoremstyle{plain}
\newtheorem{prop}{\protect\propositionname}[section]
\theoremstyle{plain}
\newtheorem{lem}{\protect\lemmaname}[section]

\usepackage{graphicx}
\usepackage{amsmath}
\usepackage{dcolumn}
\usepackage{listings}
\usepackage{color}
\usepackage{setspace}
\usepackage[normalem]{ulem}  
\definecolor{hellgelb}{rgb}{1,1,0.8}
\definecolor{colKeys}{rgb}{0,0,1}
\definecolor{colIdentifier}{rgb}{0,0,0}
\definecolor{colComments}{rgb}{1,0,0}
\definecolor{colString}{rgb}{0,0.5,0}

\usepackage{lmodern}
\usepackage[T1]{fontenc}
\usepackage{geometry}
\usepackage{fancyhdr}
\usepackage{amsthm}
\usepackage{thmtools}
\usepackage{amsmath}
\usepackage{amssymb}
\usepackage{esint}
\usepackage{multirow}
\usepackage{mathrsfs} 
\usepackage{textgreek}
\usepackage{amsfonts}
\usepackage{amssymb}
\usepackage{mathrsfs} 

\usepackage[USenglish]{isodate}
\usepackage{chngcntr}

\numberwithin{equation}{section}
\numberwithin{table}{section}
\counterwithout{figure}{section}
\counterwithout{table}{section}
\counterwithout{example}{section}

\makeatother

  \providecommand{\assumptionname}{Assumption}

  \providecommand{\corollaryname}{Corollary}
  
  \providecommand{\definitionname}{Definition}
  \providecommand{\examplename}{Example}

  \providecommand{\lemmaname}{Lemma}

  \providecommand{\propositionname}{Proposition}

  \providecommand{\theoremname}{Theorem}
 \providecommand{\corollaryname}{Corollary}
 \providecommand{\theoremname}{Theorem}
 
\usepackage{thmtools} 

\counterwithout{thm}{section}
\counterwithout{defn}{section}
\counterwithout{cor}{section}
\counterwithout{prop}{section}
\newtheoremstyle{MyTheoremstyle}
  {\topsep} 
  {\topsep} 
  {} 
  {} 
  {\bfseries} 
  {.} 
  {.90em} 
  {} 
\theoremstyle{MyTheoremstyle} 
\theoremstyle{MyTheoremstyle} 
\theoremstyle{MyTheoremstyle} 
\theoremstyle{MyTheoremstyle} 
\theoremstyle{MyTheoremstyle}

\declaretheoremstyle[
    headfont=\bfseries,
    notefont=\normalfont,
    bodyfont=\itshape,
    headpunct=\newline,
    headformat={%
        \makebox{\NAME\ \NUMBER\ }{\NOTE}%
    },
]{theorem}

\newlength{\spacelength}
\declaretheoremstyle[
    headfont=\bfseries,
    notefont=\normalfont,
    bodyfont=\itshape,
    headpunct=\newline,
    headformat={%
        \makebox[0pt][l]{\NAME\ \NUMBER\ }\hskip-\spacelength{\NOTE}%
    },
]{theore}

 \usepackage{fancyhdr}
\usepackage{booktabs}
\usepackage{float}
\usepackage{graphicx}
\usepackage{epstopdf}
\usepackage{morefloats}
\usepackage[referable]{threeparttablex}
\usepackage{footnote}

\usepackage{setspace} 
\usepackage{graphicx,color}

\usepackage{listings}
\RequirePackage[T1]{fontenc}
\RequirePackage{ae,fancyvrb}
\DefineVerbatimEnvironment{Sinput}{Verbatim}{fontshape=sl}
\DefineVerbatimEnvironment{Soutput}{Verbatim}{}
\DefineVerbatimEnvironment{Scode}{Verbatim}{fontshape=sl}

\title{\bf Identification of Causal Effects in High-Frequency Event Studies}
\author{
\textsc{\textcolor{MyBlue}{Alessandro Casini}}\thanks{Dep. of Economics and Finance, University of Rome Tor Vergata, Via Columbia 2, Rome, 00133, IT. 
Email: 
\texttt{\textcolor{MyBlue}{{alessandro.casini@uniroma2.it}}}.} 
\\
\small{\text{University of Rome Tor Vergata}}
\and
\textsc{\textcolor{MyBlue}{Adam McCloskey}}\thanks{Dep. of Economics, University of Colorado at Boulder, 256 UCB, Boulder, CO 80309, US. 
Email: 
\texttt{\textcolor{MyBlue}{\mbox{adam.mccloskey@colorado.edu}}}.} 
\\
\small{\text{University of Colorado at Boulder}}
}

\date{\small{\today} 
}



\usepackage{babel}
\addto\captionsenglish{%
}

\usepackage{color}
\usepackage{xcolor}

\definecolor{MyRed}{rgb}{0.8,0,0}
\definecolor{MyBlue}{rgb}{0,0,0.7}
\definecolor{Green}{rgb}{0,0.5,0}
\definecolor{hellgelb}{rgb}{1,1,0.8}
\definecolor{colKeys}{rgb}{0,0,1}
\definecolor{colIdentifier}{rgb}{0,0,0}
\definecolor{colComments}{rgb}{1,0,0}
\definecolor{colString}{rgb}{0,0.5,0}

\definecolor{MyLightRed}{rgb}{2.2,0.2,0.4} 
\definecolor{MyLightRed2}{rgb}{0.6,0.2,0.3} 
\definecolor{MyLightRed2temp}{rgb}{0.6,0.2,0.3}
\definecolor{MyLightRed3}{rgb}{0.8,0.1,0.1} 
\definecolor{MyRed}{rgb}{0.7,0.0,0}

\definecolor{MyLigthBlue13}{rgb}{0,0.2,0.7}
 \definecolor{MyLigthBlack}{rgb}{0.2,0.25,0.3} 

\hypersetup{%
  bookmarks=true
  pdftitle = {Title Here},
  pdfsubject = {},
  pdfkeywords = {Keywords},
  pdfauthor = {Alessandro Casini},}
\hypersetup{ 
  colorlinks = {true}, 
  citecolor = {MyBlue},
  linkcolor = {MyRed},
  filecolor= {Green},	
  urlcolor = {MyBlue},
  hyperindex = {true},
  linktocpage = {true},
  linkbordercolor ={1 0 0},
 citebordercolor ={0 1 1},
 urlbordercolor ={0 1 1},
hypertexnames=false
}

\usepackage{bookmark}

\usepackage[bottom]{footmisc}
\usepackage{multibib}
\newcites{ReferencesSupp}{References}

\makeatother

\providecommand{\assumptionname}{Assumption}
\providecommand{\corollaryname}{Corollary}
\providecommand{\definitionname}{Definition}
\providecommand{\examplename}{Example}
\providecommand{\lemmaname}{Lemma}
\providecommand{\propositionname}{Proposition}
\providecommand{\theoremname}{Theorem}

\begin{document}
\pagebreak{}

\setcounter{page}{0}

\raggedbottom
\title{\textbf{Identification, Estimation and Inference in High-Frequency
Event Study Regressions\thanks{This paper previously circulated as \textquotedblleft Identification
and Estimation of Causal Effects in High-Frequency Event Studies\textquotedblright .
We thank Chris House, Eva Janssens, Donggyu Kim, Emi Nakamura, Mikkel
Plagborg-M\o{}ller, Andres Santos, J\'{o}n Steinsson, Stepehen Terry,
Mark Watson and Kaspar W\"{u}thrich for useful comments. We thank
seminar participants at Bank of Italy, Princeton University, UCLA,
UC Riverside, UC Santa Cruz, University of Colorado and University
of Michigan. McCloskey acknowledges support from the National Science
Foundation under Grant SES-2341730. The replication code is available
on our websites.}}}
\maketitle
\begin{abstract}
{\footnotesize We consider identification, estimation and inference
in high-frequency event study regressions, which have been used widely
in the recent macroeconomics, financial economics and political economy
literatures. The high-frequency event study method regresses changes
in an outcome variable on a measure of unexpected changes in a policy
variable in a narrow time window around an event or a policy announcement
(e.g., a 30-minute window around an FOMC announcement). We show that,
contrary to popular belief, the narrow size of the window is not sufficient
for identification. Rather, the population regression coefficient
identifies a causal estimand when (i) the effect of the policy shock
on the outcome does not depend on the other variables (separability)
and (ii) the surprise component of the news or event dominates all
other variables that are present in the event window (}{\footnotesize\textit{relative
exogeneity}}{\footnotesize ). Technically, the latter condition requires
the ratio between the variance of the policy shock and that of the
other variables to be infinite in the event window. Under these conditions,
we establish the causal meaning of the event study estimand corresponding
to the regression coefficient and super-consistency of the event study
estimator with rate of convergence faster than the parametric rate.
We show the asymptotic normality of the estimator and propose bias-corrected
inference. We also provide bounds on the worst-case bias and use them
to quantify its impact on the worst-case coverage properties of confidence
intervals, as well as to construct a bias-aware critical value. Notably,
this standard linear regression estimator is robust to general forms
of nonlinearity. We apply our results to \citeauthor{nakamura/steinsson:2018}'s
(\citeyear{nakamura/steinsson:2018}) analysis of the real economic
effects of monetary policy, providing a simple empirical procedure
to analyze the extent to which the standard event study estimator
adequately estimates causal effects of interest.}{\footnotesize\par}

{\footnotesize}{\footnotesize\par}
\end{abstract}

\onehalfspacing
\thispagestyle{empty}

\pagebreak{}

\section{Introduction}

Randomized controlled experiments offer an ideal framework for identifying
causal effects. However, in macroeconomics and finance controlled
experiments cannot be run in practice. Instead, economists often search
for pseudo-experiments, that is situations in which one can extract
plausible exogenous variation in policy and use this variation to
estimate the effect of the policy on some economic outcome {[}cf.
\citet{nakamura/steinsson:2018_survey}{]}. Recently, there has been
a surge of interest in high-frequency event study regressions for
estimating causal effects in applied work in macroeconomics, financial
economics and political economy, among others. The idea behind the
event study approach based on high-frequency data is that in a narrow
time window around a policy announcement or data release {[}e.g.,
a Federal Open Market Committee (FOMC) announcement, a U.S. employment
report released by the Bureau of Labor Statistics, a GDP release report
by the Bureau of Economic Analysis, etc.{]}, one can extract the unexpected
change or surprise in the policy and regress the changes in an outcome
variable within the narrow window on the policy surprises to estimate
the causal effect of the policy.

Though this approach has become central to empirical work, there are
no corresponding theoretical results establishing identification of
causal effects via this method, and informal identification arguments
used by different authors do not always coincide.\footnote{Recent examples of informal identification arguments include \citet{bauer/swanson:2021}
and \citet{nakamura/steinsson:2018}, who discussed the exogeneity
and relevance of monetary policy surprises typically constructed around
FOMC announcements for identifying the macroeconomic effects of monetary
policy shocks.} In this paper we establish precise conditions for nonparametric identification
of causal effects by high-frequency event study regressions. We show
that, contrary to popular belief, the narrow size of the time window
the event study regression is run over is not sufficient for the identification
of causal effects. Rather, the population regression coefficient identifies
a causal estimand when (i) the effect of the policy shock on the outcome
does not depend on the other variables (separability) and (ii) the
surprise component of the news or event dominates all other variables
that are present in the event window (\textit{relative exogeneity}).
Under these conditions, we establish the causal meaning of the event
study estimand corresponding to the regression coefficient and the
consistency and asymptotic normality of the event study regression
estimator. Notably, this standard linear regression estimator is robust
to general forms of nonlinearity (e.g., between the outcome and policy
variables).

Separability holds, for example, when the model is linear, which is
often assumed in applied work. The key condition deserving of careful
scrutiny in practice, relative exogeneity, holds when the policy shock
has infinite variance while the other variables have finite variance
within the event window. More precisely, relative exogeneity holds
when the ratio between the variance of the policy shock and that of
the other variables (i.e., background noise) is infinite in the event
window. Thus, relative exogeneity also holds when the variance of
the policy shock is finite while the variance of the background noise
is vanishing. The latter variables correspond to factors that are
not specific to the announcement and may also be present in non-announcement
periods. In contrast, the policy shock occurs in a lumpy manner as
the unexpected part of the news quickly spreads among economic agents.
It is this lumpy manner in which a disproportionate amount of policy
news is revealed that can justify relative exogeneity. Even when the
policy shock does not have infinite variance, which can be difficult
to verify in practice, we show that the event study estimator has
low bias for a weighted average of causal effects when the variance
of the policy shock is large enough relative to that of other variables
in the window. In this sense, relative exogeneity can be seen as an
idealized limiting case that can serve as a good approximation to
the practically-relevant case of a very large variance ratio for the
policy shock relative to the other variables in the window.

Relative exogeneity relates to the size of the event window. As the
size of the window expands, it becomes less likely that the policy
shock dominates all other variables within the window, making it more
likely that relative exogeneity provides a poor approximation. Relative
exogeneity can also fail if there is information leakage about the
policy news or some market anticipation of the policy change. Information
leakage or frictions result in a reduction in the variance of the
policy shock relative to the other variables in the window. The finance
literature has recently documented strong evidence of information
leakage, informal communication and informed trading around policy
announcements {[}see, e.g., \citet{cieslak/morse/vissing-jorgensen:2019},
\citet{cieslak/schrimpf:2019} and \citet{lucca:moench:2015}{]}.
However, if the extent of the information leakage is small, then it
will lead only to partial market anticipation and will not prevent
the news from coming out in a lumpy manner at the release time, thereby
allowing relative exogeneity to hold.

When separability and relative exogeneity hold we show that (i) any
reverse causality from the outcome variable to the policy variable
does not generate bias since the reverse causality is dominated by
the policy shock and (ii) common unobserved factors correlated with
the policy variable do not generate omitted variables bias since they
are also dominated by the policy shock. Since the event study method
regresses changes in an outcome on the unexpected changes in a policy
variable at dates in the policy sample, the event study estimand is
equal to the corresponding population regression coefficient. We establish
that the latter identifies a weighted average of standardized marginal
causal effects (MCEs) of the policy on the outcome.\footnote{The MCE is the derivative of the potential outcome with respect to
the policy variable. This should not be confused with the marginal
treatment effect (MTE) used in microeconometrics which refers to the
treatment effect of a program or intervention for those at the margin
of participation {[}cf. \citet{heckman/vytlacil:2001}{]}.} When relative exogeneity is violated, we show that the event study
estimand can be decomposed into the same weighted average of MCEs
and a selection bias factor. The magnitude of the selection bias factor
is decreasing in the variance of the policy shock so that even when
relative exogeneity fails, the event study estimator will not have
substantial bias so long as the variance of the policy shock is relatively
large.

It is difficult to use statistical tests to verify relative exogeneity.
Existing tests for (in)finite variance {[}e.g., \citet{trapani:2016}{]}
require a large sample of the corresponding random variable to be
observed or residuals from a correctly-specified regression. Given
the potential for simultaneity and omitted variables bias in high-frequency
event study regressions, these requirements are not satisfied in the
current context. In addition, the OLS event study estimator can still
perform well when relative exogeneity holds approximately. A test
for infinite variance of the policy shocks alone, even when feasible,
is unable to detect when relative exogeneity provides a good enough
approximation for the event study estimator to perform well in practice.
Instead of trying to test for infinite variance, we introduce a simple
empirical procedure that can be used for a sensitivity analysis to
diagnose whether relative exogeneity is ``close enough'' to holding
that the OLS event study estimator should be expected to have mean-squared
error at least as small as that of an oracle estimator in a corresponding
regression with no endogeneity. This sensitivity analysis uses information
available in the relevant control sample, e.g., days when there is
no FOMC meeting, in order to obtain a proxy for the variance of the
variables that are not specific to the policy within the event window.
The procedure can accommodate nonlinearities, heteroskedasticity and
serial correlation. This procedure requires no additional data relative
to that already used by existing high-frequency event studies and
can thus be applied easily. We introduce our procedure in the context
of an empirical example aimed at assessing the causal effects of monetary
policy news on real interest rates. The empirical results show that
relative exogeneity is likely to approximately hold in the analysis
of \citet{nakamura/steinsson:2018} based on a 30-minute or 1-day
window for the outcome and policy variable.

Under separability and relative exogeneity, we establish the super-consistency
of the OLS event study estimator for a weighted average of standardized
MCEs, with a rate of convergence equal to the square root of the sample
size multiplied by the standard deviation of the policy variable.
Asymptotic bias arises when relative exogeneity is not \textquotedblleft strong
enough\textquotedblright\textemdash that is, when the standard deviation
of the policy variable diverges at the same rate or more slowly than
the square root of the sample size. Nonetheless, we show that under
the \textquotedblleft sharp null\textquotedblright{} hypothesis of
zero MCEs, it is possible to consistently estimate the asymptotic
bias and develop inference refinements based on bias correction, even
when relative exogeneity is not ``strong enough''. 

Since the bias vanishes only in the limit under relative exogeneity,
we consider methods to quantify how this bias can affect the worst-case
properties of the estimator by deriving a bound on the worst-case
bias. In particular, we use this bound to study the worst-case asymptotic
coverage of standard confidence intervals based on the OLS event study
estimator. We then propose a bias-aware critical value that accounts
for the estimator\textquoteright s potential asymptotic bias in addition
to its variance, by adjusting the standard normal critical value upward
to compensate for the bias. The resulting bias-aware inference does
not rely on the imposition of the \textquotedblleft sharp null\textquotedblright{}
hypothesis.

We use our identification framework to shed light on the recent debate
in the literature on some puzzling event study regression results
involving the causal effect of monetary policy on Blue Chip forecasts
of real GDP. We show that the estimates with signs opposite to the
predictions from standard macroeconomic models likely arise because
relative exogeneity fails in this type of regression. Since the Blue
Chip forecast revision is constructed as a one-month change while
the monetary policy surprise is constructed as a 30-minute change,
it is highly unlikely that the variance of the policy shocks dominates
the variance of the other variables that determine the Blue Chip forecast:
the latter aggregate news over a much longer time frame. Thus, any
endogeneity of the policy surprise is not likely to be drowned out
by the variation in the policy shock.

Before proceeding with the analysis, it is worth clarifying the estimation
approach we analyze in relation to other related approaches. We analyze
the widely popular event study estimand in order to determine the
causal effects under study in empirical practice when estimating high-frequency
event study regressions. Under the conditions we provide, the high-frequency
event study estimand identifies a weighted average of MCEs of the
policy variable rather than those of exogenous shocks, which are more
customary in impulse response analysis of structural vector autoregressions
(SVARs). Our approach is thus to elucidate \emph{which} causal effects
are being analyzed in high-frequency event study regressions, not
to take a stand on which type of MCEs are more or less policy-relevant.
However, we note that there are many applications for which identifying
the causal effects of the policy variable is of interest. These include,
for example, estimating the slope of demand functions, price elasticities,
and response or reaction coefficients\textemdash such as how asset
prices respond to monetary policy. High-frequency policy surprises
used in event studies are also employed as instruments for the policy
shock in SVAR contexts to identify impulse responses. This approach
corresponds to a different estimand and identified causal effect,
which are not the focus of the present paper.

Although it bares some conceptual similarity, the high-frequency event
study approach is distinct from regression discontinuity. The former
involves a single observation for each announcement or event, and
the estimand is the linear projection coefficient, given by the ratio
of the covariance between the outcome and the policy variable to the
variance of the policy variable. In contrast, regression discontinuity
relies on many observations to the left and right of a given cutoff,
and the estimand is the difference in the conditional mean of the
outcome at the cutoff. A key identification condition for regression
discontinuity is local continuity of the mean of the potential outcome
in the absence of treatment around the cutoff. The high-frequency
event study estimator does not converge to the regression discontinuity
estimand, and local continuity is therefore not sufficient to recover
a causal effect\textemdash even if one assumes the announcement window
shrinks to zero. Nevertheless, the infinite variance condition required
by relative exogeneity imposes a discontinuity in the policy variable
around the announcement, which is similar in spirit to the discontinuity
used in regression discontinuity designs. The technical difference
is that, in high-frequency identification, the bias is scaled by the
inverse of the ratio of the variance of the policy shock to the variance
of other shocks, which converges to zero. In contrast, in regression
discontinuity, the bias is eliminated through local differencing.

The rest of the paper is structured as follows. Section \ref{Section: Event Study Identification of Causal Effects}
reviews the high-frequency event study literature, presents several
empirical examples, introduces our identification results, and discusses
robustness to information leakage. Section \ref{Section: Properties-of-the OLS ES}
establishes the asymptotic properties of the OLS event study estimator.
Section \ref{Section: Bounds on bias, coverage, bias-aware} derives
bounds on the asymptotic bias, examines the worst-case asymptotic
coverage properties of confidence intervals based on the OLS event
study estimator, and proposes bias-aware inference. Section \ref{Section: Empirical Procedure Identification}
develops a sensitivity analysis to assess the identification conditions,
applies both the sensitivity analysis and our inference results to
an empirical example, and discusses identification issues in event
study regressions involving Blue Chip forecasts. Section \ref{Section Conclusions}
concludes. 

\section{Identification in High-Frequency Event Studies \label{Section: Event Study Identification of Causal Effects}}

We begin with a brief review of the event study methodology in empirical
work in Section \ref{Subsection: Event Study Design}. We then introduce
the potential outcomes framework and provide formal conditions for
nonparametric identification of causal effects via the event study
regressions in Section \ref{Subsection: Identification-Conditions}.
We present the identification results in Section \ref{Subsection: Identification-Results}
and relate them to the literature in Section \ref{Subsection: Relation to the Literature}.
Robustness to information leakage is discussed in Section \ref{Subsection: Robustness to Leakage}.

\subsection{\label{Subsection: Event Study Design}Event Study Design}

Consider a system of dynamic simultaneous equations that relate an
outcome variable $Y_{t}$ and a (measure of) policy action $D_{t}$
to each other: 
\begin{align}
Y_{t} & =\beta D_{t}+X'_{t}\theta+Z'_{t}\gamma_{1}+u_{t},\label{Eq.: Event study general model}\\
D_{t} & =\alpha Y_{t}+X'_{t}\phi+Z'_{t}\gamma_{2}+e_{t},\nonumber 
\end{align}
 where $X_{t}$ represents observed macroeconomic variables that
might influence both the outcome and policy variable, $Z_{t}$ represents
unobserved macroeconomic factors that also might affect the outcome
and policy variable and $u_{t}$ and $e_{t}$ are serially uncorrelated
shocks that are mutually uncorrelated. The parameters $\beta$ and
$\alpha$ are scalars while $\theta,\,\phi,$ $\gamma_{1}$ and $\gamma_{2}$
are finite-dimensional vectors. $X_{t}$ is a vector that could include
lags of $Y_{t}$ and $D_{t}$. The parameter of interest is $\beta$
which captures the response of the outcome variable to the policy
action. The policy variable is endogenous, i.e., $D_{t}$ reacts to
$Y_{t}$. Further, since $Z_{t}$ is not observed there is an omitted
variables problem as $Z_{t}$ and $D_{t}$ may be correlated. Hence,
the identification of $\beta$ requires one to overcome both simultaneity
and an omitted variables problem.

The event study approach based on high-frequency observations (e.g.,
weekly, daily and intradaily frequencies) of $Y_{t}$ and $D_{t}$
can be used to address these identification issues. The idea is that
in a narrow time window around a policy announcement or data release
(e.g., FOMC announcement, U.S. employment report, GDP data release,
etc.) one can extract the unexpected change (or surprise) in the policy
action to form $D_{t}$. The sample consists of observations of $Y_{t}$
and $D_{t}$ at the dates corresponding to the relevant policy announcement,
data release or event. We call this sample the policy sample and denote
it by $\mathbf{P}.$ Under the identification conditions to be discussed
below, a simple OLS regression of $Y_{t}$ on $D_{t}$ over the policy
sample (i.e, over all $t\in\mathbf{P})$ recovers causal effects of
the policy action on the outcome. 

We now present a few examples that use the model \eqref{Eq.: Event study general model}.
\begin{example}
\label{Example: Monetary policy, NK and Kuttner}{[}FOMC announcements:
\citeauthor{bauer/swanson:2022} (\citeyear{bauer/swanson:2021},
\citeyear{bauer/swanson:2022}), \citet{kuttner:2001}, \citet{nakamura/steinsson:2018}
and \citet{rigobon/sack:2004}{]} Several authors  investigated the
impact of monetary policy on the real economy using the event study
approach.\footnote{See \citet{ai/bansal:2018}, \citet{cochrane/piazzesi:2002}, \citet{cook/hahn:1989},
\citet{gurkaynak/sack/swanson:2005}, \citet{lucca:moench:2015},
\citet{bernile/hu/tang:2016}, \citet{hu/pan/wang/zhu:2022}, \citeauthor{caballero/simsek:2022}
(\citeyear{caballero/simsek:2022}, \citeyear{caballero/simsek:2023}),
\citet{cieslak/morse/vissing-jorgensen:2019}, \citet{cieslak/mcmahon:2023},
\citet{cieslak/schrimpf:2019}, \citet{hansen/mcmahon/prat:2018},
\citet{hanson/stein:2015}, \citet{jarocinski/karadi:2020}, \citet{michelacci/paciello:2020},
\citet{neuhierl/weber:2019} and \citet{swanson:2021}.} \citet{kuttner:2001} explained how to use Federal funds futures
contracts to separate changes in the Fed funds rate (i.e., the short-term
interest rate) into anticipated and unanticipated monetary policy
actions, the latter being $D_{t}$. In \citet{kuttner:2001} $D_{t}$
is the 1-day change in the spot-month Federal funds future rate and
$Y_{t}$ represents a  yield on a zero-coupon Treasury bill (or bond)
at some maturity or a change in an asset price. The policy sample
$\mathbf{P}$ collects the dates of the FOMC announcements and the
dates when the Fed funds target rate was changed (if that did not
coincide with an FOMC announcement).\\
\indent This analysis was further elaborated by, among others, \citet{nakamura/steinsson:2018}
who used intradaily data and looked at a 30-minute window surrounding
each FOMC announcement. The authors considered 30-minute changes in
the zero-coupon yields and instantaneous forward rates constructed
using Treasury Inflation Protected Security data at different maturities
and changes in survey expectations on output and inflation, as the
outcome variable $Y_{t}$. To construct the monetary policy news $D_{t}$
the authors extracted the first principle component of the unanticipated
change over 30-minute windows of five interest rates chosen among
the Federal funds futures and eurodollar futures. The latter provide
a direct measure of the unexpected component of the policy change.
They estimated the causal effect of $D_{t}$ on $Y_{t}$ by running
an event study OLS regression of $Y_{t}$ on $D_{t}$ for dates in
the policy sample: 
\begin{align}
Y_{t} & =\beta D_{t}+\widetilde{u}_{t},\qquad\qquad t\in\mathbf{P},\label{Eq. (1) Event study regression in the literature}
\end{align}
where $\widetilde{u}_{t}$ is an error term. The control variables
$X_{t}$ that could be included in this analysis are monthly releases
of major macroeconomic variables or other low-frequency variables.
For example, the consumer price index (CPI), nonfarm payrolls, producer
price index (PPI), retail sales, etc. However, in practice event study
regressions do not typically involve control variables.\footnote{See \citet{bauer/swanson:2021} and \citet{rigobon/sack:2003} for
exceptions.}\\
\indent It is not obvious how to justify the exclusion of either
the observed or unobserved factors $X_{t}$ and $Z_{t}$ from \eqref{Eq. (1) Event study regression in the literature}.
Asset prices likely react to $X_{t}$ and $Z_{t}$ even within the
event window, no matter how small the window is. If not, a recursive
argument soon would contradict asset pricing theory. Think about splitting
the regular trading hours for the U.S. stock market, 9:30am-4:00pm,
into non-overlapping small time windows of, for example, 30 minutes.
If one assumes that asset prices do not respond to observed or unobserved
macroeconomic factors over such a tight window, then applying this
argument recursively to each trading day implies that asset prices
never respond to such factors. That is, the choice of a very tight
window bracketing a policy announcement is not a sufficient condition
for precluding omitted variables bias from the regression \eqref{Eq. (1) Event study regression in the literature}
due to the correlation of $D_{t}$ with $X_{t}$ and/or $Z_{t}$.
We will show that under our identification conditions, which do not
refer explicitly to the size of the window, the OLS event study regression
\eqref{Eq. (1) Event study regression in the literature} that excludes
both the observed and unobserved factors $X_{t}$ and $Z_{t}$ can
still recover causal effects of interest. \\
\indent The recent literature documented evidence pointing to simultaneous
determination of $D_{t}$ and $Y_{t}$ in FOMC announcement applications.
See Section \ref{Subsection: Relation to the Literature} for details.
We will also show that under our identification conditions, the OLS
event study regression \eqref{Eq. (1) Event study regression in the literature}
recovers causal effects of interest in the presence of simultaneity.
\end{example}
\begin{example}
{[}Macroeconomic announcements: \citet{faust/rogers/wang/wright:2003},
\citet{gurkaynaky/kisacikoglu/wright:2020}, \citet{gurkaynak/sack/swanson:2005}
and \citeauthor{kanzig:2021} (\citeyear{kanzig:2021}; \citeyear{kanzig:2021_Carbon}){]}
Several works studied the effects of macroeconomic announcements on
changes in asset prices and exchange rates in a narrow window around
news releases such as the U.S. employment report, U.S. GDP releases,
Census reports on retail sales, and CPI and PPI data releases. \citet{gurkaynak/sack/swanson:2005}
also estimated the event study regression \eqref{Eq. (1) Event study regression in the literature}
but with a vector-valued $D_{t}$ and coefficient $\beta$, where
$Y_{t}$ is the log-return on an asset or a change in a bond yield
and $D_{t}$ is a vector of news, or unexpected, components of the
considered macroeconomic announcements and $u_{t}$ is a serially
uncorrelated error term.\footnote{Actually, \citet{gurkaynak/sack/swanson:2005} considered a vector
of asset prices $Y_{t}$. But since the cross-sectional variation
is not exploited for identification we simply take $Y_{t}$ to be
a scalar.} Unlike for the FOMC announcements, there are no traded instruments
from which to infer market expectations for macroeconomic announcements.
Thus, in order to identify surprise announcements the literature relies
instead on economists' forecasts from surveys: each element of $D_{t}$
is computed as the difference between the actual macroeconomic data
release and its market expectation obtained from the most recent survey.
These surveys are the Blue Chip Economic Indicators Survey or those
run by Action Economics, or alternatively by Bloomberg. Thus, $D_{t}$
captures the surprise component of each data release.\\
\indent Reverse causality from $Y_{t}$ to $D_{t}$ is ruled out
by the authors' identification argument that in a 20-minute window
around news releases, changes in asset prices do not cause news. For
example, the employment report that is released on the first Friday
of each month pertains to the labor market data in the previous month.
Thus, by construction changes in asset prices within the window affect
neither the news release nor the survey expectations since the latter
are collected earlier.\footnote{The amount of time between when the survey is given and the release
of the news is typically less than a week.}\\
\indent The exclusion of the  unobserved factors $Z_{t}$ that can
affect both asset prices and the unexpected component of the news
is easier to justify than in the case of FOMC announcements. Since
$D_{t}$ here involves economic data releases about the previous month
and survey expectations are collected in the days preceding the event
window, correlation between $Z_{t}$ and $D_{t}$ seems unlikely.
However, economic and business news and other events that occur in
the hours or days before the macroeconomic data release may require
time to incorporate into financial markets and may be subject to uncertainty
with respect to how they are interpreted by other market participants.
Therefore, $Z_{t}$ may contain information about other news and events
(e.g., company-specific news) that occur before the announcement window
and therefore influence survey expectations, and in turn $D_{t}$,
generating correlation between $Z_{t}$ and $D_{t}$. This issue is
independent from the size of the window used in the event study regression.
\end{example}
\begin{example}
{[}Political and war announcements: \citet{acemoglu/hassan/tahoun:2018},
\citet{dube/kaplan/naidu:2011}, \citet{garred/stickland/warrinnier:2023}
and \citet{giudolin/laferrara:2007}{]}. We briefly note that the
high-frequency event study methodology is also applied in the field
of political economy. \citet{giudolin/laferrara:2007} provided evidence
that violent conflict may be perceived by investors as beneficial
to incumbent firms. They focused on the Angolan civil war and its
effects on the industry of diamond production. They exploited the
sudden ceasefire of the civil war after the announcement of the death
of the rebels' leader, Jonas Savimbi, on February 22, 2002, showing
that international stock markets perceived Savimbi's death as bad
news for the companies in the diamond industry operating in Angola.
\end{example}

\subsection{\label{Subsection: Identification-Conditions}Nonparametric Event
Study Design and Identification Conditions}

Although we discussed examples in the high-frequency event study literature
in the context of the linear model \eqref{Eq.: Event study general model}
in the previous section, our identification results for event study
regressions actually apply to a much more general nonparametric class
of simultaneous equations models.\footnote{To simplify the exposition, we present the identification conditions
and results for the case when there are no control or pre-treatment
variables $X_{t}$ included in the design since they do not play an
important role for the identification analysis and are not typically
included in the high-frequency event study regressions, in which case
they can be considered elements of $Z_{t}$.} Unlike the linear model \eqref{Eq.: Event study general model},
this more general class of models allows for general forms of nonlinear
relationships between the variables in the system. Specifically, let
$Z_{t}$ denote an unobserved random vector and $u_{t}$ and $e_{t}$
denote scalar-valued random shocks to the outcome $Y_{t}$ and policy
$D_{t}$, respectively. When the observation belongs to the policy
sample $\mathbf{P}$, the outcome variable $Y_{t}$ is an unknown
nonparametric function of the policy variable $D_{t}$, $Z_{t}$,
$u_{t}$ and the time index $t$ while the policy variable $D_{t}$
is simultaneously an unknown nonparametric function of $Y_{t}$, $Z_{t}$,
$e_{t}$ and $t$: 
\begin{equation}
Y_{t}=\varphi_{Y}\left(D_{t},\,Z_{t},\,u_{t},\,t\right)\qquad\text{and}\qquad D_{t}=\varphi_{D}\left(Y_{t},\,Z_{t},\,e_{t},\,t\right),\label{Eq: nonparametric model}
\end{equation}
for all $t\in\mathbf{P}$. The unknown structural functions $\varphi_{Y}$
and $\varphi_{D}$ may be nonlinear in their arguments, allow for
simultaneous causality of $Y_{t}$ and $D_{t}$ and, since $Z_{t}$
is unobserved, allow for omitted variables in the system determining
$Y_{t}$ and $D_{t}$.

In Example \ref{Example: Monetary policy, NK and Kuttner}, $D_{t}$
depends on the Federal Reserve's best estimates of the strength of
the economy in the near-term and of potential inflationary pressures.
Those estimates will be influenced by the macroeconomic factors $Z_{t}$
and by changes in asset prices $Y_{t}$. The monetary policy shock
$e_{t}$ represents shifts in the preferences of individual FOMC members
or in the manner in which their views are aggregated. For example,
it could include changes to policy makers' goals and beliefs about
the economy, political factors, and the temporary pursuit of objectives
other than changes in the outcomes of interest (e.g., targeting inflation
rather than unemployment or exchange rates). The shock to the outcome
equation, $u_{t}$, captures any change in $Y_{t}$ not attributable
to the common macroeconomic factors $Z_{t}$ and policy action, $D_{t}$.
In Example \ref{Example: Monetary policy, NK and Kuttner}, $u_{t}$
is referred to as the asset price shock and is primarily driven by
shifts in investors' risk preferences.

To establish the identification of causal effects we impose some structure
on the system, in particular, a partial additive separability of the
structural function for the outcome variable.
\begin{assumption}
\label{Assumption: Structural Form}(Structural form separability)
For all $t\in\mathbf{P}$, $Y_{t}=\varphi_{Y}\left(D_{t},\,Z_{t},\,u_{t},\,t\right)=\varphi_{Y,D}\left(D_{t},\,t\right)+\varphi_{Y,u}\left(Z_{t},\,u_{t},\,t\right)$
for some functions $\varphi_{Y}$, $\varphi_{Y,D}$ and $\varphi_{Y,u}$.
\end{assumption}
If some smoothness on $\varphi_{Y,D}\left(\cdot,\,\cdot\right)$ is
assumed, then Assumption \ref{Assumption: Structural Form} implies
the following restrictions on partial effects: 
\begin{align*}
\frac{\partial^{2}Y_{t}}{\partial D_{t}\partial u_{t}}=0\qquad\text{and}\qquad\frac{\partial^{2}Y_{t}}{\partial D_{t}\partial Z_{t}}=0.
\end{align*}
This means that the marginal effect of the policy on the outcome variable
does not depend on the shock to the outcome variable or the unobserved
factors $Z_{t}$. Without this additional smoothness, Assumption \ref{Assumption: Structural Form}
implies that the effect of the policy on the outcome does not vary
with $Z_{t}$ or $u_{t}$.

Next, we assume that the simultaneous structural equations imply a
reduced-form for $D_{t}$ that is analogously additively separable
so that the effect of $e_{t}$ on $D_{t}$ is not influenced by $Z_{t}$
or $u_{t}$.
\begin{assumption}
\label{Assumption: Reduced-form}(Reduced-form separability) For all
$t\in\mathbf{P}$, $D_{t}=g_{D,e}\left(e_{t},\,t\right)+g_{D,u}\left(Z_{t},\,u_{t},\,t\right)$
for some functions $g_{D,e}$ and $g_{D,u}$.
\end{assumption}
Assumption \ref{Assumption: Reduced-form} is implied by Assumption
\ref{Assumption: Structural Form}, separability in $\varphi_{D}\left(Y_{t},\,Z_{t},\,e_{t},\,t\right)$
across $(Y_{t},\,Z_{t})$ and $e_{t}$ and standard invertibility
requirements on the system of equations. This holds trivially if
the true data-generating process is a system of linear simultaneous
equations, like \eqref{Eq.: Event study general model} under standard
rank conditions. In structural VARs, invertibility is a standard condition
that is imposed to obtain the reduced-form shocks from the structural
shocks {[}see \citet{plagborgmoller:2019} for a discussion about
invertibility of impulse responses{]}. We directly impose Assumption
\ref{Assumption: Reduced-form} on the reduced-form rather than imposing
invertibility requirements to avoid introducing further notation for
defining the partial inverse of a multivariate function. Assumptions
\ref{Assumption: Structural Form}-\ref{Assumption: Reduced-form}
are satisfied in the event study applications considered in empirical
work since the specification for the outcome and treatment variables
is typically a simultaneous linear equations model as in \eqref{Eq.: Event study general model}.
\begin{assumption}
\label{Assumption: Structural shocks}(Structural shocks) For all
$t\in\mathbf{P},$ $u_{t}$ and $e_{t}$ have zero mean and no serial
correlation, are mutually independent, and are each independent from
$Z_{t}$.
\end{assumption}
Under Assumption \ref{Assumption: Structural shocks}, $u_{t}$ and
$e_{t}$ are interpreted as structural shocks, i.e., primitive, unanticipated
impulses that are unforecastable and mutually uncorrelated. 

In the assumed specification \eqref{Eq: nonparametric model}, $Y_{t}$
and $D_{t}$ are determined simultaneously and so are endogenous.
Often endogeneity is overcome by using instrumental variables. This
leads to the identification of the local average treatment effect
via the instrumental variables estimand {[}cf. \citet{imbens/angrist:1994}{]}.
In contrast, the event study approach can identify causal effects
in the presence of endogeneity without the need to find instrumental
variables.\footnote{It is well-known that it is hard to find valid instruments in macroeconomics
applications. Considering Example \ref{Example: Monetary policy, NK and Kuttner},
it is difficult to find any instrument that would affect asset price
returns $\left(Y_{t}\right)$ without changing short-term interest
rates $\left(D_{t}\right)$ as any variable related to the macroeconomic
outlook would not satisfy this criterion. Neither would variables
related to corporate revenues and profits since they would likely
contain information about the economic outlook and be correlated with
interest rate changes.} The key idea is that in a narrow time window around a particular
event (or change in policy) the variation in the policy variable is
dominated by the variation in the policy shock. The effect of the
outcome variable on the policy variable is still present within the
time window but it is negligible relative to the effect of the policy
shock. Similarly, the effect of the omitted variables $Z_{t}$ on
$D_{t}$ is negligible relative to that of $e_{t}$. This idea is
formalized by the following assumption.
\begin{assumption}
\label{Assumption: Relative Exogeneity}(Relative exogeneity) For
all $t\in\mathbf{P}$,\\
(i) $\sigma_{e,t}^{2}=\mathrm{Var}(e_{t})\rightarrow\infty,$\\
(ii) $\mathrm{Var}(g_{D,e}\left(e_{t},\,t\right))$ is increasing
in $\sigma_{e,t}^{2},$ \\
(iii) $\mathbb{E}(\varphi_{Y,u}\left(Z_{t},\,u_{t},\,t\right)^{2})$
and $\mathbb{E}(g_{D,u}\left(Z_{t},\,u_{t},\,t\right)^{2})$ are finite.
\end{assumption}
Assumption \ref{Assumption: Relative Exogeneity} requires that within
the event window the policy shock has infinite variance (condition
(i,ii)) while the other variables have finite variance (condition
(iii)) and so the policy shock dominates the changes in the policy
variable in the  window. Assumption \ref{Assumption: Relative Exogeneity}(iii)
is implied by a correspondence between the boundness of the second
moments of $e_{t}$, $Z_{t}$ and $u_{t}$ with those of $D_{t}$
and $Y_{t}$. This easily holds for \eqref{Eq.: Event study general model}
since the second moments of $Y_{t}$ and $D_{t}$ depend on the second
moments of $e_{t}$, $Z_{t}$ and $u_{t}$ and on the moments of their
products. Altogether, Assumption \ref{Assumption: Relative Exogeneity}
implies that at periods immediately surrounding a policy announcement
(i.e., $t\in\mathbf{P}$) the policy shock $e_{t}$ dominates the
other shock $u_{t}$ and the omitted factors $Z_{t}$. The latter
involve shocks and factors that are not related to the announcement
and are also present in non-announcement periods. Hence, it is reasonable
to expect these variables to have finite variance. In contrast, the
policy news shock is much more pronounced in the announcement window
and occurs in a lumpy manner as it is completely unexpected. Consequently,
the market reacts leading to realized volatility and trading volume
to significantly decline before the  announcement and then jump at
the announcement {[}see, e.g., \citet{lucca:moench:2015} and \citet{hu/pan/wang/zhu:2022}{]}.
It is this lumpy manner in which a disproportionate amount of policy
news is revealed that can make the policy variable $D_{t}$ \textit{relatively
exogenous}. When this occurs we show that within the  event window:
(i) the reverse causality problem disappears (changes in the outcome
variable do not affect changes in the policy variable since the latter
are entirely driven by the policy shock);\footnote{The same conclusion holds when observable factors are included in
the analysis since these are typically low-frequency variables that
are dominated by the policy shock $e_{t}$.} (ii) the common unobserved factors $Z_{t}$ do not generate omitted
variables bias. Note that the endogeneity problem is overcome only
in the policy sample $\mathbf{P}$. In the control sample $\mathbf{C}$,
defined as the collection of all $t$ such that $t\notin\mathbf{P}$
(e.g., days with no FOMC announcement), the endogeneity  remains. 

Based on Assumption \ref{Assumption: Relative Exogeneity}, we establish
our identification results by taking the limits as
\begin{equation}
\sigma_{e,t}^{2}\rightarrow\infty,\label{Eq.: Relative exogeneity}
\end{equation}
 and refer to this limiting case as \emph{relative exogeneity}.\footnote{More precisely, relative exogeneity holds when the ratios between
the variance of the policy shock and the variances of the other variables
in the event window diverge. Thus, one may instead frame relative
exogeneity as the condition for which the variance of the policy shock
is finite and the variance of the background noise vanishes in the
event window in analogy with a continuous time jump-diffusion model
being sampled at high frequency, where the diffusion component corresponds
to background noise and the jump component corresponds to the policy
shock. See, e.g., \citet{bandi/nguyen:2003}. We show that the same
identification results hold under this latter condition in Appendix
\ref{Section Shrinking-Variance}.} The condition \eqref{Eq.: Relative exogeneity} makes clear one point
that has been overlooked by the empirical literature. Namely, it requires
that the ``large'' variance condition for the policy shock has to
hold for all $t\in\mathbf{P}$. For example, suppose that it is satisfied
only for a few announcements in the policy sample. Since the variance
cannot be negative, an estimate of the average variance computed in
the policy sample could still be very large. This could be misleadingly
interpreted as support for relative exogeneity. However, the endogeneity
in most of the policy sample would not be overcome and relative exogeneity
would fail. Hence, verifying that the sample variance of $D_{t}$
is large does not allow one to conclude that relative exogeneity holds.

\subsubsection{\label{Subsubsection: Potential Outcomes}Potential Outcomes Framework}

To conduct our identification analysis, we introduce the relevant
potential outcomes framework, which is useful for determining nonparametric
conditions under which the OLS event study estimands have a causal
interpretation. We define causal effects using the notion of potential
outcomes introduced by \citet{rubin:1974} and extended to time series
settings by \citet{angrist/kuersteiner:2011} and \citet{rambachan/shephard:2021}.
Potential outcomes are defined as the counterfactuals of $Y_{t}$
that would arise in response to a hypothetical value of the policy
variable $D_{t}$.
\begin{defn}
\label{Definition: Potential Outcome}The potential outcome, $Y_{t}\left(d\right)$,
is defined as the value taken by $Y_{t}$ if $D_{t}=d$. 
\end{defn}
We assume that $d\in\mathbf{D}$ for an appropriate set $\mathbf{D}$.
A potential outcome $Y_{t}\left(d\right)$ describes which value the
outcome would have taken at time $t$ under treatment value $d$.
The definition implies that the potential outcome $Y_{t}\left(d\right)$
does not depend on future treatments. \citet{rambachan/shephard:2021}
described this property as ``non-anticipating potential outcomes''.
The definition also implies that the potential outcome $Y_{t}\left(d\right)$
does not depend on past treatments (unless they are elements of $Z_{t}$).
This is realistic in the high-frequency event study setting for two
reasons. First, $Y_{t}$ typically measures a change in an asset price
or a survey forecast within a narrow window around a policy announcement.
Thus, market efficiency or rational expectations, respectively, imply
that market participants or professional forecasters use all public
information available at the start of the window and so $Y_{t}$ is
constructed by conditioning on that information. Second, for $t\in\mathbf{P}$
the latest past treatment is $D_{t-1}$ $(\{t-1\}\in\mathbf{C})$
which captures the surprise change in a policy within a 30-minute
window that does not involve any policy announcement and so the change
is very small or equal zero. The potential outcome should not be confused
with the outcome $Y_{t}=Y_{t}\left(D_{t}\right)$. Finally, note that
in terms of the structural function in \eqref{Eq: nonparametric model},
$Y_{t}(d)=\varphi_{Y}(d\,,Z_{t}\,,u_{t}\,,t)$.

The notation $Y_{t}\left(d\right)$ focuses on the effect of the current
treatment $d$ on the current outcome. In our context, the hypothesis
of no causal effects of the policy means that $Y_{t}\left(d\right)=Y_{t}\left(d'\right)$
for all $d,\,d'\in\mathbf{D}$. To analyze the causal effect of the
policy variable, it is useful to define the effect of a marginal change
in the policy variable on the potential outcome, the MCE. Define the
normalized variables $\widetilde{Y}_{t}\left(d\right)=\sigma_{D,t}^{-1}Y_{t}\left(d\right)$,
$\widetilde{D}_{t}=\sigma_{D,t}^{-1}D_{t}$ and $\widetilde{e}_{t}=\sigma_{D,t}^{-1}e_{t}$
for all $t$, where $\sigma_{D,t}^{2}=\mathrm{Var}(D_{t})$. We impose
two technical assumptions to enable MCEs to be well-defined under
relative exogeneity \eqref{Eq.: Relative exogeneity}. The first is
on the support of the normalized policy variable and the second is
on the smoothness of the normalized potential outcome process.
\begin{assumption}
\label{Assumption: Support of D}For all $t\in\mathbf{P}$, $\widetilde{D}_{t}\in\mathbf{D}=[\underline{d},\,\overline{d}]$
with $\underline{d}<\overline{d}$. 
\end{assumption}
\begin{assumption}
\label{Assumption: Differentiability of Potential Outcome}$($Differentiability$)$
For all $t\in\mathbf{P}$, $\widetilde{Y}_{t}\left(d\right)$ is continuously
differentiable in $d\in(\underline{d},\,\overline{d})$. 
\end{assumption}
Under these assumptions, the MCE of the time $t$ policy on the time
$t$ normalized potential outcome is defined as $\partial\widetilde{Y}_{t}\left(d\right)/\partial d$.
Assumptions \ref{Assumption: Support of D}-\ref{Assumption: Differentiability of Potential Outcome}
involve the normalized quantities instead of the actual quantities
since we analyze estimands and estimators under relative exogeneity
\eqref{Eq.: Relative exogeneity}, for which the support of $e_{t}$
and $D_{t}$ necessarily become unbounded. Following \citet{rambachan/shephard:2021},
to apply basic tools such as the fundamental theorem of calculus we
need the argument of the relevant function to have support on a closed
interval.

\subsection{\label{Subsection: Identification-Results}Identification Results}

The simple event study regression approach regresses the outcome variable
$Y_{t}$ on the policy variable $D_{t}$ at dates in the policy sample
$t\in\mathbf{P}$. Thus, the corresponding event study estimand at
time $t\in\mathbf{P}$ is the linear projection estimand: 
\begin{align}
\beta_{\mathrm{ES},t} & =\frac{\mathrm{Cov}\left(Y_{t},\,D_{t}\right)}{\mathrm{Var}\left(D_{t}\right)}.\label{Eq. (Event Study beta)}
\end{align}
We begin by establishing the value of this estimand in the general
nonparametric model \eqref{Eq: nonparametric model} without imposing
relative exogeneity (cf. Assumption \ref{Assumption: Relative Exogeneity})
and successively obtain corresponding results under relative exogeneity
and  then linear homogeneous treatment effects implied by \eqref{Eq.: Event study general model},
the linear model commonly assumed in practice. To establish the first
result we use the following assumption in place of Assumption \ref{Assumption: Relative Exogeneity}. 
\begin{assumption}
\label{Assumption: NO Exogeneity}For all $t\in\mathbf{P}$, $\mathbb{E}(g_{D,e}\left(e_{t},\,t\right)^{2}),$
$\mathbb{E}(\varphi_{Y,u}\left(Z_{t},\,u_{t},\,t\right)^{2})$ and
$\mathbb{E}(g_{D,u}\left(Z_{t},\,u_{t},\,t\right)^{2})$ are finite,
and $\mathbb{E}(g_{D,e}\left(e_{t},\,t\right)^{2})$ is decreasing
in $\sigma_{e,t}^{2}$. 
\end{assumption}
The event study estimand can be decomposed into a weighted average
of MCEs and a selection bias factor. 
\begin{thm}
\label{Theorem: WATE Event Study}Let Assumptions \ref{Assumption: Structural Form}-\ref{Assumption: Structural shocks}
and \ref{Assumption: Support of D}-\ref{Assumption: NO Exogeneity}
hold. Then for $t\in\mathbf{P}$, 
\begin{align*}
\beta_{\mathrm{ES},t}=\int_{\mathbf{D}}\frac{\partial\widetilde{Y}_{t}\left(d\right)}{\partial d}\mathbb{E}\left(H_{t}\left(d\right)\right)\mathrm{d}d+\Delta_{t},
\end{align*}
where $H_{t}\left(d\right)=\mathbf{1}\{d\leq\widetilde{D}_{t}\}(\widetilde{D}_{t}-\mathbb{E}(\widetilde{D}_{t}))$
with $\mathbb{E}(H_{t}\left(d\right))\geq0,$ and $\int_{\mathbf{D}}\mathbb{E}\left(H_{t}\left(d\right)\right)\mathrm{d}d=1$,
\[
\Delta_{t}=\mathbb{E}\left[\widetilde{Y}_{t}\left(\underline{d}\right)\left(\widetilde{D}_{t}-\mathbb{E}\left(\widetilde{D}_{t}\right)\right)\right]=\mathrm{Cov}\left(\widetilde{Y}_{t}\left(\underline{d}\right),\,\widetilde{D}_{t}\right),
\]
 and $\left|\Delta_{t}\right|$ is decreasing in $\sigma_{e,t}^{2}$. 
\end{thm}
The bias $\Delta_{t}$ depends on the covariance between the policy
variable and the potential outcome $Y_{t}(\underline{d})$. This bias
would be zero if the policy $D_{t}$ were randomly assigned, i.e.,
when there is no reverse causality and no omitted variables. Unfortunately,
this is quite unrealistic in practice. However in Theorem \ref{Theorem: WATE Event Study with Relative Exogeneity},
we show that as $\sigma_{e,t}^{2}$ grows large, this bias term disappears.
This implies that the event study estimand and the event study regression
estimates can have small enough bias to remain meaningful so long
as $\sigma_{e,t}^{2}$ is large enough.

The following theorem establishes that $\beta_{\mathrm{ES},t}$ identifies
a weighted average of MCEs of the time $t$ policy on the time $t$
outcome when relative exogeneity holds.
\begin{thm}
\label{Theorem: WATE Event Study with Relative Exogeneity}Let Assumptions
\ref{Assumption: Structural Form}-\ref{Assumption: Differentiability of Potential Outcome}
hold. Then for $t\in\mathbf{P}$, as $\sigma_{e,t}^{2}\rightarrow\infty$
\begin{align*}
\beta_{\mathrm{ES},t}\rightarrow\lim_{\sigma_{e,t}^{2}\rightarrow\infty}\int_{\mathbf{D}}\frac{\partial\widetilde{Y}_{t}\left(d\right)}{\partial d}\mathbb{E}\left(H_{t}\left(d\right)\right)\mathrm{d}d,
\end{align*}
 where $H_{t}\left(d\right)=\mathbf{1}\{d\leq\widetilde{D}_{t}\}(\widetilde{D}_{t}-\mathbb{E}(\widetilde{D}_{t}))$
with $\mathbb{E}(H_{t}\left(d\right))\geq0,$ $\int_{\mathbf{D}}\mathbb{E}\left(H_{t}\left(d\right)\right)\mathrm{d}d=1$. 
\end{thm}
For the weighted average of the MCEs in Theorem \ref{Theorem: WATE Event Study with Relative Exogeneity},
a higher weight 
\begin{align*}
\mathbb{E}\left(H_{t}\left(d\right)\right) & =\mathbb{E}\left(\widetilde{D}_{t}-\mathbb{E}(\widetilde{D}_{t})|\,d\leq\widetilde{D}_{t}\right)\times\mathbb{P}\left(d\leq\widetilde{D}_{t}\right)
\end{align*}
is not necessarily given to large values of $d$ because large values
of $d$ may be associated with small tail probabilities of the distribution
of $\widetilde{D}_{t}$, $\mathbb{P}(d\leq\widetilde{D}_{t})$. The
intuition on how $\beta_{\mathrm{ES},t}$ is able to recover a weighted
average of MCEs under relative exogeneity is as follows. Relative
exogeneity, in combination with Assumptions \ref{Assumption: Structural Form}-\ref{Assumption: Structural shocks},
imply that when appropriately normalized, the potential outcome $Y_{t}\left(d\right)$
and the policy variable $D_{t}$ behave as if they are uncorrelated.
Intuitively, the effect of the policy shock on $D_{t}$ is not influenced
by $Y_{t}$ (separability) and so changes in the policy are entirely
determined by changes in the policy shock (relative exogeneity). In
addition, as $\sigma_{e,t}^{2}\rightarrow\infty$ the relative bias
generated by the omitted variables $Z_{t}$ becomes negligible since
the correlation between $Z_{t}$ and $D_{t}$ is an order of magnitude
smaller than the variation in $D_{t}$ generated by $e_{t}$.

It is noteworthy that the event study estimand obtained from a standard
linear regression is able to recover a nonparametric causal effect
in this case. The event study empirical applications in the literature
most often assume a linear simultaneous equations model. Our assumptions
on the structural and reduced-form as well as the moment conditions
{[}cf. Assumptions \ref{Assumption: Structural Form}-\ref{Assumption: Relative Exogeneity}(ii,iii){]}
are easily satisfied in those contexts. Under the linear model \eqref{Eq.: Event study general model}
and relative exogeneity, Theorem \ref{Theorem: WATE Event Study with Relative Exogeneity}
implies that $\beta_{\mathrm{ES},t}\rightarrow\beta$ since $\partial\widetilde{Y}_{t}\left(d\right)/\partial d=\beta.$

The key identification condition is relative exogeneity, which also
relates to the size of the time window surrounding an announcement
or event. As the size of the window expands, it becomes less likely
that the policy shock dominates all other variables within the window.
Given that the narrow size of the window is not a sufficient condition
for identification, our results suggest that empirical work using
the high-frequency event study regression approach should be very
careful to isolate the surprise component of the policy news so that
the policy shock dominates the other variables in the event window.
Hence, our theoretical results support the concerns expressed recently
by \citeauthor{bauer/swanson:2022} (\citeyear{bauer/swanson:2021},
\citeyear{bauer/swanson:2022}) on the credibility of some high-frequency
event study estimates when the outcome variable is the Blue Chip forecast
revision that is based on a one-month window.

Note that the high-frequency event study method is different from
heteroskedasticity-based identification {[}cf. \citet{lewis:2020}
and \citet{rigobon:2003}{]}, sometimes referred to as Rigobon's method.
The latter also uses information from the control sample, which includes
windows (either 30-minute or 1-day windows) that do not bracket an
announcement or event. The main heteroskedasticity-based identification
condition is that the volatility of the policy shock is larger in
the event windows than in the control windows. This is different from
relative exogeneity for two reasons. First, the required increase
in volatility of the policy shock is not relative to the other variables
in the event window but relative to the policy shock in the control
windows. Second, the increase in volatility need not be infinite.
Thus, heteroskedasticity-based identification requires neither stronger
nor weaker conditions than the relative exogeneity condition. In addition,
Rigobon's estimator is an instrumental variables estimator and therefore
different from the OLS high-frequency event study estimator. Since
these two estimators correspond to different estimands, they identify
different causal effect in general.

\subsection{\label{Subsection: Relation to the Literature}Relation of the Theoretical
Results with the Empirical Literature}

There are essentially no formal identification results about high-frequency
event studies in the literature. Empirical works have often mentioned
that changes in the policy variable in the event windows are dominated
by the information about future monetary policy contained in the FOMC
announcements {[}see, e.g., \citet{nakamura/steinsson:2018}{]}. However,
precise conditions have not been provided. \citet{rigobon/sack:2004}
noted that the bias of the OLS event study estimator disappears in
a stylized linear model when relative exogeneity holds. However, this
result does not suffice to prove identification of a causal estimand.
Further, the model they analyzed is admittedly a ``clear oversimplification''.
As such, this stylized model does not enable practically-relevant
identification analysis or econometric results. Nevertheless, we
credit Rigobon and Sack's heuristic analysis as an early insight into
understanding the validity of the OLS event study estimator. \citet{gurkaynak/wright:2013}
surveyed the empirical literature and presented useful discussions.
They also argued that the event study regression works even when the
variance of the policy shock is not large relative to the variance
of the other variables because in a narrow window around an FOMC announcement
the policy news can depend on lagged changes in asset returns but
not on contemporaneous changes. That is, they wrote $D_{t}=\alpha Y_{t-j}+Z'_{t-j}\gamma_{2}+e_{t}$,
where $j\geq1$. However, as we explain in Example \ref{Example: Monetary policy, NK and Kuttner},
the choice of a narrow window is not sufficient for ruling out omitted
variables bias. Further, when $Y_{t}$ is the change in an asset price
and $D_{t}$ is, for example, the price change in the Federal funds
or eurodollar futures, assuming that $Y_{t}$ does not contemporaneously
affect $D_{t}$ constitutes a strong empirical restriction.

In recent work, \citeauthor{bauer/swanson:2022} (\citeyear{bauer/swanson:2021},
\citeyear{bauer/swanson:2022}), \citet{cieslak:2018} and \citet{mirandaagrippino/ricco:2021}
provided empirical evidence for some predictability of $D_{t}$ with
publicly available macroeconomic or financial market information that
predates the FOMC announcement. This implies that $D_{t}$ does not
correctly isolate the unexpected component of the policy surprise
and may be in fact simultaneously determined with the outcome variable
$Y_{t}$.\footnote{This simultaneity follows from the plausible correlation between the
publicly available macroeconomic and financial market information
prior to the announcement and the macroeconomic factors at the time
of the announcement that may affect $Y_{t}$.} \citet{bauer/swanson:2021} proposed to take the residuals from a
regression of those surprises on the economic and financial variables
that predate the announcements. So in \citet{bauer/swanson:2021}
$D_{t}$ is actually the orthogonalized monetary policy surprise rather
than the surprise itself. However, \citet{bauer/swanson:2021} showed
that the orthogonalized policy variables yield the same results as
the unadjusted policy variables when $Y_{t}$ is measured as the change
in a 30-minute window, as for the case of asset prices or Treasury
yields. This corroborates our result that some endogeneity of $D_{t}$
does not preclude the validity of the approach if the variance of
the unadjusted policy shocks is much larger than that of the other
variables in the system.

Evidence of nonlinearities is often documented in the empirical literature
{[}cf. \citet{bauer/swanson:2021}{]}. Our results establish the causal
meaning of the event study estimand when the relationship between
the outcome and the policy variable is potentially nonlinear as long
as the additive separability conditions hold.

As we explain more in detail below, the finance literature has recently
documented strong evidence of information leakage, informal communication
and informed trading around policy announcements {[}\citet{bernile/hu/tang:2016},
\citet{cieslak/morse/vissing-jorgensen:2019}, \citet{cieslak/mcmahon:2023},
\citet{cieslak/schrimpf:2019} and \citet{lucca:moench:2015}{]}.
For example, government agencies routinely allow pre-release access
to information to accredited news agencies under embargo agreements.
\citet{bernile/hu/tang:2016} found evidence consistent with informed
trading during embargoes of the FOMC announcements. They documented
significant abnormal order imbalances that are in the direction of
the subsequent policy surprises and showed that the information contained
in lockup-related trading activity (i.e., the window immediately before
the scheduled release) predicts the market reaction to the actual
FOMC announcement. Here the information leakage may arise from the
news media with pre-release access or from other FOMC insiders with
incentives to mimic such behavior.

\citet{lucca:moench:2015} documented large average excess returns
on U.S. equities in anticipation of monetary policy decisions made
at scheduled FOMC meetings.\footnote{More specifically, \citet{lucca:moench:2015} looked at unconditional
excess returns in the twenty-four hours before scheduled FOMC announcements
while \citet{hu/pan/wang/zhu:2022} looked at the overnight excess
returns before the same announcements.} This pre-FOMC drift is not found for fixed-income assets. They noted
that the pre-FOMC drift cannot be explained by changes in the public
information set in the twenty-four hours ahead of the FOMC meeting
as FOMC members refrain from providing monetary policy information
through speeches and interviews in the week before FOMC meetings.
They were more inclined to attribute the pre-FOMC drift to informational
frictions. This is empirically supported by \citet{cieslak/morse/vissing-jorgensen:2019}
who showed that large pre-FOMC drift is the result of news leakage
prior to the announcement of unexpectedly accommodating monetary policy.
They provided evidence of systematic informal communication, including
both outright leaks emerging in the media and private newsletters
and systematic preferential access to the Fed enjoyed by some private
financial institutions. The subsequent literature {[}see, e.g., \citet{hu/pan/wang/zhu:2022}{]}
found that other major U.S. macroeconomic news announcements give
rise to pre-announcement drifts in excess returns. Here the sources
of the leakage depend on the specific context. Another implication
of leakage, serial dependence, is also documented empirically {[}see,
e.g., \citet{bernile/hu/tang:2016}, \citet{cieslak/morse/vissing-jorgensen:2019}
and \citet{lucca:moench:2015}{]}. Overall, the literature contains
substantive evidence of information leakage and informal public communication.

As we discuss below, the leakage documented in the literature does
not necessarily imply that relative exogeneity is violated. Intuitively,
as long as the key news is revealed with the actual announcements,
the high-frequency event study is still characterized by the lumpy
manner with which a disproportionate amount of information is unveiled
to the public.\footnote{This argument applies to the effects of President Trump's tweets
that criticize the Federal Reserve on financial markets documented
by \citet{bianchi/kind/kung:2020}. They showed that those tweets
had a negative effect on the expected Fed funds rate with the magnitude
growing by horizon. If the tweet occurs in between the time the survey
is collected and the scheduled macroeconomic announcement, then $D_{t}$
does not isolate the expected component of the news correctly. This
challenges relative exogeneity. However, if the key macroeconomic
data information is revealed in the announcement, then $D_{t}$ is
primarily driven by the infinite variance policy shock $e_{t}$ and
so missing the effect of the tweets on the updates of the expectations
is negligible.}

With regards to the choice of the length of the event window, the
early literature commonly used a 1-day window while the more recent
literature recommended to use narrower windows with the goal of reducing
the background noise. In addition, it is common in the literature
to use the same window for both $Y_{t}$ and $D_{t}$. \citet{nakamura/steinsson:2018}
is an exception as they used a 1-day window for $Y_{t}$ and a 30-minute
window for $D_{t}$.

Although a narrower window is associated with a smaller probability
of including news about events other than the policy announcement,
it also has disadvantages relative to using a longer window for $Y_{t}$.
First, with some information leakage, asset prices $Y_{t}$ may respond
before the announcement is actually made. Second, with learning or
sluggish market adjustments, asset prices may take some time to incorporate
the news and so changes in $Y_{t}$ may occur also in the hours after
the announcement.\footnote{Note that the complications arising from the effect of information
leakage and learning on $Y_{t}$ are different from those we discuss
in Section \ref{Subsection: Robustness to Leakage} for proper construction
of $D_{t}$. Intuitively, the impact of information leakage on $Y_{t}$
depends on the potential causal effect of the policy news on $Y_{t}$.
In the linear model \eqref{Eq.: Event study general model} this is
captured by $\beta$. If $\beta=0$, then information leakage has
no effect on $Y_{t}$ while it does complicate the proper construction
of the surprise $D_{t}$ irrespective of the value of $\beta$.} Our analysis shows that the key identification condition (relative
exogeneity) does not explicitly refer to the size of the window, it
only requires that, whatever window length is chosen, the policy shock
dominates any other shock within that window.

Theorem \ref{Theorem: WATE Event Study} allows us to provide a formal
explanation for a recent debate on some puzzling event study regression
results documented in the literature. It has been shown that regressions
of private-sector macroeconomic forecast revisions on monetary policy
surprises often produce coefficients with signs opposite to those
of standard macroeconomic models.\footnote{For example, a surprise monetary policy tightening is associated with
a statistically significant upward revision in the Blue Chip consensus
forecasts for real GDP growth. This is inconsistent with the standard
macroeconomic view that a monetary policy tightening should cause
future GDP to fall.} \citet{campbell/evans/fisher/justiniano:2012}, \citet{nakamura/steinsson:2018}
and \citet{romer/romer:2000} argued in favor of the ``Fed information
effect'' for which these puzzling results are due to monetary policy
surprises revealing private information held by the Federal Reserve.
\citet{bauer/swanson:2021} challenged these views, arguing that these
event study estimates suffer from omitted variables bias. In Section
\ref{Section: Empirical Procedure Identification}, we analyze this
puzzle in detail and show that in the event study regressions involving
the Blue Chip forecasts, it is unlikely that relative exogeneity holds
because the forecast revisions are evaluated at a much lower frequency
than the policy variable. Intuitively, while the policy surprise
$D_{t}$ is constructed as a 30-minute change, $Y_{t}$ is the one-month
change in the Blue Chip forecasts and so the latter likely has a large
variance relative to the former as it aggregates all news and factors
that are relevant over the month. Thus, it becomes important to control
for macroeconomic and financial variables that predate the announcements
in this context. This explains why using the orthogonalized shocks
indeed allowed \citet{bauer/swanson:2021} to overturn the documented
puzzling estimates.  In contrast, when $Y_{t}$ is a 30-minute or
1-day change in an asset price or Treasury yield, the variance of
$D_{t}$ is much larger in relative terms and can eliminate the endogeneity
arising from omitted variables that predate the FOMC announcement.

\subsection{\label{Subsection: Robustness to Leakage}Robustness to Information
Leakage}

It is interesting to analyze when relative exogeneity may or may not
fail due to information leakage about the policy news or some market
anticipation of the policy change. Leakage is highly relevant in applications
as documented recently in the finance literature discussed above.
Leakage has the following empirical features. First, it leads to market
anticipation of the news that is to be revealed at the event time.
This may reduce the variance of $e_{t}$ substantially for $t\in\mathbf{P}$,
possibly making the policy shock on the same order of magnitude as
the other random variables in the system, failing to dominate them
in the event window. Second, leakage can be associated with learning
and sluggish market adjustments since the news may initially reach
a small number of market participants and, through their reactions,
may slowly spread into the market. This is likely to generate serial
dependence in the policy shock over the hours or days prior to the
scheduled announcement. We introduce these features into the model
and discuss how this can alter the identification results. The introduction
of leakage into the model depends on the specific context of the event
under consideration. Here we focus on FOMC announcements.

Consider two successive FOMC announcement dates $T_{0},\,T_{1}\in\mathbf{P}$
where $T_{0}<T_{1}$. There are usually six weeks between any two
successive FOMC announcements.\footnote{To be precise, there are eight regularly scheduled FOMC announcements
per year that are spaced roughly six to eight weeks apart.} Consider splitting the regular trading hours within these six weeks
into non-overlapping 30-minute windows indexed by $t=T_{0}+1,\,\ldots,\,T_{1}$.
The following model for the policy shock is useful for describing
some of the empirical features of leakage: 
\begin{align}
e_{t} & =\phi_{t}\sigma_{t}v_{t}+\left(1-\phi_{t}\right)\left(\vartheta_{1}v_{t-1}+\ldots+\vartheta_{q}v_{t-q}\right),\label{Eq. (e_t model with leakage)}
\end{align}
where $v_{t}$ is a white noise process with zero mean and unit variance,
$|\vartheta_{j}|<\infty$ for all $j=1,\ldots,\,q$ and $q>0$ is
a finite integer. Assume that $\sigma_{t}^{2}\rightarrow\infty$ if
$t\in\mathbf{P}$ and $|\sigma_{t}^{2}|<\infty$ if $t\in\mathbf{C}$.
The parameter $\phi_{t}$ is the leakage parameter: 
\begin{align*}
\phi_{t} & =\begin{cases}
1, & \text{no leakage}\\
\sigma_{t}^{-1}, & \text{leakage}
\end{cases}.
\end{align*}
When $\phi_{T_{1}}=1$ there is no leakage as we have $e_{T_{1}}=\sigma_{T_{1}}v_{T_{1}}$,
$\mathbb{E}\left(e_{T_{1}}\right)=0$, $\mathrm{Var}(e_{T_{1}})=\sigma_{e,T_{1}}^{2}=\sigma_{T_{1}}^{2}$
and $\mathbb{E}\left(e_{t}e_{t-j}\right)=0$ for $j>0.$ Then, relative
exogeneity holds ($\sigma_{e,T_{1}}^{2}\rightarrow\infty$ since $T_{1}\in\mathbf{P}$)
and the identification results of Section \ref{Subsection: Identification-Results}
apply. On the other hand, when $\phi_{T_{1}}=\sigma_{T_{1}}^{-1}$
there is leakage and relative exogeneity fails. To see this, note
that 
\[
e_{T_{1}}=v_{T_{1}}+(1-\sigma_{T_{1}}^{-1})\left(\vartheta_{1}v_{T_{1}-1}+\ldots+\vartheta_{q}v_{T_{1}-q}\right),
\]
$\mathbb{E}\left(e_{T_{1}}\right)=0$, $\mathrm{Var}(e_{T_{1}})<\infty$
since $1-\sigma_{T_{1}}^{-1}\rightarrow1$, $|\vartheta_{j}|<\infty$
for all $j=1,\ldots,\,q$, and $\mathbb{E}\left(e_{t}e_{t-j}\right)\neq0$
for $j=1,\ldots,\,q$ and $\mathbb{E}\left(e_{t}e_{t-j}\right)=0$
for $j>q.$ Thus, when $\phi_{T_{1}}=\sigma_{T_{1}}^{-1}$ the variance
of the policy shock $e_{T_{1}}$ is not an order of magnitude larger
than the variances of $Z_{t}$ and $u_{t}.$ Intuitively, the market
has anticipated the content of the FOMC announcement. Further, $e_{T_{1}}$
exhibits serial dependence up to $q$ lags. This captures the idea
that information leakage is associated with learning and sluggish
market adjustments so that the information content of the announcement
can be predicted.\footnote{We use a moving-average specification for $e_{t}$ here because the
dependence that stems from leakage is limited to a few periods prior
to the FOMC announcement {[}see, e.g., \citet{lucca:moench:2015}{]}.} The model also implies that the leakage cannot start in the $q$
periods following an FOMC announcement, i.e., $\phi_{t}=1$ for $t=T_{0}+1,\ldots,\,T_{0}+q$.
Otherwise, the information leakage would be correlated with the news
from the previous announcement, contradicting the idea behind leakage.

The model \eqref{Eq. (e_t model with leakage)} suggests several testable
empirical implications. Relative exogeneity implies that the variance
of the policy variable is unbounded at each announcement window. Statistically,
this corresponds to a jump in $D_{t}$ at the time of the announcement.
Preliminary inspection of the high-frequency time series data on $D_{t}$
can be useful. A formal test involves testing for infinite variance
or testing for jumps {[}see, e.g., \citet{trapani:2016} and \citet{li/todorov/tauchen:2017}{]}.
Apart from the assumptions involved, a limitation of these tests for
our purposes is that they do not provide information on whether a
particular application may be characterized by a value of $\sigma_{e,t}^{2}$
that, although possibly finite, is relatively large enough to imply
low bias in the event study estimand (see Theorem \ref{Theorem: WATE Event Study}).
We introduce a simple procedure in Section \ref{Section: Empirical Procedure Identification}
to diagnose whether relative exogeneity is ``close enough'' to holding
that the event study estimator should be expected to perform well
in practice.

A second implication of leakage is the serial dependence in $e_{t}$
prior to the announcement. Unfortunately, $e_{t}$ is not observable
and is not recoverable in the presence of endogeneity. However, serial
dependence in $e_{t}$ implies serial dependence in $D_{t}$, which
is observed. A test for leakage can thus be obtained from testing
for autocorrelation in $D_{t}$ in sub-samples close to the announcement.

If the extent of the information leakage is small, then it will lead
only to partial market anticipation and will not prevent the news
from coming out in a lumpy manner at the release time. Thus, relative
exogeneity can hold in this case. However, $Y_{t}$ will not capture
the overall effect of the policy news as asset prices respond also
during the lockup window (i.e., the window immediately before the
scheduled release), leading to attenuation bias in the event study
estimator. One way to address this issue would be to take $Y_{t}$
to be the change in the relevant asset price in a wider window (e.g.,
a 1-hour or 1-day window) around the FOMC announcement than in the
30-minute window used for $D_{t}$.

Finally, information leakage could also have negative consequences
for obtaining good measures of market expectations about the policy
news thereby making it difficult to accurately construct the surprise
component of the policy action $D_{t}$. Failure to isolate the expected
component of the news leads to attenuation bias as asset prices have
already responded to the expected part of the policy news. Whether
this attenuation bias is important or not depends on how large the
information leakage is relative to the unexpected part of the policy
news that is driven by $e_{t}$ in the event window. When the leakage
is small, one does not need to extend the event window to bracket
the pre-release embargo when computing $D_{t}$ since the policy shock
during the 30-minute window is still an order of magnitude larger
than the other variables.

\section{\label{Section: Properties-of-the OLS ES}Properties of the OLS Event
Study Estimator}

In this section we establish the asymptotic properties of the OLS
event study estimator. Theorem \ref{Theorem: WATE Event Study with Relative Exogeneity}
shows that a weighted average of MCEs can be identified by the ratio
of the covariance between $Y_{t}$ and $D_{t}$ and the variance of
$D_{t}$. This is exactly what the OLS event study estimator estimates
under homogeneous treatment effects and covariance-stationarity of
the normalized processes.\footnote{Technically speaking, $D_{t}$ and $Y_{t}$ cannot be said to be covariance-stationary
because their second moments may diverge under relative exogeneity.
In contrast, it is meaningful to say that the normalized processes
are covariance-stationary.} However, when $\beta_{\mathrm{ES},t}$ varies with $t$, it is infeasible
to estimate each $\beta_{\mathrm{ES},t}$ separately. In this section,
we seek to obtain the limiting behavior of the OLS event study estimator
generally, without imposing homogeneous treatment effects or covariance-stationarity.
We show that the OLS event study estimator estimates a time-average
of the $\beta_{\mathrm{ES},t}$'s under relative exogeneity. Let $T_{P}$
denote the number of observations in $\mathbf{P}$. The event study
estimator is defined as 
\begin{align*}
\widehat{\beta}_{\mathrm{ES}} & =\frac{\sum_{t=1}^{T_{P}}\left(D_{t}-\overline{D}\right)\left(Y_{t}-\overline{Y}\right)}{\sum_{t=1}^{T_{P}}\left(D_{t}-\overline{D}\right)^{2}},
\end{align*}
where $\overline{D}=T_{P}^{-1}\sum_{t=1}^{T_{P}}D_{t}$ and $\overline{Y}=T_{P}^{-1}\sum_{t=1}^{T_{P}}Y_{t}$,
and an intercept is added to the regression.

Let $\sigma_{D}^{2}=\lim_{T_{P}\rightarrow\infty}T_{P}^{-1}\sum_{t=1}^{T_{P}}\mathrm{Var}\left(D_{t}\right)$,
$D_{t}^{*}=\sigma_{D}^{-1}D_{t}$, $Y_{t}^{*}=\sigma_{D}^{-1}Y_{t}$
and $Y_{t}^{*}\left(d\right)=\sigma_{D}^{-1}Y_{t}\left(d\right)$
for all $t\in\mathbf{P}.$ We make the following assumption in order
to study the asymptotic properties of $\widehat{\beta}_{\mathrm{ES}}$.
\begin{assumption}
\label{Assumption LLN}As $T_{P}\rightarrow\infty$ we have\\
 (i) $T_{P}^{-1}\sum_{t=1}^{T_{P}}(D_{t}^{*}-\overline{D}^{*})(Y_{t}^{*}-\overline{Y}^{*})\overset{\mathbb{P}}{\rightarrow}\int_{0}^{1}c(D^{*},\,Y^{*},\,s)\mathrm{d}s,$\\
 (ii) $T_{P}^{-1}\sum_{t=1}^{T_{P}}(D_{t}^{*}-\overline{D}^{*})^{2}\overset{\mathbb{P}}{\rightarrow}1,$\\
 where $\overline{D}^{*}=T_{P}^{-1}\sum_{t=1}^{T_{P}}D_{t}^{*},$
$\overline{Y}^{*}=T_{P}^{-1}\sum_{t=1}^{T_{P}}Y_{t}^{*},$ and $c(D^{*},\,Y^{*},\,s)=\lim_{T_{P}\rightarrow\infty}\mathrm{Cov}(D_{\left\lfloor T_{P}s\right\rfloor }^{*},\,Y_{\left\lfloor T_{P}s\right\rfloor }^{*})$.
\end{assumption}
\noindent Assumption \ref{Assumption LLN} requires that a law of
large numbers holds in an infill asymptotic embedding where the observations
originally defined on the time span $t=1,\ldots,\,T_{P}$ are mapped
into the unit interval $\left[0,\,1\right]$ through $s=t/T_{P}$.
We refer to the index $s\in\left[0,\,1\right]$ as the rescaled time
index. This is a mild assumption. It allows the observations to be
heterogeneous, i.e., to have time-varying moments. If one assumes
covariance-stationarity, then $c\left(D^{*},\,Y^{*},\,s\right)=c\left(D^{*},\,Y^{*}\right)$
and the infill asymptotic embedding is no longer required. Additionally,
$D_{t}^{*}=\widetilde{D}_{t}$, $Y_{t}^{*}=\widetilde{Y}_{t}$ and
$c(D^{*},\,Y^{*})=c(\widetilde{D},\,\widetilde{Y})$ since $\sigma_{D,t}^{2}=\sigma_{D}^{2}$
for all $t\in\mathbf{P}$ by covariance-stationarity of the normalized
processes. 
\begin{thm}
\label{Theorem: OLS ES consistency +  delta}Let Assumptions \ref{Assumption: Structural Form}-\ref{Assumption: Structural shocks}
and \ref{Assumption: Support of D}-\ref{Assumption LLN} hold. As
$T_{P}\rightarrow\infty$ we have $\widehat{\beta}_{\mathrm{ES}}\overset{\mathbb{P}}{\rightarrow}\beta_{\mathrm{ES}}$
where 
\begin{align*}
\beta_{\mathrm{ES}} & =\lim_{T_{P}\rightarrow\infty}\int_{0}^{1}\int_{\mathbf{D}}\frac{\partial Y_{\left\lfloor T_{P}s\right\rfloor }^{*}\left(d\right)}{\partial d}h\left(d,\,s\right)\mathrm{d}d\mathrm{d}s+\Delta,\qquad\qquad\mathrm{and}\\
h\left(d,\,s\right) & =\lim_{T_{P}\rightarrow\infty}\mathbb{E}\left(\mathbf{1}\{d\leq D_{\left\lfloor T_{P}s\right\rfloor }^{*}\}(D_{\left\lfloor T_{P}s\right\rfloor }^{*}-\mathbb{E}(D_{\left\lfloor T_{P}s\right\rfloor }^{*}))\right)
\end{align*}
with $h\left(d,\,s\right)\geq0$, $\int_{\mathbf{D}}h\left(d,\,s\right)\mathrm{d}d=1$
and $|\Delta|=|\lim_{T_{P}\rightarrow\infty}\int_{0}^{1}\mathbb{E}(Y_{\left\lfloor T_{P}s\right\rfloor }^{*}(\underline{d})(D_{\left\lfloor T_{P}s\right\rfloor }^{*}-\mathbb{E}(D_{\left\lfloor T_{P}s\right\rfloor }^{*})))ds|$
is decreasing in $\lim_{T_{P}\rightarrow\infty}\sigma_{e,\left\lfloor T_{P}s\right\rfloor }^{2}$
for all $s\in\left[0,\,1\right]$. 
\end{thm}
\noindent Theorem \ref{Theorem: OLS ES consistency +  delta} shows
that the OLS event study estimator is consistent for a weighted average
of standardized MCEs, plus the bias term $\Delta$. The former is
characterized by two types of averaging. First, there is averaging
over time for a given treatment $d$. Second, there is averaging over
different treatments for a given rescaled time $s.$ These treatment
effects are said to be standardized because they involve the standardized
outcome $Y_{\left\lfloor T_{p}s\right\rfloor }^{*}$ rather than the
original outcome $Y_{\left\lfloor T_{p}s\right\rfloor }$.

When $Y_{t}$ is the change in an asset price, it is interesting to
note that in the special case for which relative exogeneity fails,
$Z_{t}$ is absent and $u_{t}$ is driven solely by financial microstructure
noise, the OLS event study estimator suffers from attenuation bias
so that it can be used to bound a true causal effect from below. Microstructure
noise typically appears in ultra high-frequency data (e.g., 5 minutes
and less). Since the common size of the window in event study regressions
is 20 or 30 minutes, microstructure noise may be less relevant in
this context, though its presence ultimately depends on the liquidity
of the asset under consideration.

The following theorem establishes the consistency of the OLS event
study estimator under relative exogeneity (cf. Assumption \ref{Assumption: Relative Exogeneity}).
Moreover, since $|\Delta|$ is decreasing in $\sigma_{e,t}^{2}$,
the large-sample bias of $\widehat{\beta}_{\mathrm{ES}}$ for estimating
the weighted average of standardized MCEs is small when $\sigma_{e,t}^{2}$
is large. For technical reasons inherent to the proof, to establish
this result we take the limits as $T_{P}\rightarrow\infty$ and $\min_{t\in\mathbf{P}}\sigma_{e,t}^{2}\rightarrow\infty$
sequentially. 
\begin{thm}
\label{Theorem: OLS consistency}Let Assumptions \ref{Assumption: Structural Form}-\ref{Assumption: Differentiability of Potential Outcome}
and \ref{Assumption LLN} hold. Then as $T_{P}\rightarrow\infty$,
then $\min_{t\in\mathbf{P}}\sigma_{e,t}^{2}\rightarrow\infty$, 
\begin{align}
\widehat{\beta}_{\mathrm{ES}}\overset{\mathbb{P}}{\rightarrow}\lim_{\min_{t\in\mathbf{P}}\sigma_{e,t}^{2}\rightarrow\infty}\lim_{T_{P}\rightarrow\infty}\int_{0}^{1}\int_{\mathbf{D}}\frac{\partial Y_{\left\lfloor T_{P}s\right\rfloor }^{*}\left(d\right)}{\partial d}h\left(d,\,s\right)\mathrm{d}d\mathrm{d}s.\label{Eq. (OLS ES Consistency)}
\end{align}
\end{thm}
Under covariance-stationarity the event study estimand on the right-hand
side of \eqref{Eq. (OLS ES Consistency)} reduces to the estimand
in Theorem \ref{Theorem: WATE Event Study with Relative Exogeneity}:
\begin{align*}
\lim_{\min_{t\in\mathbf{P}}\sigma_{e,t}^{2}\rightarrow\infty}\lim_{T_{P}\rightarrow\infty}\int_{\mathbf{D}}h\left(d\right)\int_{0}^{1}\frac{\partial\widetilde{Y}_{\left\lfloor T_{P}s\right\rfloor }\left(d\right)}{\partial d}\mathrm{d}s\mathrm{d}d=\lim_{\sigma_{e,t}^{2}\rightarrow\infty}\int_{\mathbf{D}}\frac{\partial\widetilde{Y}_{t}\left(d\right)}{\partial d}\mathbb{E}\left(H_{t}\left(d\right)\right)\mathrm{d}d,
\end{align*}
since $h\left(d,\,s\right)=\mathbb{E}\left(H_{t}\left(d\right)\right)$,
$Y_{t}^{*}=\widetilde{Y}_{t}$ and $\partial\widetilde{Y}_{t}\left(d\right)/\partial d$
is invariant to $t$ by stationarity.

Finally, we analyze the asymptotic distribution of the event study
estimator. 
\begin{assumption}
\label{Assumption CLT}Let $\varepsilon_{t}=Y_{t}^{*}-\overline{Y}^{*}-\beta_{\mathrm{ES}}(D_{t}^{*}-\overline{D}^{*})$.
As $T_{P}\rightarrow\infty$ we have 
\begin{align*}
\frac{\sigma_{D}}{\sqrt{T_{P}}}\sum_{t=1}^{T_{P}}\left(D_{t}^{*}-\overline{D}^{*}\right)\varepsilon_{t}\overset{d}{\rightarrow} & \mathscr{N}\left(0,\,J\right),
\end{align*}
where 
\begin{align*}
J & =\int_{0}^{1}J\left(s\right)\mathrm{d}s,\qquad\mathrm{and}\qquad J\left(s\right)=\lim_{T_{P}\rightarrow\infty}\mathbb{E}\left[\varepsilon_{\left\lfloor T_{P}s\right\rfloor }^{2}\left(D_{\left\lfloor T_{P}s\right\rfloor }^{*}-\mathbb{E}(D_{\left\lfloor T_{P}s\right\rfloor }^{*})\right)^{2}\right].
\end{align*}
\end{assumption}
Assumption \ref{Assumption CLT} requires a central limit theorem
to hold in an infill asymptotic embedding. The asymptotic variance
$J$ allows for heteroskedasticity but not serial correlation. This
is implied by Assumption \ref{Assumption: Structural shocks} and
no serial correlation in $Z_{t}$. The assumption of no serial correlation
in high-frequency event studies is standard since the observations
are in first-differences and the original series are often asset prices.
When there is serial correlation in either $u_{t}$, $e_{t}$ or $Z_{t}$
the asymptotic variance in the assumption should be modified to 
\begin{align*}
J & =\lim_{T_{P}\rightarrow\infty}\frac{1}{T_{P}}\sum_{t=1}^{T_{P}}\sum_{l=1}^{T_{P}}\mathbb{E}\left[\varepsilon_{t}\varepsilon_{l}\left(D_{t}^{*}-\mathbb{E}(D_{t}^{*})\right)\left(D_{l}^{*}-\mathbb{E}(D_{l}^{*})\right)\right]\\
 & =\int_{0}^{1}\int_{0}^{1}\lim_{T_{P}\rightarrow\infty}\mathbb{E}\left[\varepsilon_{\left\lfloor T_{P}s\right\rfloor }\varepsilon_{\left\lfloor T_{P}r\right\rfloor }\left(D_{\left\lfloor T_{P}s\right\rfloor }^{*}-\mathbb{E}(D_{\left\lfloor T_{P}s\right\rfloor }^{*})\right)\left(D_{\left\lfloor T_{P}r\right\rfloor }^{*}-\mathbb{E}(D_{\left\lfloor T_{P}r\right\rfloor }^{*})\right)\right]\mathrm{d}s\mathrm{d}r.
\end{align*}

\begin{thm}
\label{Theorem: CLT}Let Assumptions \ref{Assumption: Structural Form}-\ref{Assumption: Structural shocks}
and \ref{Assumption: Support of D}-\ref{Assumption CLT} hold. As
$T_{P}\rightarrow\infty$ we have 
\begin{align*}
\sqrt{T_{P}\sigma_{D}^{2}}\left(\widehat{\beta}_{\mathrm{ES}}-\beta_{\mathrm{ES}}\right) & \overset{d}{\rightarrow}\mathscr{N}\left(0,\,J\right).
\end{align*}
\end{thm}
Theorem \ref{Theorem: CLT} shows that heteroskedasticity-robust standard
errors suffice for valid inference when there is no serial correlation
in $u_{t}$, $e_{t}$ and $Z_{t}$. Under covariance-stationarity
$J\left(s\right)$ does not depend on $s,$ and so $J=\mathbb{E}[\varepsilon_{t}^{2}\left(D_{t}^{*}-\mathbb{E}(D_{t}^{*})\right)^{2}]$.
However, heteroskedasticity-robust standard errors are still required
as $\varepsilon_{t}$ may not be homoskedastic.

Theorem \ref{Theorem: CLT} establishes the asymptotic normality of
$\hat{\beta}_{\mathrm{ES}}$ as an estimator for the biased quantity
$\beta_{\mathrm{ES}}$. To conduct unbiased inference on the weighted
average of standardized MCEs in \eqref{Eq. (OLS ES Consistency)}
using $\widehat{\beta}_{\mathrm{ES}}$, we require a strengthening
of relative exogeneity as formalized in the following corollary. 
\begin{cor}
\label{Corollary: unbiased OLS anorm}Let Assumptions \ref{Assumption: Structural Form}-\ref{Assumption: Differentiability of Potential Outcome}
and \ref{Assumption LLN}-\ref{Assumption CLT} hold. Then if $\sqrt{T_{P}}/\sigma_{D}\rightarrow c\in[0,\infty)$
as $T_{P}\rightarrow\infty$, 
\begin{align*}
\sqrt{T_{P}\sigma_{D}^{2}}\left(\widehat{\beta}_{\mathrm{ES}}-\left(\beta_{\mathrm{ES}}-\Delta(T_{P})\right)\right) & \overset{d}{\rightarrow}\mathscr{N}\left(\mathrm{aBias}\left(\widehat{\beta}_{\mathrm{ES}}\right),\,J\right)
\end{align*}
as $T_{P}\rightarrow\infty$, where $\Delta(T_{P})\equiv\int_{0}^{1}\mathbb{E}[Y_{\lfloor T_{P}s\rfloor}^{*}(\underline{d})(D_{\lfloor T_{P}s\rfloor}^{*}-\mathbb{E}(D_{\lfloor T_{P}s\rfloor}^{*}))]\mathrm{d}s\rightarrow\Delta$
and
\begin{align*}
\mathrm{aBias}\left(\widehat{\beta}_{\mathrm{ES}}\right) & =c\cdot\lim_{T_{P}\rightarrow\infty}\int_{0}^{1}\mathbb{E}\left[\varphi_{Y,u}\left(Z_{\lfloor T_{P}s\rfloor},\,u_{\lfloor T_{P}s\rfloor},\,\left\lfloor T_{P}s\right\rfloor \right)\right.\\
 & \quad\times\left.\left(g_{D,u}\left(Z_{\lfloor T_{P}s\rfloor},\,u_{\lfloor T_{P}s\rfloor},\,\left\lfloor T_{P}s\right\rfloor \right)-\mathbb{E}\left(g_{D,u}\left(Z_{\lfloor T_{P}s\rfloor},\,u_{\lfloor T_{P}s\rfloor},\,\left\lfloor T_{P}s\right\rfloor \right)\right)\right)\right]\mathrm{d}s.
\end{align*}
\end{cor}
Clearly, if relative exogeneity is ``strong enough'' in the sense
that $c=0$, $\mathrm{aBias}(\widehat{\beta}_{\mathrm{ES}})=0$ and
$\widehat{\beta}_{\mathrm{ES}}$ is asymptotically unbiased. In addition,
$\widehat{\beta}_{\mathrm{ES}}$ is super-consistent, with a faster
than standard $\sqrt{T_{P}}$ rate of convergence since $\sigma_{D}\rightarrow\infty$
under relative exogeneity. Under the conditions of the corollary,
it is also possible to bias-correct $\widehat{\beta}_{\mathrm{ES}}$
in case $c\neq0$ under the ``sharp null'' of zero MCEs at all time
periods, providing a further inference refinement. Specifically, if
$Y_{t}(d)$ does not depend upon $d$, Assumption \ref{Assumption: Structural Form}
implies $Y_{t}=\varphi_{Y,u}(Z_{t},\,u_{t},\,t)$ so that 
\[
\mathbb{E}\left[Y_{t}\left(D_{t}-\mathbb{E}(D_{t})\right)\right]=\mathbb{E}\left[\varphi_{Y,u}(Z_{t},\,u_{t},\,t)\left(g_{D,u}\left(Z_{t},\,u_{t},\,t\right)-\mathbb{E}\left(g_{D,u}(Z_{t},\,u_{t},\,t)\right)\right)\right],
\]
by Assumptions \ref{Assumption: Reduced-form}-\ref{Assumption: Structural shocks}.
Therefore, $\mathrm{aBias}(\widehat{\beta}_{\mathrm{ES}})$ can be
consistently estimated by 
\[
\sqrt{\frac{T_{P}}{T_{P}^{-1}\sum_{t=1}^{T_{P}}\left(D_{t}-\overline{D}\right)^{2}}}T_{P}^{-1}\sum_{t=1}^{T_{P}}Y_{t}\left(D_{t}-\overline{D}\right)
\]
under Assumption \ref{Assumption LLN}. In the following section,
we discuss bias-aware inference that does not rely upon the imposition
of the ``sharp null''.

\section{\label{Section: Bounds on bias, coverage, bias-aware}Bias Bounds,
Worst-Case Coverage and Bias-Aware Inference}

The bias in the high-frequency event study estimator vanishes in the
limit under relative exogeneity. We consider methods to quantify how
this bias can affect the worst-case properties of the estimator using
a bound on the bias, and we propose a bias-aware critical value that
accounts for the potential asymptotic bias of the estimator in addition
to its variance.

\subsection{Worst-Case Bias}

We begin with deriving the worst-case (unscaled) bias $\Delta$ from
Theorem \ref{Theorem: OLS ES consistency +  delta}. The following
proposition produces a useful expression for this.
\begin{prop}
\label{Proposition: Worst-case bias}Under Assumptions \ref{Assumption: Structural Form}-\ref{Assumption: Structural shocks}
and \ref{Assumption: Support of D}-\ref{Assumption LLN}, 
\begin{align}
\Delta & =\frac{1}{\sigma_{D}^{2}}\lim_{T_{P}\rightarrow\infty}\int_{0}^{1}\sqrt{\mathrm{Var}\left(\overline{\varphi}_{Y,u}\left(s\right)\right)}\sqrt{\mathrm{Var}\left(\overline{g}_{D,u}\left(s\right)\right)}\rho_{Zu,T_{P}}\left(s\right)ds,\label{Eq. Bias expression}
\end{align}
 where $\overline{\varphi}_{Y,u}\left(s\right)=\varphi_{Y,u}(Z_{\left\lfloor T_{P}s\right\rfloor },\,u_{\left\lfloor T_{P}s\right\rfloor },\,\left\lfloor T_{P}s\right\rfloor )$,
$\overline{g}_{D,u}\left(s\right)=g_{D,u}(Z_{\left\lfloor T_{P}s\right\rfloor },\,u_{\left\lfloor T_{P}s\right\rfloor },\,\left\lfloor T_{P}s\right\rfloor )$
and $\rho_{Zu,T_{P}}\left(s\right)=\mathrm{Corr}(\varphi_{Y,u}(Z_{\left\lfloor T_{P}s\right\rfloor },\,u_{\left\lfloor T_{P}s\right\rfloor },\,\left\lfloor T_{P}s\right\rfloor ),\,g_{D,u}(Z_{\left\lfloor T_{P}s\right\rfloor },\,u_{\left\lfloor T_{P}s\right\rfloor },\,\left\lfloor T_{P}s\right\rfloor ))$
is the correlation coefficient between the two terms in the parentheses.
\end{prop}
If the two variances on right-hand side of \eqref{Eq. Bias expression}
were known, we could obtain an upper bound on the magnitude of the
asymptotic bias by bounding $|\rho_{Zu,T_{P}}\left(s\right)|\leq\rho_{\mathrm{max}}\leq1$
for all $s\in\left[0,\,1\right]$: 
\[
|\Delta|\leq\frac{\rho_{\mathrm{max}}}{\sigma_{D}^{2}}\lim_{T_{P}\rightarrow\infty}\int_{0}^{1}\sqrt{\mathrm{Var}\left(\overline{\varphi}_{Y,u}\left(s\right)\right)}\sqrt{\mathrm{Var}\left(\overline{g}_{D,u}\left(s\right)\right)}\mathrm{d}s.
\]
With this result in hand, we can immediately derive the worst-case
scaled bias under the same strengthening of relative exogeneity as
that in Corollary \ref{Corollary: unbiased OLS anorm}:~under Assumptions
\ref{Assumption: Structural Form}-\ref{Assumption LLN} and $\sqrt{T_{p}}/\sigma_{D}\rightarrow c\in[0,\infty)$
as $T_{P}\rightarrow\infty$, 
\begin{align}
\sup_{|\rho_{Zu,T_{P}}\left(s\right)|\leq\rho_{\mathrm{max}}\,\forall s\in[0,1]} & \sqrt{T_{P}\sigma_{D}^{2}}\left|\mathbb{E}\left(\widehat{\beta}_{\mathrm{ES}}\right)-\left(\beta_{\mathrm{ES}}-\Delta(T_{P})\right)\right|\label{equation: bias-bound}\\
 & \rightarrow c\rho_{\mathrm{max}}\lim_{T_{P}\rightarrow\infty}\int_{0}^{1}\sqrt{\mathrm{Var}\left(\overline{\varphi}_{Y,u}\left(s\right)\right)}\sqrt{\mathrm{Var}\left(\overline{g}_{D,u}\left(s\right)\right)}\mathrm{d}s.\nonumber 
\end{align}
 The worst-case bias depends on $c\rho_{\mathrm{max}}$ and on the
variances of components that are functions of the shock $u_{t}$ and
omitted variables $Z_{t}$.

\subsection{\label{Subsection: Worst-Case-Asymptotic-Coverage}Worst-Case Asymptotic
Coverage}

We use the worst-case scaled bias to study the worst-case asymptotic
coverage of the standard confidence intervals: 
\begin{align*}
\mathrm{CI}\left(\widehat{\beta}_{\mathrm{ES}}\right) & =\left[\widehat{\beta}_{\mathrm{ES}}\pm z_{1-a/2}\sqrt{J/(T_{P}\sigma_{D}^{2})}\right],
\end{align*}
where $z_{1-a/2}$ is the $1-a/2$ quantile of the standard normal
distribution. The asymptotic variance $J$ can be estimated consistently
as suggested in Section \ref{Section: Properties-of-the OLS ES},
without affecting the results.
\begin{cor}
\label{Corollary: Worst-Case Asymptotic Coverage}Let Assumptions
\ref{Assumption: Structural Form}-\ref{Assumption: Differentiability of Potential Outcome}
and \ref{Assumption LLN}-\ref{Assumption CLT} hold, $\sqrt{T_{P}}/\sigma_{D}\rightarrow c\in\left[0,\,\infty\right)$
and $\mathcal{Z}\sim\mathscr{N}(0,1)$. Then: 
\begin{align*}
\inf_{|\rho_{Zu,T_{P}}\left(s\right)|\leq\rho_{\mathrm{max}}\,\forall s\in[0,1]} & \mathbb{P}\left(\left(\beta_{\mathrm{ES}}-\Delta(T_{P})\right)\in\mathrm{CI}\left(\widehat{\beta}_{\mathrm{ES}}\right)\right)\\
 & \rightarrow\mathbb{P}\left(\left|\mathcal{Z}+\frac{c\rho_{\mathrm{max}}}{\sqrt{J}}\lim_{T_{P}\rightarrow\infty}\int_{0}^{1}\sqrt{\mathrm{Var}\left(\overline{\varphi}_{Y,u}\left(s\right)\right)}\sqrt{\mathrm{Var}\left(\overline{g}_{D,u}\left(s\right)\right)}\mathrm{d}s\right|\leq z_{1-a/2}\right).
\end{align*}
\end{cor}
We can study how the worst-case asymptotic coverage varies as $c\rho_{\mathrm{max}}$
and the ratio involving $\mathrm{Var}(\overline{\varphi}_{Y,u}\left(s\right))$,
$\mathrm{Var}(\overline{g}_{D,u}\left(s\right))$ and $\sqrt{J}$
change. The asymptotic variance $J$ can be estimated consistently.
To estimate $\mathrm{Var}(\overline{\varphi}_{Y,u}\left(s\right))$
and $\mathrm{Var}(\overline{g}_{D,u}\left(s\right))$ we assume stationarity,
under which these expressions reduce to $\mathrm{Var}(\overline{\varphi}_{Y,u})$
and $\mathrm{Var}(\overline{g}_{D,u})$, respectively. Since $\overline{\varphi}_{Y,u}$
and $\overline{g}_{D,u}$ are functions of the shock $u_{t}$ and
of omitted variables $Z_{t}$\textemdash both unobserved factors that
are also present in non-announcement windows\textemdash we propose
estimating $\mathrm{Var}(\overline{\varphi}_{Y,u})$ and $\mathrm{Var}(\overline{g}_{D,u})$
using the sample variance of the outcome and policy variables in the
control sample. Specifically, we use $\widehat{\sigma}_{Y,C}^{2}$
for the outcome variable and $\widehat{\sigma}_{D,C}^{2}$ for the
policy variable, where the control sample includes time windows on
days without announcements. Under stationarity, it is reasonable to
assume that the variance of $Y_{t}$ on control days provides an upper
bound for $\mathrm{Var}(\overline{\varphi}_{Y,u})$, and similarly,
the variance of $D_{t}$ on control days provides an upper bound for
$\mathrm{Var}(\overline{g}_{D,u})$. See Section \ref{Section: Empirical Procedure Identification}
for further details and for an application of Corollary \ref{Corollary: Worst-Case Asymptotic Coverage}
to the setting in \citet{nakamura/steinsson:2018}.

\subsection{\label{Subsection: Bias-Aware-Inference}Bias-Aware Inference}

An alternative way to deal with a vanishing bias is to adjust the
critical value upward to compensate for this bias, as suggested by
\citet{armstrong/kolesar:2021}. Using the bound from Proposition
\ref{Proposition: Worst-case bias}, we define the bias-aware confidence
interval as 

\begin{align*}
\mathrm{CI}_{\mathrm{BA}}\left(\widehat{\beta}_{\mathrm{ES}},\,c\rho_{\mathrm{max}}\right) & =\left[\widehat{\beta}_{\mathrm{ES}}\pm\mathrm{cv}_{1-a/2}\left(c\rho_{\mathrm{max}}\frac{\lim_{T_{P}\rightarrow\infty}\int_{0}^{1}\sqrt{\mathrm{Var}\left(\overline{\varphi}_{Y,u}\left(s\right)\right)}\sqrt{\mathrm{Var}\left(\overline{g}_{D,u}\left(s\right)\right)}\mathrm{d}s}{\sqrt{J}}\right)\sqrt{J/(T\sigma_{D}^{2})}\right],
\end{align*}
where $\mathrm{cv}_{1-a/2}\left(\mathcal{B}\right)$ is the bias-aware
critical value defined as the number such that $\mathbb{P}(|\mathcal{Z}+\mathcal{B}|\leq\mathrm{cv}_{1-a/2}\left(\mathcal{B}\right))=1-a$.
By construction the bias-aware confidence interval has correct asymptotic
coverage probability but it can be conservative.
\begin{cor}
\label{Corollary: Bias-aware Inference}Let Assumptions \ref{Assumption: Structural Form}-\ref{Assumption: Differentiability of Potential Outcome}
and \ref{Assumption LLN}-\ref{Assumption CLT} hold and $\sqrt{T_{P}}/\sigma_{D}\rightarrow c\in\left[0,\,\infty\right)$.
We have: 
\begin{align*}
\underset{T_{P}\rightarrow\infty}{\lim}\inf_{|\rho_{Zu,T_{P}}\left(s\right)|\leq\rho_{\mathrm{max}}\,\forall s\in[0,1]}\mathbb{P}\left(\left(\beta_{\mathrm{ES}}-\Delta(T_{P})\right)\in\mathrm{CI}_{\mathrm{BA}}\left(\widehat{\beta}_{\mathrm{ES}},\,c\rho_{\mathrm{max}}\right)\right) & =1-a.
\end{align*}
 
\end{cor}

\section{\label{Section: Empirical Procedure Identification}Empirical Analysis
of Identification and Inference }

In Section \ref{Subsection: Response-of-Interest} we present a simple
procedure that can be used as a sensitivity analysis to determine
whether relative exogeneity provides a good approximation in practice
and apply it to the analysis of interest rate responses to monetary
policy shocks. In Section \ref{Subsection: Application Worst-Case-Coverage and bias-aware CI}
we examine the worst-case asymptotic coverage of the standard confidence
interval and compare the length of the bias-aware confidence interval
to that of the conventional one. In Section \ref{Subsection: Response-of-Blue}
we discuss some identification issues in the event study regressions
that involve Blue Chip forecasts about real GDP growth.

\subsection{\label{Subsection: Response-of-Interest}Response of Interest Rates
to Monetary Policy News}

We consider the regression of \citet{nakamura/steinsson:2018} which
is a special case of \eqref{Eq.: Event study general model}, 
\begin{align}
Y_{t} & =\beta D_{t}+\widetilde{u}_{t},\qquad\qquad t\in\mathbf{P},\label{Eq. (1) Note}
\end{align}
where $Y_{t}$ is the 1-day change in the 2-Year or 5-Year U.S. Treasury
instantaneous real forward rate, $D_{t}$ is the policy news surprise
constructed by the authors as a change over a 30-minute window (see
Example \ref{Example: Monetary policy, NK and Kuttner} above) and
$\widetilde{u}_{t}$ is an error term. The policy variable can be
endogenous, i.e., $\mathbb{E}\left(\widetilde{u}_{t}|\,D_{t}\right)\neq0$.
However, Theorem \ref{Theorem: OLS consistency} implies that if relative
exogeneity \eqref{Eq.: Relative exogeneity} holds then the OLS event
study estimator $\widehat{\beta}_{\mathrm{ES}}$ in \eqref{Eq. (1) Note}
is consistent for $\beta.$ Moreover, even if $\sigma_{e,t}^{2}$
is finite but much larger than the variance of $\widetilde{u}_{t}$,
Theorem \ref{Theorem: OLS ES consistency +  delta} implies that $\widehat{\beta}_{\mathrm{ES}}$
has low bias. 

We propose a simple empirical procedure to check if $\sigma_{e,t}^{2}$
is relatively large enough that $\widehat{\beta}_{\mathrm{ES}}$ has
low bias. We apply our analysis to the setting of \citet{nakamura/steinsson:2018}
for concreteness, noting that it can be straightforwardly applied
to any other event study regression.\footnote{It can also be applied in settings where the researcher believes there
is a specific nonlinear relationship between the economic variables.
In such cases, one would need to consider the appropriate nonlinear
regression model in place of \eqref{Eq. (1) Note}. } While it is possible to estimate $\mathrm{Var}\left(D_{t}\right)$
using the sample variance of $D_{t}$ in the policy sample, this could
be very large even when relative exogeneity is satisfied only for
a few announcements  since the variance cannot be negative, as discussed
above. Additionally, it is more difficult to estimate $\sigma_{\widetilde{u},t}^{2}$
since $\widetilde{u}_{t}$ is not observed and the OLS residuals may
not be close in probability to the corresponding true errors $\widetilde{u}_{t}$
given that $\mathbb{E}\left(\widetilde{u}_{t}|\,D_{t}\right)\neq0$.
The key idea to our  procedure is that since $\widetilde{u}_{t}$
includes macroeconomic news or factors that are present even when
there is no announcement, the order of magnitude of its variance can
be retrieved from the order of magnitude of the variance of $Y_{t}$
in the control sample $\mathbf{C}$. In fact, in the control sample
the variance of $D_{t}$ and $\widetilde{u}_{t}$ have the same order
of magnitude and given that $\left|\beta\right|<\infty$, the order
of magnitude of the variance of $Y_{t}$ for $t\in\mathbf{C}$ is
the same as that of $\widetilde{u}_{t}$ so long as \eqref{Eq. (1) Note}
also holds for $t\in\mathbf{C}$. Thus, we can proxy the average variance
of $\widetilde{u}_{t}$ for $t\in\mathbf{P}$ by the average variance
of $Y_{t}$ for $t\in\mathbf{C}$.

We initially simulate the regression \eqref{Eq. (1) Note} calibrated
to the corresponding regression in \citet{nakamura/steinsson:2018}
for the control sample, making the draws of $D_{t}$ and $\widetilde{u}_{t}$
independent, and in each draw we estimate $\beta$ by OLS. We repeat
this many times and compute bias, mean-absolute error (MAE), and mean-squared
error (MSE) for this idealized OLS estimator. We name it the ``oracle''
estimator since it is obtained in the absence of endogeneity and it
is efficient by the Gauss-Markov Theorem.

The performance of the oracle estimator is compared to that of the
corresponding OLS event study estimator when we allow for correlation
between $D_{t}$ and $\widetilde{u}_{t}$. In the latter regression,
we successively increase the variance of $D_{t}$ and record how much
the performance of the event study estimator approaches that of the
oracle estimator in the idealized regression. In particular, we determine
the threshold value for the variance of $D_{t}$ that ensures that
the event study estimator performs as well as the oracle. Lastly,
we verify whether the sample variance of $D_{t}$ in \citet{nakamura/steinsson:2018}
is larger than this threshold. If it is, relative exogeneity likely
holds and we can interpret $\sigma_{e,t}^{2}$ as being large enough
relative to the variance of $\widetilde{u}_{t}$ for the OLS event
study estimator to have low enough bias to perform well in practice.

Let $T_{C}$ denote the sample size of the control sample $t\in\mathbf{C}$.
The policy sample consists of all regularly scheduled FOMC meeting
days from 1/1/2000 to 3/19/2014. The control sample includes all Tuesdays
and Wednesdays that are not FOMC meeting days over the same period.
This yields $T_{P}=74$ and $T_{C}=762$ when the dependent variable
is the 2-Year real yield, and $T_{P}=106$ and $T_{C}=1130$ when
it is the 5-Year real yield.

Consider the following steps: 
\begin{enumerate}
\item Estimate the mean, variance and autoregressive coefficient of order
1 from all $D_{t}$ in the control sample ($t\in\mathbf{C}$). Denote
these by $\overline{D}_{C}$, $\widehat{\sigma}_{D,C}^{2}$ and $\widehat{\rho}_{D,C,1}$.
Similarly, obtain $\widehat{\sigma}_{Y,C}^{2}$ and $\widehat{\rho}_{Y,C,1}$
from all $Y_{t}$ in the control sample ($t\in\mathbf{C}$). Test
whether $\widehat{\rho}_{D,C,1}$ and $\widehat{\rho}_{Y,C,1}$ are
statistically significant different from zero.
\item Obtain the sample variances, $\widehat{\sigma}_{D,P}^{2}$ and $\widehat{\sigma}_{Y,P}^{2}$,
from all $D_{t}$ and $Y_{t}$ in the policy sample ($t\in\mathbf{P}$).
\item If $Y_{t}$ is constructed as a change over a substantially longer
frequency than $D_{t}$, compute $\widehat{\sigma}_{\widetilde{u}}^{2}=\widehat{\sigma}_{Y,P}^{2}\widehat{\sigma}_{D,C}^{2}/\widehat{\sigma}_{D,P}^{2}$.
Note that $\widehat{\sigma}_{\widetilde{u}}^{2}$ is a proxy for the
variance of $Y_{t}$ in the control sample. Since $Y_{t}$ is the
1-day change in the forward rate while $D_{t}$ is the change in a
30-minute window, the variance of $D_{t}$ is substantially smaller
than that of $Y_{t}$ in the control sample and the variance of $Y_{t}$
and $\widetilde{u}_{t}$ are very similar. So we choose $\widehat{\sigma}_{\widetilde{u}}^{2}$
such that the ratio of the variances of $Y_{t}$ in the policy and
control sample is the same as that of $D_{t}$, i.e., $\widehat{\sigma}_{Y,P}^{2}/\widehat{\sigma}_{\widetilde{u}}^{2}=\widehat{\sigma}_{D,P}^{2}/\widehat{\sigma}_{D,C}^{2}.$\footnote{Note that $\widehat{\sigma}_{\widetilde{u}}^{2}$ approximates the
variance of $\widetilde{u}_{t}$ for $t\in\mathbf{P}$ when $D_{t}$
and $\widetilde{u}_{t}$ are independent and the variance of $\widetilde{u}_{t}$
is similar across the policy and control samples.} Otherwise, compute $\widehat{\sigma}_{\widetilde{u}}^{2}=\widehat{\sigma}_{Y,C}^{2}-\beta^{2}\widehat{\sigma}_{D,C}^{2}$
with $\beta$ set equal to 0.99, which is the OLS estimate in \citet{nakamura/steinsson:2018}.\footnote{We could choose any other value for $\beta$, the intuition behind
the procedure and the results would not change.}
\item For $t=1,\ldots,\,T_{P}$ generate $\widetilde{D}_{t}\sim i.i.d.\,\mathscr{N}(\overline{D}_{C},\,\widehat{\sigma}_{D,C}^{2})$
if $\widehat{\rho}_{D,C,1}$ is not statistically different from zero.
Otherwise, generate $\widetilde{D}_{t}=(1-\widehat{\rho}_{D,C,1})\overline{D}_{C}+\widehat{\rho}_{D,C,1}\widetilde{D}_{t-1}+u_{D,t}$
where $u_{D,t}\sim i.i.d.\,\mathscr{N}(0,\,(1-\widehat{\rho}_{D,C,1}^{2})\widehat{\sigma}_{D,C}^{2})$
for $t=1,\ldots,\,T_{P}$ and $\widetilde{D}_{0}=0$.
\item For $t=1,\ldots,\,T_{P}$ generate $\widetilde{u}_{t}\sim i.i.d.\,\mathscr{N}(0,\,\widehat{\sigma}_{\widetilde{u}}^{2})$
if $\widehat{\rho}_{Y,C,1}$ is not statistically different from zero.
Otherwise generate $\widetilde{u}_{t}$ according to $\widetilde{u}_{t}=\widehat{\rho}_{Y,C,1}\widetilde{u}_{t-1}+\widetilde{v}_{t}$
where $\widetilde{v}_{t}\sim i.i.d.\,\mathscr{N}(0,\,(1-\widehat{\rho}_{Y,C,1}^{2})\widehat{\sigma}_{\widetilde{u}}^{2})$
and $\widetilde{u}_{0}=0$.
\item Generate $\widetilde{Y}_{t}=\beta\widetilde{D}_{t}+\widetilde{u}_{t}$
for $t=1,\ldots,\,T_{P}$ where $\widetilde{D}_{t}$ is generated
in 4., $\widetilde{u}_{t}$ in 5. and $\beta$ is set equal to 0.99
as explained in 3. Run OLS for the regression of $\widetilde{Y}_{t}$
on $\widetilde{D}_{t}$ to obtain the OLS estimator $\widehat{\beta}_{\mathrm{orcale}}$. 
\item Repeat 4-6. 5,000 times and compute the bias, MAE and MSE of $\widehat{\beta}_{\mathrm{orcale}}$.
\item Repeat the regression in 6. but now allow for correlation $\rho$
between $\widetilde{u}_{t}$ and $\widetilde{D}_{t}$. For a given
$\rho\in\left[-1,\,1\right]$, generate $\widetilde{u}_{t}=\rho(\widehat{\sigma}_{\widetilde{u}}/\widehat{\sigma}_{D,C})(\widetilde{D}_{t}-\overline{D}_{C})+\sqrt{1-\rho^{2}}\eta_{t}$,
where $\eta_{t}\sim i.i.d.\,\mathscr{N}(0,\,\widehat{\sigma}_{\widetilde{u}}^{2})$
for $t=1,\ldots,\,T_{P}$.\footnote{Note that the correlation between $\widetilde{u}_{t}$ and $\widetilde{D}_{t}$
is equal to $\rho$ and $\widetilde{u}_{t}$ has variance $\widehat{\sigma}_{\widetilde{u}}^{2}$.} The case $\rho=0$ corresponds to the regression in 6. Generate $\widetilde{Y}_{t}=\beta\widetilde{D}_{t}+\widetilde{u}_{t}$
for $t=1,\ldots,\,T_{P}$ with $\beta$ as in 6. and $\widetilde{D}_{t}$
as in 4. Run OLS for the regression of $\widetilde{Y}_{t}$ on $\widetilde{D}_{t}$
and label this estimator $\widehat{\beta}_{\mathrm{ES}}\left(\rho\right)$.
Repeat this 5,000 times and compute its bias, MAE and MSE.
\item Consider the regression in 8. but start to raise the variance of $\widetilde{D}_{t}$.
That is, define $\widetilde{D}_{\delta,t}=\overline{D}_{C}+(\widetilde{D}_{t}-\overline{D}_{C})\sqrt{1+\delta}$
for some $\delta\geq0$ where $\widetilde{D}_{t}$ is generated in
4.\footnote{Note that $\widetilde{D}_{\delta,t}$ is a rescaled version of $\widetilde{D}_{t}$
with variance $\widehat{\sigma}_{D,C}^{2}\left(1+\delta\right)$.} Generate $\widetilde{Y}_{t}=\beta\widetilde{D}_{\delta,t}+\widetilde{u}_{t}$
for $t=1,\ldots,\,T_{P}$ with $\beta$ and $\widetilde{u}_{t}$ as
in 8. Run OLS for the regression of $\widetilde{Y}_{t}$ on $\widetilde{D}_{\delta,t}$
and label this estimator $\widehat{\beta}_{\mathrm{ES}}\left(\rho,\,\delta\right)$.\footnote{$\widehat{\beta}_{\mathrm{ES}}\left(\rho\right)$ in 8. corresponds
to $\widehat{\beta}_{\mathrm{ES}}\left(\rho,\,0\right)$.} Repeat this 5,000 times and compute its bias, MAE and MSE.
\item Find the value of $\delta$ such that $\mathrm{MSE}(\widehat{\beta}_{\mathrm{ES}}\left(\rho,\,\delta\right))=\mathrm{MSE}(\widehat{\beta}_{\mathrm{orcale}})$
{[}or $\mathrm{MAE}(\widehat{\beta}_{\mathrm{ES}}\left(\rho,\,\delta\right))=\mathrm{MAE}(\widehat{\beta}_{\mathrm{orcale}})${]}
and label it $\delta^{*}\left(\rho\right)$. Define $\widehat{\sigma}_{D,*}^{2}\left(\rho\right)=\widehat{\sigma}_{D,C}^{2}\left(1+\delta^{*}\left(\rho\right)\right).$
\item Repeat 8.-10. for a grid of $\rho$ values in $[-1,\,1]$. 
\end{enumerate}
Steps 1.-11. provide a detailed sensitivity analysis that computes
the values of the variance of $D_{t}$ that allows the event study
estimator to perform as well as the oracle estimator over a range
of values of endogeneity $\rho$, namely $\widehat{\sigma}_{D,*}^{2}\left(\rho\right)$.
If $\widehat{\sigma}_{D,P}^{2}\geq\widehat{\sigma}_{D,*}^{2}\left(\rho\right)$
one can be confident that the variance of the policy shock in \citet{nakamura/steinsson:2018}
is relatively large enough for a degree of endogeneity that satisfies
$|\mathrm{Corr}\left(\widetilde{u}_{t},\,D_{t}\right)|\leq|\rho|$
for the event study estimator to have low enough bias for its MSE
(MAE) to be as low as in the oracle regression with no endogeneity.
If one is not interested in the full sensitivity analysis, but rather
just in evaluating the values of $\rho$ for which the policy shock
in \citet{nakamura/steinsson:2018} is large enough to produce a credible
event study estimator, one could  replace $1+\delta$ in 9. by $\widehat{\sigma}_{D,P}^{2}/\widehat{\sigma}_{D,C}^{2}$
and instead of 10., find the value of $\rho$ such that $\mathrm{MSE}(\widehat{\beta}_{\mathrm{ES}}(\rho,\,\widehat{\sigma}_{D,P}^{2}/\widehat{\sigma}_{D,C}^{2}-1))=\mathrm{MSE}(\widehat{\beta}_{\mathrm{orcale}})$
{[}or $\mathrm{MAE}(\widehat{\beta}_{\mathrm{ES}}(\rho,\,\widehat{\sigma}_{D,P}^{2}/\widehat{\sigma}_{D,C}^{2}-1))=\mathrm{MAE}(\widehat{\beta}_{\mathrm{orcale}})${]}.

The validity of the sensitivity analysis relies on the following two
properties: (i) the model in \eqref{Eq. (1) Note} is correct also
in the control sample, i.e., for $t\in\mathbf{C}$ and (ii) the order
of magnitude of $\widehat{\sigma}_{\widetilde{u}}^{2}$ is a good
proxy for the order of magnitude of $T_{P}^{-1}\sum_{t\in\mathbf{P}}\sigma_{\widetilde{u},t}^{2}$.
Property (i) is implicitly assumed in typical VAR estimation of simultaneous
equations models since typically the relationship between monetary
policy and real economic variables is estimated with VARs at monthly,
quarterly, yearly frequencies and each observation aggregates both
FOMC announcement and non-announcement days. Property (ii) is reasonable
since $\widetilde{u}_{t}$ for $t\in\mathbf{P}$ represents news and
latent factors that are not specific to the announcement and would
be present even if there were no FOMC announcements. Thus, its variation
is expected to be similar to that in non-announcement days which in
turn has the same order of magnitude as that of $Y_{t}$ in non-announcement
days. 

Table \ref{Table Bias, MAE and MSE} shows the bias, MAE and MSE of
$\widehat{\beta}_{\mathrm{orcale}}$ and $\widehat{\beta}_{\mathrm{ES}}\left(\rho,\,\delta\right)$
for different values of $\rho$ and $\delta$. We first compare $\widehat{\beta}_{\mathrm{orcale}}$
and $\widehat{\beta}_{\mathrm{ES}}\left(\rho,\,\delta\right)$ for
$\rho=0,\,0.25,\ldots,\,1$ and $\delta=0$. As we raise the endogeneity
parameter $\rho$ from 0 to 1 the bias, MAE and MSE of $\widehat{\beta}_{\mathrm{ES}}$
increases substantially.

\begin{table}[H]
\caption{\label{Table Bias, MAE and MSE}Bias, MAE and MSE of $\widehat{\beta}_{\mathrm{oracle}}$
and $\widehat{\beta}_{\mathrm{ES}}$}

\smallskip{}

\begin{centering}
{\footnotesize{}%
\begin{tabular}{ccccccccccccc}
\hline 
\multicolumn{13}{c}{}\tabularnewline
\multicolumn{13}{c}{{\footnotesize 2-Year U.S. Treasury instantaneous real forward rates}}\tabularnewline
\hline 
\hline 
 & \multicolumn{3}{c}{{\footnotesize$\widehat{\beta}_{\mathrm{orcale}}$}} & \multicolumn{3}{c}{{\footnotesize$\widehat{\beta}_{\mathrm{ES}}\left(\rho,\,\delta=0\right)$}} & \multicolumn{3}{c}{{\footnotesize$\widehat{\beta}_{\mathrm{ES}}\left(\rho,\,\delta=1\right)$}} & \multicolumn{3}{c}{{\footnotesize$\widehat{\beta}_{\mathrm{ES}}\left(\rho,\,\delta=8\right)$}}\tabularnewline
{\footnotesize$\rho$} & {\footnotesize Bias} & {\footnotesize MAE} & {\footnotesize MSE} & {\footnotesize Bias} & {\footnotesize MAE} & {\footnotesize MSE} & {\footnotesize Bias} & {\footnotesize MAE} & {\footnotesize MSE} & {\footnotesize Bias} & {\footnotesize MAE} & {\footnotesize MSE}\tabularnewline
\hline 
{\footnotesize 0.} & {\footnotesize -0.004} & {\footnotesize 0.158} & {\footnotesize 0.039} & {\footnotesize 0.000} & {\footnotesize 0.155} & {\footnotesize 0.037} & {\footnotesize -0.004} & {\footnotesize 0.126} & {\footnotesize 0.025} & {\footnotesize 0.000} & {\footnotesize 0.069} & {\footnotesize 0.008}\tabularnewline
{\footnotesize 0.25} &  &  &  & {\footnotesize 0.294} & {\footnotesize 0.302} & {\footnotesize 0.128} & {\footnotesize 0.278} & {\footnotesize 0.282} & {\footnotesize 0.101} & {\footnotesize 0.177} & {\footnotesize 0.179} & {\footnotesize 0.039}\tabularnewline
{\footnotesize 0.50} &  &  &  & {\footnotesize 0.580} & {\footnotesize 0.580} & {\footnotesize 0.367} & {\footnotesize 0.550} & {\footnotesize 0.550} & {\footnotesize 0.322} & {\footnotesize 0.349} & {\footnotesize 0.349} & {\footnotesize 0.128}\tabularnewline
{\footnotesize 0.75} &  &  &  & {\footnotesize 0.869} & {\footnotesize 0.869} & {\footnotesize 0.777} & {\footnotesize 0.826} & {\footnotesize 0.826} & {\footnotesize 0.696} & {\footnotesize 0.524} & {\footnotesize 0.524} & {\footnotesize 0.278}\tabularnewline
{\footnotesize 1} &  &  &  & {\footnotesize 1.167} & {\footnotesize 1.163} & {\footnotesize 1.362} & {\footnotesize 1.098} & {\footnotesize 1.098} & {\footnotesize 1.211} & {\footnotesize 0.699} & {\footnotesize 0.699} & {\footnotesize 0.490}\tabularnewline
\hline 
 & \multicolumn{3}{c}{{\footnotesize$\widehat{\beta}_{\mathrm{ES}}\left(\rho,\,\delta=16\right)$}} & \multicolumn{3}{c}{{\footnotesize$\widehat{\beta}_{\mathrm{ES}}\left(\rho,\,\delta=32\right)$}} & \multicolumn{3}{c}{{\footnotesize$\widehat{\beta}_{\mathrm{ES}}\left(\rho,\,\delta=40\right)$}} & \multicolumn{3}{c}{{\footnotesize$\widehat{\beta}_{\mathrm{ES}}\left(\rho,\,\delta=48\right)$}}\tabularnewline
{\footnotesize$\rho$} & {\footnotesize Bias} & {\footnotesize MAE} & {\footnotesize MSE} & {\footnotesize Bias} & {\footnotesize MAE} & {\footnotesize MSE} & {\footnotesize Bias} & {\footnotesize MAE} & {\footnotesize MSE} & {\footnotesize Bias} & {\footnotesize MAE} & {\footnotesize MSE}\tabularnewline
\hline 
{\footnotesize 0} & {\footnotesize 0.000} & {\footnotesize 0.052} & {\footnotesize 0.004} & {\footnotesize 0.000} & {\footnotesize 0.038} & {\footnotesize 0.002} & {\footnotesize -0.001} & {\footnotesize 0.034} & {\footnotesize 0.002} & {\footnotesize 0.000} & {\footnotesize 0.031} & {\footnotesize 0.002}\tabularnewline
{\footnotesize 0.25} & {\footnotesize 0.133} & {\footnotesize 0.134} & {\footnotesize 0.022} & {\footnotesize 0.098} & {\footnotesize 0.099} & {\footnotesize 0.012} & {\footnotesize 0.091} & {\footnotesize 0.091} & {\footnotesize 0.000} & {\footnotesize 0.081} & {\footnotesize 0.082} & {\footnotesize 0.008}\tabularnewline
{\footnotesize 0.50} & {\footnotesize 0.267} & {\footnotesize 0.267} & {\footnotesize 0.074} & {\footnotesize 0.197} & {\footnotesize 0.179} & {\footnotesize 0.041} & {\footnotesize 0.177} & {\footnotesize 0.177} & {\footnotesize 0.032} & {\footnotesize 0.163} & {\footnotesize 0.164} & {\footnotesize 0.028}\tabularnewline
{\footnotesize 0.75} & {\footnotesize 0.401} & {\footnotesize 0.401} & {\footnotesize 0.163} & {\footnotesize 0.296} & {\footnotesize 0.296} & {\footnotesize 0.088} & {\footnotesize 0.266} & {\footnotesize 0.266} & {\footnotesize 0.072} & {\footnotesize 0.245} & {\footnotesize 0.245} & {\footnotesize 0.061}\tabularnewline
{\footnotesize 1} & {\footnotesize 0.534} & {\footnotesize 0.534} & {\footnotesize 0.285} & {\footnotesize 0.398} & {\footnotesize 0.398} & {\footnotesize 0.155} & {\footnotesize 0.355} & {\footnotesize 0.355} & {\footnotesize 0.126} & {\footnotesize 0.326} & {\footnotesize 0.326} & {\footnotesize 0.107}\tabularnewline
\hline 
 &  &  &  &  &  &  &  &  &  &  &  & \tabularnewline
\multicolumn{13}{c}{{\footnotesize 5-Year U.S. Treasury instantaneous real forward rates}}\tabularnewline
\hline 
\hline 
 & \multicolumn{3}{c}{{\footnotesize$\widehat{\beta}_{\mathrm{orcale}}$}} & \multicolumn{3}{c}{{\footnotesize$\widehat{\beta}_{\mathrm{ES}}\left(\rho,\,\delta=0\right)$}} & \multicolumn{3}{c}{{\footnotesize$\widehat{\beta}_{\mathrm{ES}}\left(\rho,\,\delta=1\right)$}} & \multicolumn{3}{c}{{\footnotesize$\widehat{\beta}_{\mathrm{ES}}\left(\rho,\,\delta=8\right)$}}\tabularnewline
{\footnotesize$\rho$} & {\footnotesize Bias} & {\footnotesize MAE} & {\footnotesize MSE} & {\footnotesize Bias} & {\footnotesize MAE} & {\footnotesize MSE} & {\footnotesize Bias} & {\footnotesize MAE} & {\footnotesize MSE} & {\footnotesize Bias} & {\footnotesize MAE} & {\footnotesize MSE}\tabularnewline
\hline 
{\footnotesize 0} & {\footnotesize 0.0017} & {\footnotesize 0.1171} & {\footnotesize 0.0217} & {\footnotesize -0.002} & {\footnotesize 0.119} & {\footnotesize 0.022} & {\footnotesize 0.000} & {\footnotesize 0.095} & {\footnotesize 0.014} & {\footnotesize -0.002} & {\footnotesize 0.052} & {\footnotesize 0.004}\tabularnewline
{\footnotesize 0.25} &  &  &  & {\footnotesize 0.268} & {\footnotesize 0.272} & {\footnotesize 0.093} & {\footnotesize 0.249} & {\footnotesize 0.250} & {\footnotesize 0.076} & {\footnotesize 0.159} & {\footnotesize 0.159} & {\footnotesize 0.029}\tabularnewline
{\footnotesize 0.50} &  &  &  & {\footnotesize 0.529} & {\footnotesize 0.530} & {\footnotesize 0.299} & {\footnotesize 0.502} & {\footnotesize 0.502} & {\footnotesize 0.263} & {\footnotesize 0.319} & {\footnotesize 0.319} & {\footnotesize 0.108}\tabularnewline
{\footnotesize 0.75} &  &  &  & {\footnotesize 0.796} & {\footnotesize 0.796} & {\footnotesize 0.646} & {\footnotesize 0.750} & {\footnotesize 0.751} & {\footnotesize 0.572} & {\footnotesize 0.478} & {\footnotesize 0.478} & {\footnotesize 0.231}\tabularnewline
{\footnotesize 1} &  &  &  & {\footnotesize 1.061} & {\footnotesize 1.061} & {\footnotesize 1.131} & {\footnotesize 1.002} & {\footnotesize 1.002} & {\footnotesize 1.006} & {\footnotesize 0.638} & {\footnotesize 0.637} & {\footnotesize 0.407}\tabularnewline
\hline 
 & \multicolumn{3}{c}{{\footnotesize$\widehat{\beta}_{\mathrm{ES}}\left(\rho,\,\delta=16\right)$}} & \multicolumn{3}{c}{{\footnotesize$\widehat{\beta}_{\mathrm{ES}}\left(\rho,\,\delta=32\right)$}} & \multicolumn{3}{c}{{\footnotesize$\widehat{\beta}_{\mathrm{ES}}\left(\rho,\,\delta=40\right)$}} & \multicolumn{3}{c}{{\footnotesize$\widehat{\beta}_{\mathrm{ES}}\left(\rho,\,\delta=48\right)$}}\tabularnewline
{\footnotesize$\rho$} & {\footnotesize Bias} & {\footnotesize MAE} & {\footnotesize MSE} & {\footnotesize Bias} & {\footnotesize MAE} & {\footnotesize MSE} & {\footnotesize Bias} & {\footnotesize MAE} & {\footnotesize MSE} & {\footnotesize Bias} & {\footnotesize MAE} & {\footnotesize MSE}\tabularnewline
{\footnotesize 0} & {\footnotesize 0.000} & {\footnotesize 0.040} & {\footnotesize 0.003} & {\footnotesize 0.000} & {\footnotesize 0.029} & {\footnotesize 0.002} & {\footnotesize 0.000} & {\footnotesize 0.025} & {\footnotesize 0.001} & {\footnotesize 0.000} & {\footnotesize 0.023} & {\footnotesize 0.000}\tabularnewline
{\footnotesize 0.25} & {\footnotesize 0.123} & {\footnotesize 0.123} & {\footnotesize 0.017} & {\footnotesize 0.090} & {\footnotesize 0.090} & {\footnotesize 0.009} & {\footnotesize 0.081} & {\footnotesize 0.081} & {\footnotesize 0.008} & {\footnotesize 0.074} & {\footnotesize 0.074} & {\footnotesize 0.006}\tabularnewline
{\footnotesize 0.50} & {\footnotesize 0.243} & {\footnotesize 0.243} & {\footnotesize 0.061} & {\footnotesize 0.179} & {\footnotesize 0.179} & {\footnotesize 0.033} & {\footnotesize 0.162} & {\footnotesize 0.162} & {\footnotesize 0.027} & {\footnotesize 0.149} & {\footnotesize 0.149} & {\footnotesize 0.023}\tabularnewline
{\footnotesize 0.75} & {\footnotesize 0.367} & {\footnotesize 0.364} & {\footnotesize 0.134} & {\footnotesize 0.269} & {\footnotesize 0.269} & {\footnotesize 0.073} & {\footnotesize 0.243} & {\footnotesize 0.243} & {\footnotesize 0.059} & {\footnotesize 0.223} & {\footnotesize 0.223} & {\footnotesize 0.050}\tabularnewline
{\footnotesize 1} & {\footnotesize 0.487} & {\footnotesize 0.487} & {\footnotesize 0.237} & {\footnotesize 0.359} & {\footnotesize 0.359} & {\footnotesize 0.129} & {\footnotesize 0.324} & {\footnotesize 0.324} & {\footnotesize 0.105} & {\footnotesize 0.297} & {\footnotesize 0.297} & {\footnotesize 0.088}\tabularnewline
\hline 
\end{tabular}}{\footnotesize\par}
\par\end{centering}
{\scriptsize{}%
\noindent\begin{minipage}[t]{1\columnwidth}%
{\scriptsize The bias, MAE and MSE of $\widehat{\beta}_{\mathrm{orcale}}$
and $\widehat{\beta}_{\mathrm{ES}}\left(\rho,\,\delta\right)$. In
the top panel, the dependent variable in each regression is calibrated
to the 2-Year real forward rate and $T_{P}=74$. In bottom panel,
the dependent variable in each regression is calibrated to the 5-Year
real forward rate and $T_{P}=106$. The number of replications is
5,000.}%
\end{minipage}}{\scriptsize\par}
\end{table}

To provide numerical support for the theoretical results of Theorem
\ref{Theorem: WATE Event Study}-\ref{Theorem: OLS consistency},
we look at the change in the bias, MAE and MSE of $\widehat{\beta}_{\mathrm{ES}}\left(\rho,\,\delta\right)$
as $\delta$ increases for a given $\rho\neq0$ when looking at the
2-Year real yields. The results show that these summary statistics
decrease quickly as $\delta$ grows. For example, for a small degree
of endogeneity $\rho=0.25$, it is sufficient to increase the variance
of the policy variable in the announcement windows nine-fold $\left(\delta=8\right)$
for the event study estimator to exhibit an MSE that is as small as
that of the oracle estimator. A similar feature applies to MAE, though
a slightly larger $\delta$ is needed. As we raise $\delta$ further,
the MAE and MSE of $\widehat{\beta}_{\mathrm{ES}}$ become smaller
and smaller relative to those of $\widehat{\beta}_{\mathrm{oracle}}$.
In particular, both the MAE and MSE of $\widehat{\beta}_{\mathrm{ES}}$
converge to zero quickly as $\delta$ increases for $\rho=0.25$.

For a larger value of $\rho$ it takes larger values of $\delta$
to decrease the MAE and MSE of $\widehat{\beta}_{\mathrm{ES}}$ to
those of $\widehat{\beta}_{\mathrm{oracle}}$. For example, for $\rho=0.5$
increasing the variance of the policy variable in the policy sample
nine-fold is not sufficient, while increasing it roughly thirty three
times is. Similar features are found when the dependent variable
is calibrated to the 5-Year real yield.

The bias of $\widehat{\beta}_{\mathrm{oracle}}$ is always smaller
than that of $\widehat{\beta}_{\mathrm{ES}}$. This follows because
for $\widehat{\beta}_{\mathrm{oracle}}$ positive and negative deviations
from the true value $\beta$ cancel each other across simulation replications,
while for $\widehat{\beta}_{\mathrm{ES}}$ these deviations tend to
be positive since $\rho>0$. On the other hand, these deviations for
$\widehat{\beta}_{\mathrm{ES}}$ tend to be smaller in magnitude,
making its variance smaller than that of $\widehat{\beta}_{\mathrm{oracle}}$
for large values of $\delta$. Focusing only on bias does not account
for the higher precision of $\widehat{\beta}_{\mathrm{ES}}$. A user
concerned about both accuracy and precision should rather focus on
performance measures such as MSE or MAE, which incorporate both bias
and variance.

We now turn to the most important step of our sensitivity analysis.
Given that the regression is calibrated to that in \citet{nakamura/steinsson:2018},
we can verify up to which degree of endogeneity $\rho$ the variance
of the policy shock is large enough for the event study estimator
to be expected to perform well in this setting. This entails finding
for each $\rho>0$ the value $\delta^{*}\left(\rho\right)$ such that
$\mathrm{MSE}(\widehat{\beta}_{\mathrm{oracle}})=\mathrm{MSE}(\widehat{\beta}_{\mathrm{ES}}(\rho,\,\delta^{*}\left(\rho\right)))$
and then determining whether the estimate $\widehat{\sigma}_{D,P}^{2}$
from \citet{nakamura/steinsson:2018} is larger than $\widehat{\sigma}_{D,*}^{2}\left(\rho\right)=\widehat{\sigma}_{D,C}^{2}\left(1+\delta^{*}\left(\rho\right)\right).$

We report this information in Table \ref{Table Delta*}. As $\rho$
rises the value of $\delta^{*}$ increases since stronger endogeneity
requires higher variance in the policy variable to make the event
study estimator less biased. From \citet{nakamura/steinsson:2018}
$\widehat{\sigma}_{D,P}^{2}=32.43\widehat{\sigma}_{D,C}^{2}$. Thus,
when $Y_{t}$ is the 2-Year real yield we have $\widehat{\sigma}_{D,P}^{2}\approx\widehat{\sigma}_{D,*}^{2}\left(0.50\right)$
while when $Y_{t}$ is the 5-Year real yield we have $\widehat{\sigma}_{D,P}^{2}\approx\widehat{\sigma}_{D,*}^{2}\left(0.40\right)$.
These imply that the event study estimator can be expected to perform
well for all degrees of endogeneity no larger than roughly $\rho=0.50$
($\rho=0.40$) for the 2-Year (5-Year) real yield.

Values of $\rho=0.50$ and 0.40 represent quite strong empirical contemporaneous
correlation for any pair of 30-minute or 1-day changes in common macroeconomic
and financial variables. These high-frequency contemporaneous correlations
should not be that large in practice for two reasons. First, it is
well-known that taking first-differences of trending variables reduces
their variability and their contemporaneous correlation is smaller
than that corresponding to the series in levels. Second, many relationships
between macroeconomic and financial variables are in the form of lead-lag
which implies that significant portions of dependence between any
two variables is not contemporaneous. We verify this empirically by
computing the pairwise contemporaneous correlations between all the
time series used by \citet{nakamura/steinsson:2018} and the main
macroeconomic time series available from FRED. Even though these pairwise
correlations are typically larger than zero in absolute value, they
never exceed 0.5 and they average about 0.2.

\begin{table}[t]
\caption{\label{Table Delta*} Values of $\delta^{*}$ for each $\rho$}

\smallskip{}

\begin{centering}
{\footnotesize{}%
\begin{tabular}{ccc}
\hline 
 & {\footnotesize 2-Year real forward} & {\footnotesize 5-Year real forward}\tabularnewline
{\footnotesize$\rho$} & {\footnotesize$\delta^{*}$} & {\footnotesize$\delta^{*}$}\tabularnewline
\hline 
\hline 
{\footnotesize 0.10} & {\footnotesize 0.78} & {\footnotesize 1.50}\tabularnewline
{\footnotesize 0.15} & {\footnotesize 2.50} & {\footnotesize 4.00}\tabularnewline
{\footnotesize 0.20} & {\footnotesize 4.60} & {\footnotesize 7.30}\tabularnewline
{\footnotesize 0.25} & {\footnotesize 7.60} & {\footnotesize 11.9}\tabularnewline
{\footnotesize 0.30} & {\footnotesize 11.40} & {\footnotesize 17.80}\tabularnewline
{\footnotesize 0.35} & {\footnotesize 15.80} & {\footnotesize 24.80}\tabularnewline
{\footnotesize 0.40} & {\footnotesize 21.00} & {\footnotesize 32.40}\tabularnewline
{\footnotesize 0.45} & {\footnotesize 26.80} & {\footnotesize 41.20}\tabularnewline
{\footnotesize 0.50} & {\footnotesize 33.50} & {\footnotesize 50.80}\tabularnewline
{\footnotesize 0.55} & {\footnotesize 40.80} & {\footnotesize 61.50}\tabularnewline
{\footnotesize 00.60} & {\footnotesize 49.30} & {\footnotesize 73.50}\tabularnewline
{\footnotesize 0.65} & {\footnotesize 56.80} & {\footnotesize 86.50}\tabularnewline
{\footnotesize 0.70} & {\footnotesize 66.00} & {\footnotesize 100.50}\tabularnewline
{\footnotesize 0.75} & {\footnotesize 75.90} & {\footnotesize 115.30}\tabularnewline
{\footnotesize 0.80} & {\footnotesize 86.50} & {\footnotesize 132.10}\tabularnewline
{\footnotesize 0.85} & {\footnotesize 97.80} & {\footnotesize 148.20}\tabularnewline
{\footnotesize 0.90} & {\footnotesize 109.7} & {\footnotesize 166.50}\tabularnewline
{\footnotesize 0.95} & {\footnotesize 122.3} & {\footnotesize 186.70}\tabularnewline
{\footnotesize 1} & {\footnotesize 135.5} & {\footnotesize 207.50}\tabularnewline
\hline 
\end{tabular}}{\footnotesize\par}
\par\end{centering}
{\small ~~~~~~~~~~~~~~~~~~~~~~~~~~~~~~~~~~}{\scriptsize{}%
\begin{minipage}[t]{0.5\columnwidth}%
{\scriptsize The Values of $\delta^{*}$ for each $\rho=0.10,\,\ldots,\,1.$.
The dependent variable is the 2-Year real forward rate (first column)
and the 5-Year real forward rate (second column). The number of replications
is 5,000.}%
\end{minipage}}{\scriptsize\par}
\end{table}

To analyze whether the variance of the policy shock is relatively
large only for some announcement days and not for others, we compute
the sample variance of the policy variable $\mathrm{\widehat{Var}}\left(D_{t}\right)$
over disjoint sub-samples. We consider the full sample from 1/1/2000
to 3/19/2014 and the sample post-2004.\footnote{We consider the sample after 2004 because the data for 2-year forward
rates, are available from 2004 onward.} The sample sizes are $T=T_{P}+T_{C}=1,236$ $\left(T_{P}=106,\,T_{C}=1130\right)$
and $T=T_{P}+T_{C}=836$ $\left(T_{P}=74,\,T_{C}=762\right)$, respectively.
For each sample we construct several sub-samples with different window
lengths according to the rule $n_{P}=\left\lfloor T_{P}^{4/5}\right\rfloor $
and $m_{P}=\left\lfloor T_{P}/n_{P}\right\rfloor $ where $n_{P}$
is the number of observations in each window and $m_{P}$ is the number
of windows in the policy sample.\footnote{The available data for the control sample is up to and including 2012.
Thus, in constructing the sub-samples we consider all announcement
and non-announcement days until the end of 2012.}\footnote{The choice of the window length $n_{P}=\left\lfloor T_{P}^{4/5}\right\rfloor $
is optimal for nonparametric smoothing under an MSE criterion.} The number of windows in the control sample is set equal to that
in the policy sample, i.e., $m_{C}=m_{P}$, and we set $n_{C}=\left\lfloor T_{C}/m_{C}\right\rfloor $
so that each corresponding window in the policy and control sample
brackets the same period.

The evidence in Table \ref{Table Robustness} shows that there is
some time-variation in $\mathrm{\widehat{Var}}\left(D_{t}\right)$
in both the policy and control sample, though it does not deviate
much from the average value computed over the policy and control sample,
respectively. In particular, the ratio of $\mathrm{\widehat{Var}}\left(D_{t}\right)$
in the corresponding policy and control sub-samples displays some
time-variation even though it does not fall substantially below the
ratio computed over the full policy and control sample. For example,
for the period 2004-2012 the rule selects two windows. In the first
window, the ratio of $\mathrm{\widehat{Var}}\left(D_{t}\right)$ in
the policy and control sample is 26.27. This is not much smaller 
than that corresponding to the sample 2004-2012, 36.72.

We also compute the variance of the policy variable in the sub-sample
that includes the last two non-announcement days prior to an FOMC
meeting that are available from the control sample. That is, we estimate
the variance of the 30-minute changes in the policy variable across
all Tuesdays and Wednesdays of the week before the announcement, two
weeks before the announcement and three weeks before the announcement,
labeling these as $\mathrm{\widehat{Var}}(\overleftarrow{D}_{t,1})$,
$\mathrm{\widehat{Var}}(\overleftarrow{D}_{t,2})$ and $\mathrm{\widehat{Var}}(\overleftarrow{D}_{t,3})$.
The results do not show any significant evidence of leakage occurring
in the 30-minute window of the three preceding weeks of an FOMC announcement:
$\mathrm{\widehat{Var}}(\overleftarrow{D}_{t,1})$ is even smaller
than $\mathrm{\widehat{Var}}(\overleftarrow{D}_{t,2})$ and $\mathrm{\widehat{Var}}(\overleftarrow{D}_{t,3})$.
Of course, this does not exclude the possibility that there is some
leakage outside the 30-minute window {[}2:05pm-2:35pm{]} or in the
other days that precede the announcement. This can be analyzed by
applying the same approach to additional high-frequency data (e.g.,
for Thursdays, Fridays and Mondays). Here we only consider the data
from \citet{nakamura/steinsson:2018}.

\begin{table}[t]
\caption{\label{Table Robustness}{\footnotesize Estimates of $\mathrm{\mathrm{Var}}\left(D_{t}\right)$
and of $\rho_{j}$}}

\smallskip{}

\begin{centering}
{\footnotesize{}%
\begin{tabular}{ccccccc}
\hline 
{\footnotesize\textit{Panel A}}{\footnotesize . } & {\footnotesize 2004-2014} & {\footnotesize 2000-2014} &  &  & \multicolumn{1}{c}{{\footnotesize 2004-2012}} & \multicolumn{1}{c}{{\footnotesize 2000-2012}}\tabularnewline
\hline 
\hline 
 & {\footnotesize$\mathrm{\widehat{Var}}\left(D_{t}\right)$} & {\footnotesize$\mathrm{\widehat{Var}}\left(D_{t}\right)$} &  &  & {\footnotesize$\mathrm{\widehat{Var}}\left(D_{t}\right)$} & {\footnotesize$\mathrm{\widehat{Var}}\left(D_{t}\right)$}\tabularnewline
{\footnotesize Policy } & {\footnotesize 0.000808} & {\footnotesize 0.001200} &  & {\footnotesize Policy, 1st window} & {\footnotesize 0.000725} & {\footnotesize 0.002011}\tabularnewline
{\footnotesize Control } & {\footnotesize 0.000022} & {\footnotesize 0.000037} &  & {\footnotesize Control, 1st window} & {\footnotesize 0.000028} & {\footnotesize 0.000054}\tabularnewline
{\footnotesize Ratio } & {\footnotesize 36.72} & {\footnotesize 32.43} &  & {\footnotesize Policy, 2nd window} & {\footnotesize 0.001374} & {\footnotesize 0.001259}\tabularnewline
\multicolumn{3}{c}{} &  & {\footnotesize Control, 2nd window} & {\footnotesize 0.000016} & {\footnotesize 0.000020}\tabularnewline
 &  &  &  & {\footnotesize Ratio 1st window} & {\footnotesize 26.27} & {\footnotesize 37.24}\tabularnewline
 &  &  &  & {\footnotesize Ratio 2nd window} & {\footnotesize 87.85} & {\footnotesize 62.95}\tabularnewline
\multicolumn{7}{c}{}\tabularnewline
\hline 
\multicolumn{3}{c}{{\footnotesize\textit{Panel B}}{\footnotesize . 2000-2014}} &  & {\footnotesize\textit{Panel C.}} & {\footnotesize 30-Minute} & {\footnotesize 1-Day}\tabularnewline
\hline 
\hline 
{\footnotesize$\mathrm{\widehat{Var}}\left(\overleftarrow{D}_{t,1}\right)$} & {\footnotesize$\mathrm{\widehat{Var}}\left(\overleftarrow{D}_{t,2}\right)$} & {\footnotesize$\mathrm{\widehat{Var}}\left(\overleftarrow{D}_{t,3}\right)$} &  & {\footnotesize$\widehat{\rho}_{1}$} & {\footnotesize 0.14} & {\footnotesize 0.03}\tabularnewline
{\footnotesize 0.000025} & {\footnotesize 0.000052} & {\footnotesize 0.000040} &  &  & {\footnotesize (0.84)} & {\footnotesize (0.18)}\tabularnewline
 &  &  &  & {\footnotesize$\widehat{\rho}_{2}$} & {\footnotesize 0.05} & {\footnotesize -0.19}\tabularnewline
 &  &  &  &  & {\footnotesize (0.12)} & {\footnotesize (0.15)}\tabularnewline
 &  &  &  & {\footnotesize$\widehat{\rho}_{3}$} & {\footnotesize 0.18{*}{*}} & {\footnotesize 0.03}\tabularnewline
 &  &  &  &  & {\footnotesize 0.08} & {\footnotesize (0.11)}\tabularnewline
 &  &  &  & {\footnotesize$\widehat{\rho}_{4}$} & {\footnotesize -0.22} & {\footnotesize 0.07}\tabularnewline
 &  &  &  &  & {\footnotesize (0.19)} & {\footnotesize (0.12)}\tabularnewline
 &  &  &  & {\footnotesize$\widehat{\rho}_{5}$} & {\footnotesize 0.12} & {\footnotesize -0.11}\tabularnewline
 &  &  &  &  & {\footnotesize (0.18)} & {\footnotesize 0.16}\tabularnewline
 &  &  &  & {\footnotesize$\widehat{\rho}_{6}$} & {\footnotesize 0.07} & {\footnotesize -0.01}\tabularnewline
 &  &  &  &  & {\footnotesize (0.05)} & {\footnotesize (0.12)}\tabularnewline
\hline 
\end{tabular}}{\footnotesize\par}
\par\end{centering}
\begin{singlespace}
\noindent\begin{minipage}[t]{1\columnwidth}%
{\scriptsize The estimates of the average variance of the policy variable
$D_{t}$ in different samples or sub-samples (Panel A and B), and
the estimates of $\rho_{r}$ for $r=1,\ldots,\,6$ (Panel C). The
samples considered are from 1/1/2004 to 3/19/2014 and from 1/1/2000
to 3/19/2014. The sub-samples are constructed within each of the latter
two samples using window lengths according to the MSE criterion. For
the sample 2004-2014, $m_{P}=2$, $n_{P}=27$ and $n_{C}=381$. For
the sample 2000-2014, $m_{P}=2$, $n_{P}=38$ and $n_{C}=565$.}%
\end{minipage}
\end{singlespace}
\end{table}

As discussed in Section \ref{Subsection: Robustness to Leakage},
an implication of leakage is serial dependence in $D_{t}$ in the
days prior to the FOMC announcement. We evaluate this and consider
policy variables constructed both as 30-minute and 1-day changes.
For any given announcement day, we estimate the autoregressive coefficients
in the regressions, 
\begin{align*}
D_{t-j} & =c+\rho_{j+1}D_{t-j-1}+v_{t-j},\qquad\qquad t\in\mathbf{P}\,\qquad\mathrm{and}\qquad\,j=0,\ldots,\,5.
\end{align*}
The results in Table \ref{Table Robustness} show that only $\widehat{\rho}_{2}$
is significantly different from zero when $D_{t}$ is the 1-day change.
For the 30-minute change, none of the $\widehat{\rho}_{j}$'s are
statistically significant. Overall, there is little evidence of leakage
in the available data. Again, this does not exclude the possibility
that there is some leakage on the Thursdays, Fridays and Mondays that
precede an FOMC meeting.

We conclude with a final remark. The sensitivity analysis discussed
above allows for endogeneity in the form of correlation between $\widetilde{D}_{t}$
and $\widetilde{u}_{t}$. Although this is a natural specification,
it is possible that $\mathrm{corr}(\widetilde{D}_{t},\,\widetilde{u}_{t})=0$,
yet $\mathbb{E}(\widetilde{u}_{t}|\,\widetilde{D}_{t})\neq0$.\footnote{Of course this would require non-normality of either $\widetilde{u}_{t}$
or $\widetilde{D}_{t}$.} This is possible with a nonlinear relationship between $\widetilde{D}_{t}$
and $\widetilde{u}_{t}$. To accommodate this, one could change step
6. above. For example, one could specify $\widetilde{u}_{t}=\rho(\widehat{\sigma}_{\widetilde{u}}/\widehat{\sigma}_{D,C})(\widetilde{D}_{t}-\overline{D}_{C})^{2}+\sqrt{1-\rho^{2}}\eta_{t}$
with $\rho\in\left(-1,\,1\right)$ so that $\mathrm{Cov}(\widetilde{D}_{t},\,\widetilde{u}_{t})=0$
and $\mathbb{E}(\widetilde{u}_{t}|\,\widetilde{D}_{t})\neq0$. Then,
one could proceed with the other steps as above where now higher values
of $\rho$ correspond to stronger nonlinear relationship between $\widetilde{u}_{t}$
and $\widetilde{D}_{t}$, and would require a larger $\delta$ for
relative exogeneity to hold. The case of zero correlation and nonlinear
dependence is likely extreme in practice, so the original sensitivity
analysis above should suffice for most empirical applications.

\subsection{\label{Subsection: Application Worst-Case-Coverage and bias-aware CI}Worst-Case
Asymptotic Coverage and Bias-Aware Confidence Intervals}

We consider the high-frequency event study regression of the 2-Year
real forward rates on 30-minute monetary policy surprises. We first
use the bound on the bias from Proposition \ref{Proposition: Worst-case bias}
to assess the worst-case asymptotic coverage of standard confidence
intervals for the causal effect of policy surprises on real forward
rates.

\begin{table}[H]
\caption{\label{Table Worst-case Coverage}Worst-case asymptotic coverage}

\smallskip{}

\begin{centering}
{\footnotesize{}%
\begin{tabular}{cccccccc}
\hline 
{\footnotesize$a=0.90$} & \multicolumn{7}{c}{{\footnotesize$\sqrt{\mathrm{Var}\left(\overline{\varphi}_{Y,u}\right)\mathrm{Var}\left(\overline{g}_{D,u}\right)/J}$}}\tabularnewline
{\footnotesize$M$} & {\footnotesize 0} & {\footnotesize 0.00001} & {\footnotesize 0.0001} & {\footnotesize 0.001} & {\footnotesize 0.01} & {\footnotesize 0.1} & {\footnotesize 1}\tabularnewline
\hline 
\hline 
{\footnotesize 0.0001} & {\footnotesize 0.900} & {\footnotesize 0.900} & {\footnotesize 0.900} & {\footnotesize 0.900} & {\footnotesize 0.900} & {\footnotesize 0.900} & {\footnotesize 0.900}\tabularnewline
{\footnotesize 0.1} & {\footnotesize 0.900} & {\footnotesize 0.900} & {\footnotesize 0.900} & {\footnotesize 0.900} & {\footnotesize 0.900} & {\footnotesize 0.900} & {\footnotesize 0.898}\tabularnewline
{\footnotesize 0.5} & {\footnotesize 0.900} & {\footnotesize 0.900} & {\footnotesize 0.900} & {\footnotesize 0.900} & {\footnotesize 0.900} & {\footnotesize 0.900} & {\footnotesize 0.858}\tabularnewline
{\footnotesize 1} & {\footnotesize 0.900} & {\footnotesize 0.900} & {\footnotesize 0.900} & {\footnotesize 0.900} & {\footnotesize 0.900} & {\footnotesize 0.898} & {\footnotesize 0.737}\tabularnewline
{\footnotesize 10} & {\footnotesize 0.900} & {\footnotesize 0.900} & {\footnotesize 0.900} & {\footnotesize 0.900} & {\footnotesize 0.898} & {\footnotesize 0.737} & {\footnotesize 0.000}\tabularnewline
{\footnotesize 100} & {\footnotesize 0.900} & {\footnotesize 0.900} & {\footnotesize 0.900} & {\footnotesize 0.898} & {\footnotesize 0.737} & {\footnotesize 0.000} & {\footnotesize 0.000}\tabularnewline
{\footnotesize 300} & {\footnotesize 0.900} & {\footnotesize 0.900} & {\footnotesize 0.900} & {\footnotesize 0.885} & {\footnotesize 0.088} & {\footnotesize 0.000} & {\footnotesize 0.000}\tabularnewline
{\footnotesize 1000} & {\footnotesize 0.900} & {\footnotesize 0.900} & {\footnotesize 0.898} & {\footnotesize 0.737} & {\footnotesize 0.000} & {\footnotesize 0.000} & {\footnotesize 0.000}\tabularnewline
\hline 
\end{tabular}}{\footnotesize\par}
\par\end{centering}
{\small ~~~~~~~~~~~~~~~~~~~~~~~~~}{\scriptsize{}%
\begin{minipage}[t]{0.65\columnwidth}%
{\scriptsize Worst-case asymptotic coverage probability of the 90\%
confidence interval. The outcome variable is the 2-Year U.S. Treasury
instantaneous real forward rate and the policy variable is the 30-minute
policy news surprise.}%
\end{minipage}}{\scriptsize\par}
\end{table}

Assuming stationarity, Table \ref{Table Worst-case Coverage} reports
the worst-case asymptotic coverage probability of the confidence interval
$\mathrm{CI}(\widehat{\beta}_{\mathrm{ES}})$ across values of $M=c\rho_{\mathrm{max}}$
and of the variance ratio $\sqrt{\mathrm{Var}\left(\overline{\varphi}_{Y,u}\right)}\times$
$\sqrt{\mathrm{Var}\left(\overline{g}_{D,u}\right)/J}$. As $M$
or the variance ratio increases, the coverage probability deteriorates.
For the regression under consideration with $\rho_{\mathrm{max}}=1$,
we have $M\approx\sqrt{T_{P}}/\widehat{\sigma}_{D,P}=303$ and variance
ratio $\widehat{\sigma}_{Y,C}\widehat{\sigma}_{D,C}/\sqrt{\widehat{J}}=0.00013,$
which corresponds to a coverage probability of 0.8998. This suggests
that the asymptotic bias in this high-frequency regression is small
enough that it does not significantly distort the coverage probability
of the conventional confidence interval, even under a worst-case scenario.
It would seem to take an unrealistically large bias to reduce the
worst-case coverage probability significantly below the nominal level
in this application.

Next, we analyze the relative length of the bias-aware confidence
interval compared to the conventional confidence interval in the high-frequency
event study regression of 2-Year real forward rates on 30-minute policy
surprises. Table \ref{Table: Lenght CI standard and bias aware} shows
that the bias-aware and conventional confidence intervals often have
the same length. The bias-aware interval is wider only for a few combinations
of $M$ and the variance ratio. It would take an unrealistically large
asymptotic bias for the bias-aware confidence interval to be significantly
wider than the conventional one. For the values $M=\sqrt{T_{P}}/\widehat{\sigma}_{D,P}=303$
and variance ratio $\widehat{\sigma}_{Y,C}\widehat{\sigma}_{D,C}/\sqrt{\widehat{J}}=0.00013,$
which are relevant in this application, the two confidence intervals
have the same length.

\begin{table}[H]
\caption{\label{Table: Lenght CI standard and bias aware}Relative length of
$\mathrm{CI}_{\mathrm{BA}}$ versus $\mathrm{CI}$}

\smallskip{}

\begin{centering}
{\footnotesize{}%
\begin{tabular}{cccccccc}
\hline 
 & \multicolumn{7}{c}{{\footnotesize$\sqrt{\mathrm{Var}\left(\overline{\varphi}_{Y,u}\right)\mathrm{Var}\left(\overline{g}_{D,u}\right)/J}$}}\tabularnewline
{\footnotesize$M$} & {\footnotesize 0} & {\footnotesize 0.00001} & {\footnotesize 0.0001} & {\footnotesize 0.001} & {\footnotesize 0.01} & {\footnotesize 0.1} & {\footnotesize 1}\tabularnewline
\hline 
\hline 
{\footnotesize 0.0001} & {\footnotesize 1.000} & {\footnotesize 1.000} & {\footnotesize 1.000} & {\footnotesize 1.000} & {\footnotesize 1.000} & {\footnotesize 1.000} & {\footnotesize 1.000}\tabularnewline
{\footnotesize 0.1} & {\footnotesize 1.000} & {\footnotesize 1.000} & {\footnotesize 1.000} & {\footnotesize 1.000} & {\footnotesize 1.000} & {\footnotesize 1.000} & {\footnotesize 1.005}\tabularnewline
{\footnotesize 0.5} & {\footnotesize 1.000} & {\footnotesize 1.000} & {\footnotesize 1.000} & {\footnotesize 1.000} & {\footnotesize 1.000} & {\footnotesize 1.001} & {\footnotesize 1.118}\tabularnewline
{\footnotesize 1} & {\footnotesize 1.000} & {\footnotesize 1.000} & {\footnotesize 1.000} & {\footnotesize 1.000} & {\footnotesize 1.000} & {\footnotesize 1.005} & {\footnotesize 1.389}\tabularnewline
{\footnotesize 10} & {\footnotesize 1.000} & {\footnotesize 1.000} & {\footnotesize 1.000} & {\footnotesize 1.000} & {\footnotesize 1.005} & {\footnotesize 1.389} & {\footnotesize 6.869}\tabularnewline
{\footnotesize 100} & {\footnotesize 1.000} & {\footnotesize 1.000} & {\footnotesize 1.000} & {\footnotesize 1.005} & {\footnotesize 1.389} & {\footnotesize 6.869} & {\footnotesize 61.581}\tabularnewline
{\footnotesize 300} & {\footnotesize 1.000} & {\footnotesize 1.000} & {\footnotesize 1.000} & {\footnotesize 1.044} & {\footnotesize 2.603} & {\footnotesize 19.027} & {\footnotesize 183.161}\tabularnewline
{\footnotesize 1000} & {\footnotesize 1.000} & {\footnotesize 1.000} & {\footnotesize 1.005} & {\footnotesize 1.389} & {\footnotesize 6.869} & {\footnotesize 61.581} & {\footnotesize 608.693}\tabularnewline
\hline 
\end{tabular}}{\footnotesize\par}
\par\end{centering}
{\small ~~~~~~~~~~~~~~~~~~~~~~~~~}{\scriptsize{}%
\begin{minipage}[t]{0.65\columnwidth}%
{\scriptsize Relative length of bias-aware confidence interval versus
standard confidence interval. Significance level is $a=0.10$. The
outcome variable is the 2-Year U.S. Treasury instantaneous real forward
rate and the policy variable is the 30-minute policy news surprise.}%
\end{minipage}}{\scriptsize\par}
\end{table}

\subsection{\label{Subsection: Response-of-Blue}Response of Blue Chip Forecasts
on Output to Monetary Policy News}

We consider the high-frequency event study regression in \citet{nakamura/steinsson:2018}:
\begin{align}
BCrev_{t} & =\beta_{0}+\beta D_{t}+\widetilde{u}_{t},\label{Eq. (Reg BCrev)}
\end{align}
where $BCrev_{t}$ is the monthly change in Blue Chip survey expectations
about real GDP and $D_{t}$ is the policy news shock that occurs in
that month. See \citet{bauer/swanson:2021} and \citet{nakamura/steinsson:2018}
for details on how to construct $BCrev_{t}$ and $D_{t}$. It is likely
that $D_{t}$ is endogenous for several reasons. As shown by \citet{bauer/swanson:2021}
$D_{t}$ is correlated with publicly known macroeconomic and financial
market data, say $X_{t}$, that predate the FOMC announcement.\footnote{The index $t$ of $X_{t}$ should not create confusion. Since these
data releases occur before the FOMC announcement, $X_{t}$ collects
information that is known before date $t$ but is still observable
at date $t$ or is correlated with variables or news that are realized
or occur at time $t$.} Since $X_{t}$ is omitted from the regressors, $\widetilde{u}_{t}$
and $D_{t}$ are correlated. Nevertheless, the event study regression
is valid provided that relative exogeneity holds. The latter requires
the policy shock to dominate any other variable in the event window.
This means that the variance of $D_{t}$ cannot be an order of magnitude
smaller than the variance of $BCrev_{t}.$ $BCrev_{t}$ is constructed
as the change in the average of the 1-, 2-, and 3-quarter ahead consensus
forecasts. We denote the latter as $BCrev_{t}\mathrm{-1q}$, $BCrev_{t}\mathrm{-2q}$
and $BCrev_{t}\mathrm{-3q}$, respectively.

Table \ref{Table BCrev} reports the sample variances of $D_{t}$,
$BCrev_{t}$, $BCrev_{t}\mathrm{-1q}$, $BCrev_{t}\mathrm{-2q}$ and
$BCrev_{t}\mathrm{-3q}$, and of the Treasury yields over the full-sample
1995-2014 as well as over the sub-samples 2000-2014, 2000-2007 and
1995-2000. Strikingly, the variance of $D_{t}$ is much smaller than
that of $BCrev_{t}$. For example, in the sample 2000-2014 the variances
of $BCrev_{t}$ and $BCrev_{t}\mathrm{-1q}$ are thirteen and thirty
four times larger than the variance of $D_{t}$. This is likely due
to the fact that the event window for the dependent variable is one
month but it is 30 minutes for the policy variable. Intuitively, while
the policy surprise $D_{t}$ is constructed as a 30-minute change,
when $Y_{t}$ is the one-month change in the Blue Chip forecasts,
it may have too large a variance relative to the policy shock as it
aggregates all news and factors that are relevant over the month.\footnote{An alternative explanation could be that relative exogeneity holds
and $\beta$ is very large in absolute value. We rule out this possibility
as $\beta$ would need to be implausibly large in order to generate
such a large difference.} This implies that relative exogeneity does not provide a good approximation
when $Y_{t}$ is a much lower-frequency change than $D_{t}$, and
any correlation between $D_{t}$ and $\widetilde{u}_{t}$ is not overwhelmed
by the high variance of the policy shock.

\begin{table}[H]
\caption{\label{Table BCrev}The sample variances of $D_{t}$, $BCrev_{t}$
and Treasury yields}

\smallskip{}

\begin{centering}
{\footnotesize{}%
\begin{tabular}{ccccc}
\hline 
 & \multicolumn{1}{c}{{\footnotesize 1995-2014}} & \multicolumn{1}{c}{{\footnotesize 2000-2014}} & \multicolumn{1}{c}{{\footnotesize 2000-2007}} & \multicolumn{1}{c}{{\footnotesize 1995-2000}}\tabularnewline
\hline 
\hline 
{\footnotesize$D_{t}$} & {\footnotesize 0.0012} & {\footnotesize 0.0012} & {\footnotesize 0.0017} & {\footnotesize 0.0012}\tabularnewline
{\footnotesize$BCrev_{t}$} & {\footnotesize 0.0134} & {\footnotesize 0.0158} & {\footnotesize 0.0132} & {\footnotesize 0.0095}\tabularnewline
{\footnotesize$BCrev_{t}\mathrm{-1q}$} & {\footnotesize 0.0334} & {\footnotesize 0.0411} & {\footnotesize 0.0346} & {\footnotesize 0.0207}\tabularnewline
{\footnotesize$BCrev_{t}\mathrm{-2q}$} & {\footnotesize 0.0152} & {\footnotesize 0.0176} & {\footnotesize 0.0147} & {\footnotesize 0.0116}\tabularnewline
{\footnotesize$BCrev_{t}\mathrm{-3q}$} & {\footnotesize 0.0096} & {\footnotesize 0.0089} & {\footnotesize 0.0084} & {\footnotesize 0.0105}\tabularnewline
{\footnotesize 2-Year Forward} &  & {\footnotesize 0.0065} & {\footnotesize 0.0051} & \tabularnewline
{\footnotesize 5-Year Forward} &  & {\footnotesize 0.0054} & {\footnotesize 0.0026} & \tabularnewline
\hline 
\end{tabular}}{\footnotesize\par}
\par\end{centering}
{\small ~~~~~~~~~~~~~~~~~~~~~~~~~}{\scriptsize{}%
\begin{minipage}[t]{0.65\columnwidth}%
{\scriptsize The sample variance of $D_{t}$, $BCrev_{t}$ and Treasury
yields over different sub-samples. $BCrev_{t}\mathrm{-1q}$, $BCrev_{t}\mathrm{-2q}$
and $BCrev_{t}\mathrm{-3q}$ denote the 1-, 2-, and 3-quarter ahead
Blue Chip forecast revisions about real GDP, respectively. For the
Treasury yields the sample starts in January 2004.}%
\end{minipage}}{\scriptsize\par}
\end{table}

This explanation is consistent with the evidence in \citet{bauer/swanson:2021}
who showed that once macroeconomic and financial data that predate
the FOMC announcement are controlled for, the coefficient estimates
revert back to having signs consistent with standard macroeconomic
models. The authors attributed this to the correlation between $D_{t}$
and $X_{t}$. To see this, assume that the true model is linear and
the treatment effect is homogeneous. Then, standard macroeconomic
theory suggests that $\beta<0$ when $Y_{t}$ is the one-month change
in the Blue Chip forecasts for real GDP. The correlation estimates
in \citet{bauer/swanson:2021} suggest that the omitted economic news
that predate the announcement are positively correlated with $D_{t}$.
Given that the variance of $D_{t}$ is not substantially larger than
the variance of $Y_{t}$, the bias $\Delta_{t}$ is positive and so
the resulting event study estimate $\widehat{\beta}_{\mathrm{ES}}$
is an upward biased estimate of the causal effect of $D_{t}$ on real
GDP forecast revisions up to even having the wrong sign.

This argument should also apply to the event study regression with
Treasury yields as dependent variable. However, \citet{bauer/swanson:2021}
found that in regressions where the dependent variable is the change
in an asset price or Treasury yield, controlling for macroeconomic
and financial data that predate the FOMC announcement does not change
the point estimates and their  statistical significance. This difference
likely arises because the changes in Treasury yields are constructed
using a 30-minute or 1-day window around the announcement so that
relative exogeneity is likely to provide a good approximation. As
can be seen from Table \ref{Table BCrev}, the changes in Treasury
yields based on a 1-day window have an order of magnitude similar
to those of the policy variable. Since relative exogeneity provides
a good approximation in this case, controlling for $X_{t}$ does not
result in a change in the point estimates even though $D_{t}$ and
$X_{t}$ are correlated. The same omitted economic news that predate
the announcement do not generate bias when $Y_{t}$ is the change
of an asset price or Treasury yield over a similarly-sized narrow
window used to construct the policy surprise. The key point here is
that the validity of the event study approach does not require the
absence of endogeneity. Rather, it requires that the policy shock
dominates any other variable that is present in the event window.
When this condition holds, identification of causal effects from an
event study does not require the inclusion of controls in the event
study regression. It is interesting to note that this implies that,
when the identification conditions are met, event study regressions
are immune to forms of p-hacking that involve searching through different
control specifications, a potential strength of the high-frequency
event study method.

This discussion suggests that the event study approach is more credible
when a narrow window is used for both the dependent and independent
variable. Both 30-minute and 1-day windows are good choices as relative
exogeneity is more likely to  hold. This was also informally discussed
on p. 1289 in \citet{nakamura/steinsson:2018}. If longer windows
(e.g., one-month windows) are used to form the outcome variable, as
is the case for Blue Chip forecasts, then relative exogeneity is less
likely to hold and the researcher should make more effort to appropriately
control for omitted variables and simultaneity. For example, orthogonalizing
surprises rather than using the original surprises as suggested by
\citet{bauer/swanson:2021} could provide a solution in these contexts.

\section{\label{Section Conclusions}Conclusions}

We establish nonparametric conditions for identification of casual
effects in high-frequency event studies. We show that identification
can be achieved via a separability condition on the policy shock from
the other variables present in the window, and relative exogeneity
which refers to the variance of the policy shock being an order of
magnitude larger than that of the other variables. Under these conditions
we establish the causal meaning of the event study estimand, the super-consistency
and asymptotic distribution of the event study estimator and its robustness
to nonlinearities. We provide bounds on the bias and use them to study
the worst-case coverage properties of standard confidence intervals
and to construct bias-aware inference procedures. We propose a simple
procedure that can be used to assess relative exogeneity as an approximation
and apply it to \citeauthor{nakamura/steinsson:2018}'s (\citeyear{nakamura/steinsson:2018})
analysis on the real effects of monetary policy.

\newpage{}

\newpage{}

\bibliographystyle{econometrica}
\bibliography{References}
\addcontentsline{toc}{section}{References}

\newpage{}

\clearpage 
\pagenumbering{arabic}
\renewcommand*{\thepage}{A-\arabic{page}}
\appendix
\begin{center}
{\LARGE\textbf{Supplemental Appendix}}{\LARGE\par}
\par\end{center}

\section{\label{Section Mathematical-Appendix}Mathematical Proofs}

We begin with the following lemma.
\begin{lem}
\label{Lemma: Independence Y(d) and D}Let Assumptions \ref{Assumption: Structural Form}-\ref{Assumption: Relative Exogeneity}
hold. Then, for all $t\in\mathbf{P}$ and all $d\in\mathbf{D}$, 
\begin{align*}
\mathbb{E}\left(\widetilde{Y}_{t}\left(d\right)\left(\widetilde{D}_{t}-\mathbb{E}\left(\widetilde{D}_{t}\right)\right)\right)\qquad\text{and}\qquad\mathbb{E}\left(Y_{t}^{*}\left(d\right)\left(D_{t}^{*}-\mathbb{E}\left(D_{t}^{*}\right)\right)\right)
\end{align*}
are monotonically decreasing to zero as $\sigma_{e,t}^{2}\rightarrow\infty$,
where ${Y}_{t}^{*}(d)=\sigma_{D}^{-1}Y_{t}(d)$ with $\sigma_{D}^{2}=\lim_{T_{P}\rightarrow\infty}T_{P}^{-1}$
$\sum_{t=1}^{T_{P}}\mathrm{Var}\left(D_{t}\right)$ and $T_{P}$ is
the number of observations in the policy sample.
\end{lem}
\noindent{\textit{Proof}.} We only provide the proof for $\mathbb{E}(\widetilde{Y}_{t}\left(d\right)(\widetilde{D}_{t}-\mathbb{E}(\widetilde{D}_{t})))$
since the proof for $\mathbb{E}\left(Y_{t}^{*}\left(d\right)\left(D_{t}^{*}-\mathbb{E}(D_{t}^{*})\right)\right)$
is nearly identical. Using the structural and reduced forms for $Y_{t}$
and $D_{t}$ in Assumption \ref{Assumption: Structural Form}-\ref{Assumption: Reduced-form},
we have 
\begin{align}
\mathbb{E} & \left(\widetilde{Y}_{t}\left(d\right)\left(\widetilde{D}_{t}-\mathbb{E}(\widetilde{D}_{t})\right)\right)\nonumber \\
 & =\mathbb{E}\left(\sigma_{D,t}^{-1}\left(\varphi_{Y,D}\left(d,\,t\right)+\varphi_{Y,u}\left(Z_{t},\,u_{t},\,t\right)\right)\left(\widetilde{D}_{t}-\mathbb{E}(\widetilde{D}_{t})\right)\right)\nonumber \\
 & =\sigma_{D,t}^{-2}\mathbb{E}\left[\left(\varphi_{Y,D}\left(d,\,t\right)+\varphi_{Y,u}\left(Z_{t},\,u_{t},\,t\right)\right)\right.\nonumber \\
 & \times\left.\left(g_{D,e}\left(e_{t},\,t\right)+g_{D,u}\left(Z_{t},\,u_{t},\,t\right)-\mathbb{E}\left(g_{D,e}\left(e_{t},\,t\right)+g_{D,u}\left(Z_{t},\,u_{t},\,t\right)\right)\right)\right]\nonumber \\
 & =\sigma_{D,t}^{-2}\varphi_{Y,D}\left(d,\,t\right)\mathbb{E}\left(g_{D,e}\left(e_{t},\,t\right)+g_{D,u}\left(Z_{t},\,u_{t},\,t\right)-\mathbb{E}\left(g_{D,e}\left(e_{t},\,t\right)+g_{D,u}\left(Z_{t},\,u_{t},\,t\right)\right)\right)\nonumber \\
 & \quad+\sigma_{D,t}^{-2}\mathbb{E}\left(\varphi_{Y,u}\left(Z_{t},\,u_{t},\,t\right)\left(g_{D,e}\left(e_{t},\,t\right)+g_{D,u}\left(Z_{t},\,u_{t},\,t\right)-\mathbb{E}\left(g_{D,e}\left(e_{t},\,t\right)+g_{D,u}\left(Z_{t},\,u_{t},\,t\right)\right)\right)\right)\nonumber \\
 & =0+\sigma_{D,t}^{-2}\mathbb{E}\left(\varphi_{Y,u}\left(Z_{t},\,u_{t},\,t\right)\left(g_{D,e}\left(e_{t},\,t\right)-\mathbb{E}\left(g_{D,e}\left(e_{t},\,t\right)\right)\right)\right)\label{Eq. (0) Proof Lemma}\\
 & \quad+\sigma_{D,t}^{-2}\mathbb{E}\left(\varphi_{Y,u}\left(Z_{t},\,u_{t},\,t\right)\left(g_{D,u}\left(Z_{t},\,u_{t},\,t\right)-\mathbb{E}\left(g_{D,u}\left(Z_{t},\,u_{t},\,t\right)\right)\right)\right).\nonumber 
\end{align}
By Assumption \ref{Assumption: Structural shocks}, $e_{t}$ is independent
of $\left(Z_{t},\,u_{t}\right)$ and so 
\begin{align}
\sigma_{D,t}^{-2}\mathbb{E}\left(\varphi_{Y,u}\left(Z_{t},\,u_{t},\,t\right)\left(g_{D,e}\left(e_{t},\,t\right)-\mathbb{E}\left(g_{D,e}\left(e_{t},\,t\right)\right)\right)\right) & =0.\label{Eq. (1) Poof Lemma}
\end{align}
By Assumption \ref{Assumption: Relative Exogeneity}(iii), 
\begin{align}
\sigma_{D,t}^{-2} & \mathbb{E}\left(\varphi_{Y,u}\left(Z_{t},\,u_{t},\,t\right)\left(g_{D,u}\left(Z_{t},\,u_{t},\,t\right)-\mathbb{E}\left(g_{D,u}\left(Z_{t},\,u_{t},\,t\right)\right)\right)\right)\label{Eq. (2) Proof Lemma}\\
 & =\sigma_{D,t}^{-2}C,\nonumber 
\end{align}
for some $C<\infty.$ Thus, \eqref{Eq. (0) Proof Lemma}-\eqref{Eq. (2) Proof Lemma}
imply the statement of the lemma since $\sigma_{D,t}^{2}\rightarrow\infty$
as $\sigma_{e,t}^{2}\rightarrow\infty$ by Assumption \ref{Assumption: Relative Exogeneity}(ii).
$\square$

\subsection{Proof of Theorem \ref{Theorem: WATE Event Study}}

\noindent Recall that $\widetilde{Y}_{t}=\widetilde{Y}_{t}(\widetilde{D}_{t})$.
By the fundamental theorem of calculus and Assumption \ref{Assumption: Support of D}-\ref{Assumption: Differentiability of Potential Outcome},
we have 
\begin{align*}
\widetilde{Y}_{t} & =\widetilde{Y}_{t}\left(\underline{d}\right)+\int_{\underline{d}}^{\widetilde{D}_{t}}\frac{\partial\widetilde{Y}_{t}\left(d\right)}{\partial d}\mathrm{d}d\\
 & =\widetilde{Y}_{t}\left(\underline{d}\right)+\int_{\underline{d}}^{\overline{d}}\frac{\partial\widetilde{Y}_{t}\left(d\right)}{\partial d}\mathbf{1}\left\{ d\leq\widetilde{D}_{t}\right\} \mathrm{d}d.
\end{align*}
Using this and the fact that $\mathrm{Var}(\widetilde{D}_{t})=1$,
we have 
\begin{align*}
\beta_{\mathrm{ES},t} & =\frac{\mathrm{Cov}(\widetilde{Y}_{t},\widetilde{D}_{t})}{\mathrm{Var}(\widetilde{D}_{t})}=\mathrm{Cov}(\widetilde{Y}_{t},\widetilde{D}_{t})\\
 & =\mathrm{Cov}(\widetilde{Y}_{t}\left(\underline{d}\right),\widetilde{D}_{t})+\mathbb{E}\left(\int_{\underline{d}}^{\overline{d}}\frac{\partial\widetilde{Y}_{t}\left(d\right)}{\partial d}\mathbf{1}\left\{ d\leq\widetilde{D}_{t}\right\} \left(\widetilde{D}_{t}-\mathbb{E}(\widetilde{D}_{t})\right)\mathrm{d}d\right)\\
 & =\Delta_{t}+\int_{\underline{d}}^{\overline{d}}\mathbb{E}\left(\frac{\partial\widetilde{Y}_{t}\left(d\right)}{\partial d}\mathbf{1}\left\{ d\leq\widetilde{D}_{t}\right\} \mathrm{d}d\left(\widetilde{D}_{t}-\mathbb{E}(\widetilde{D}_{t})\right)\right)\\
 & =\Delta_{t}+\int_{\underline{d}}^{\overline{d}}\frac{\partial\widetilde{Y}_{t}\left(d\right)}{\partial d}\mathbb{E}\left(\mathbf{1}\left\{ d\leq\widetilde{D}_{t}\right\} \left(\widetilde{D}_{t}-\mathbb{E}\left(\widetilde{D}_{t}\right)\right)\right)\mathrm{d}d,
\end{align*}
where the fourth equality holds by Fubini's Theorem and Assumption
\ref{Assumption: Support of D}-\ref{Assumption: Differentiability of Potential Outcome}
and the final equality follows from Assumption \ref{Assumption: Structural Form}
since $\partial\widetilde{Y}_{t}\left(d\right)/\partial d=\partial\varphi_{Y,D}(d,\,t)/\partial d$,
proving the first statement of the theorem.

To see that the weights are non-negative, note that for $d\in[\underline{d},\,\overline{d}]$
we have 
\begin{align*}
\mathbb{E} & \left(\mathbf{1}\left\{ d\leq\widetilde{D}_{t}\right\} \left(\widetilde{D}_{t}-\mathbb{E}(\widetilde{D}_{t})\right)\right)\\
 & =\mathbb{E}\left(\mathbf{1}\left\{ d\leq\widetilde{D}_{t}\right\} \widetilde{D}_{t}\right)-\mathbb{E}\left(\mathbf{1}\left\{ d\leq\widetilde{D}_{t}\right\} \right)\mathbb{E}\left(\widetilde{D}_{t}\right)\\
 & =\mathbb{E}\left(\widetilde{D}_{t}|\,d\leq\widetilde{D}_{t}\right)\mathbb{E}\left(\mathbf{1}\left\{ d\leq\widetilde{D}_{t}\right\} \right)-\mathbb{E}\left(\mathbf{1}\left\{ d\leq\widetilde{D}_{t}\right\} \right)\mathbb{E}\left(\widetilde{D}_{t}\right)\\
 & =\left(\mathbb{E}\left(\widetilde{D}_{t}|\,d\leq\widetilde{D}_{t}\right)-\mathbb{E}\left(\widetilde{D}_{t}\right)\right)\mathbb{P}\left(d\leq\widetilde{D}_{t}\right)\geq0,
\end{align*}
since $\mathbb{E}(\widetilde{D}_{t}|\,d\leq\widetilde{D}_{t})-\mathbb{E}(\widetilde{D}_{t})\geq0$
for $d\in[\underline{d},\,\overline{d}]$. To see that the weights
integrate to one, note that 
\begin{align*}
\widetilde{D}_{t} & =\underline{d}+\int_{\underline{d}}^{\widetilde{D}_{t}}\mathrm{d}\widetilde{d}=\underline{d}+\int_{\underline{d}}^{\overline{d}}\mathbf{1}\left\{ \widetilde{d}\leq\widetilde{D}_{t}\right\} \mathrm{d}\widetilde{d}.
\end{align*}
Using this we have, 
\begin{align*}
1=\mathrm{Var}\left(\widetilde{D}_{t}\right) & =\mathbb{E}\left[\left(\widetilde{D}_{t}-\underline{d}\right)\left(\widetilde{D}_{t}-\mathbb{E}\left(\widetilde{D}_{t}\right)\right)\right]\\
 & \quad-\mathbb{E}\left[\mathbb{E}\left(\int_{\underline{d}}^{\overline{d}}\mathbf{1}\left\{ d\leq\widetilde{D}_{t}\right\} \mathrm{d}d\right)\left(\widetilde{D}_{t}-\mathbb{E}\left(\widetilde{D}_{t}\right)\right)\right]\\
 & =\mathbb{E}\left[\left(\widetilde{D}_{t}-\underline{d}\right)\left(\widetilde{D}_{t}-\mathbb{E}\left(\widetilde{D}_{t}\right)\right)\right]\\
 & \quad-\mathbb{E}\left(\int_{\underline{d}}^{\overline{d}}\mathbf{1}\left\{ d\leq\widetilde{D}_{t}\right\} \mathrm{d}\widetilde{d}\right)\mathbb{E}\left(\widetilde{D}_{t}-\mathbb{E}\left(\widetilde{D}_{t}\right)\right)\\
 & =\mathbb{E}\left[\left(\widetilde{D}_{t}-\underline{d}\right)\left(\widetilde{D}_{t}-\mathbb{E}\left(\widetilde{D}_{t}\right)\right)\right]\\
 & =\mathbb{E}\left[\int_{\underline{d}}^{\overline{d}}\mathbf{1}\left\{ d\leq\widetilde{D}_{t}\right\} \mathrm{d}\widetilde{d}\left(\widetilde{D}_{t}-\mathbb{E}\left(\widetilde{D}_{t}\right)\right)\right]\\
 & =\int_{\underline{d}}^{\overline{d}}\mathbb{E}\left(\mathbf{1}\left\{ d\leq\widetilde{D}_{t}\right\} \left(\widetilde{D}_{t}-\mathbb{E}\left(\widetilde{D}_{t}\right)\right)\right)\mathrm{d}d.
\end{align*}
Finally, the fact that $\Delta_{t}$ is decreasing in $\sigma_{e,t}^{2}$
follows directly from \eqref{Eq. (2) Proof Lemma} which continues
to hold when Assumption \ref{Assumption: Relative Exogeneity} is
replaced by Assumption \ref{Assumption: NO Exogeneity}. $\square$

\subsection{Proof of Theorem \ref{Theorem: WATE Event Study with Relative Exogeneity}}

The result follows directly from Theorem \ref{Theorem: WATE Event Study}
and Lemma \ref{Lemma: Independence Y(d) and D}. $\square$

\subsection{Proof of Theorem \ref{Theorem: OLS ES consistency +  delta}}

We have 
\begin{align*}
\widehat{\beta}_{\mathrm{ES}} & =\frac{\sigma_{D}^{-2}T_{P}^{-1}\sum_{t=1}^{T_{P}}\left(D_{t}-\overline{D}\right)\left(Y_{t}-\overline{Y}\right)}{\sigma_{D}^{-2}T_{P}^{-1}\sum_{t=1}^{T_{P}}\left(D_{t}-\overline{D}\right)^{2}}\\
 & =\frac{T_{P}^{-1}\sum_{t=1}^{T_{P}}\left(D_{t}^{*}-\overline{D}^{*}\right)\left(Y_{t}^{*}-\overline{Y}^{*}\right)}{T_{P}^{-1}\sum_{t=1}^{T_{P}}\left(D_{t}^{*}-\overline{D}^{*}\right)^{2}}\\
 & \overset{\mathbb{P}}{\rightarrow}\int_{0}^{1}c\left(D^{*},\,Y^{*},\,s\right)\mathrm{d}s
\end{align*}
by Assumption \ref{Assumption LLN}. Given the definition of $c\left(D^{*},\,Y^{*},\,s\right)$,
the statements of the theorem then follow from the same arguments
as in the proof of Theorem \ref{Theorem: WATE Event Study}. $\square$

\subsection{Proof of Theorem \ref{Theorem: OLS consistency}}

Note that Theorem \ref{Theorem: OLS ES consistency +  delta} is obtained
under Assumption \ref{Assumption: NO Exogeneity} while the current
theorem uses Assumption \ref{Assumption: Relative Exogeneity}. The
difference between the two assumptions is that $\mathbb{E}(g_{D,e}\left(e_{t},\,t\right)^{2})<\infty$
in Assumption \ref{Assumption: NO Exogeneity} while $\mathbb{E}(g_{D,e}\left(e_{t},\,t\right)^{2})\rightarrow\infty$
as $\sigma_{e,t}^{2}\rightarrow\infty$ in Assumption \ref{Assumption: Relative Exogeneity}.
Given that we take the limits sequentially, we can use the limiting
result from Theorem \ref{Theorem: OLS ES consistency +  delta} before
taking the limit as $\sigma_{e,t}^{2}\rightarrow\infty$. Thus, the
theorem follows directly from Theorem \ref{Theorem: OLS ES consistency +  delta}
and Lemma \ref{Lemma: Independence Y(d) and D}. $\square$


\subsection{Proof of Theorem \ref{Theorem: CLT}}

We have 
\begin{align*}
\widehat{\beta}_{\mathrm{ES}}-\beta_{\mathrm{ES}} & =\frac{T_{P}^{-1}\sum_{t=1}^{T_{P}}\left(D_{t}^{*}-\overline{D}^{*}\right)\left(Y_{t}^{*}-\overline{Y}^{*}-\beta_{\mathrm{ES}}\left(D_{t}^{*}-\overline{D}^{*}\right)\right)}{T_{P}^{-1}\sum_{t=1}^{T_{P}}\left(D_{t}^{*}-\overline{D}^{*}\right)^{2}}=\frac{T_{P}^{-1}\sum_{t=1}^{T_{P}}\left(D_{t}^{*}-\overline{D}^{*}\right)\varepsilon_{t}}{T_{P}^{-1}\sum_{t=1}^{T_{P}}\left(D_{t}^{*}-\overline{D}^{*}\right)^{2}}.
\end{align*}
Using Assumption \ref{Assumption LLN}(ii) and Assumption \ref{Assumption CLT},
we obtain the statement of the theorem. $\square$

\subsection{Proof of Corollary \ref{Corollary: unbiased OLS anorm}}

Note that $\sqrt{T_{P}\sigma_{D}^{2}}(\widehat{\beta}_{\mathrm{ES}}-\beta_{\mathrm{ES}})\overset{d}{\rightarrow}\mathscr{N}(0,J)$
by the same argument as in the proof of Theorem \ref{Theorem: CLT}.
The statement of the corollary then follows after noting 
\begin{align*}
\sqrt{T_{P}\sigma_{D}^{2}}\Delta(T_{P}) & =\frac{\sqrt{T_{P}}}{\sigma_{D}}\int_{0}^{1}\mathbb{E}\left[\varphi_{Y,u}\left(Z_{\lfloor T_{P}s\rfloor},u_{\lfloor T_{P}s\rfloor},\lfloor T_{P}s\rfloor\right)\right.\\
 & \times\left.\left(g_{D,u}\left(Z_{\lfloor T_{P}s\rfloor},u_{\lfloor T_{P}s\rfloor},\lfloor T_{P}s\rfloor\right)-\mathbb{E}\left(g_{D,u}\left(Z_{\lfloor T_{P}s\rfloor},u_{\lfloor T_{P}s\rfloor},\lfloor T_{P}s\rfloor\right)\right)\right)\right]\mathrm{d}s
\end{align*}
by the same arguments used in Lemma \ref{Lemma: Independence Y(d) and D}.
$\square$

\subsection{Proof of Proposition \ref{Proposition: Worst-case bias}}

Using Assumptions \ref{Assumption: Structural Form}-\ref{Assumption: Reduced-form},
\begin{align*}
 & \mathbb{E}\left[Y_{t}^{*}\left(\underline{d}\right)\left(D_{t}^{*}-\mathbb{E}\left(D_{t}^{*}\right)\right)\right]\\
 & =\frac{1}{\sigma_{D}^{2}}\mathbb{E}\left[Y_{t}\left(\underline{d}\right)\left(g_{D,e}\left(e_{t},t\right)+g_{D,u}\left(Z_{t},\,u_{t},t\right)-\mathbb{E}\left(g_{D,e}\left(e_{t},t\right)+g_{D,u}\left(Z_{t},u_{t},t\right)\right)\right)\right]\\
 & =\frac{1}{\sigma_{D}^{2}}\frac{\sqrt{\mathrm{Var}\left(\varphi_{Y,u}\left(Z_{t},u_{t},t\right)\right)}}{\sqrt{\mathrm{Var}\left(\varphi_{Y,u}\left(Z_{t},u_{t},t\right)\right)}}\mathbb{E}\left[\left(\varphi_{Y,u}\left(Z_{t},u_{t},t\right)\right)\left(g_{D,u}\left(Z_{t},u_{t},t\right)-\mathbb{E}\left(g_{D,u}\left(Z_{t},u_{t},t\right)\right)\right)\right]\\
 & =\frac{1}{\sigma_{D}^{2}}\sqrt{\mathrm{Var}\left(\varphi_{Y,u}\left(Z_{t},u_{t},t\right)\right)}\sqrt{\mathrm{Var}\left(g_{D,u}\left(Z_{t},u_{t},t\right)\right)}\frac{\mathbb{E}\left[\left(\varphi_{Y,u}\left(Z_{t},u_{t},t\right)\right)\left(g_{D,u}\left(Z_{t},u_{t},t\right)-\mathbb{E}\left(g_{D,u}\left(Z_{t},u_{t},t\right)\right)\right)\right]}{\sqrt{\mathrm{Var}\left(\varphi_{Y,u}\left(Z_{t},u_{t},t\right)\right)}\sqrt{\mathrm{Var}\left(g_{D,u}\left(Z_{t},u_{t},t\right)\right)}}\\
 & =\frac{1}{\sigma_{D}^{2}}\sqrt{\mathrm{Var}\left(\varphi_{Y,u}\left(Z_{t},u_{t},t\right)\right)}\sqrt{\mathrm{Var}\left(g_{D,u}\left(Z_{t},u_{t},t\right)\right)}\rho_{Zu,T_{P}}\left(t/T_{P}\right).
\end{align*}
The result follows by the expression for $\Delta$ in Theorem \ref{Theorem: OLS ES consistency +  delta}.
$\square$

\subsection{Proof of Corollary \ref{Corollary: Worst-Case Asymptotic Coverage}}

By construction, 
\begin{align*}
\mathbb{P}\left(\left(\beta_{\mathrm{ES}}-\Delta(T_{P})\right)\in\mathrm{CI}\left(\widehat{\beta}_{\mathrm{ES}}\right)\right) & =\mathbb{P}\left(\widehat{\beta}_{\mathrm{ES}}-z_{1-a/2}\sqrt{\frac{J}{T_{P}\sigma_{D}^{2}}}\leq\beta_{\mathrm{ES}}-\Delta(T_{P})\leq\widehat{\beta}_{\mathrm{ES}}+z_{1-a/2}\sqrt{\frac{J}{T_{P}\sigma_{D}^{2}}}\right)\\
 & =\mathbb{P}\left(\frac{|\sqrt{T_{P}\sigma_{D}^{2}}(\widehat{\beta}_{\mathrm{ES}}-(\beta_{\mathrm{ES}}-\Delta(T_{P})))|}{\sqrt{J}}\leq z_{1-a/2}\right)\\
 & \rightarrow\mathbb{P}\left(\left|\mathcal{Z}+\frac{\lim_{T_{P}\rightarrow\infty}\sqrt{T_{P}}\sigma_{D}\Delta(T_{P})}{\sqrt{J}}\right|\leq z_{1-a/2}\right),
\end{align*}
where $\mathcal{Z}\sim\mathscr{N}\left(0,\,1\right)$ and the convergence
follows from the same arguments as in the proof of Theorem \ref{Theorem: CLT}.
The statement of the corollary then follows from $\sqrt{T_{P}}/\sigma_{D}\rightarrow c$,
the same arguments in the proof of Proposition \ref{Proposition: Worst-case bias}
and the fact that $\mathbb{P}(\left|\mathcal{Z}+r\right|\leq x)$
is decreasing in $r$.

\subsection{Proof of Corollary \ref{Corollary: Bias-aware Inference}}

It follows by direct analogy with the proof of Corollary \ref{Corollary: Worst-Case Asymptotic Coverage}.
$\square$ 

\section{Identification Under Shrinking Variance of Background Noise}

\label{Section Shrinking-Variance}

In this appendix, we briefly discuss why the identification results
of Section \ref{Subsection: Identification-Results} continue to hold
under the alternative framing of relative exogeneity as a policy shock
with finite variance and vanishing variance of the other variables
in the event window. Replace Assumption \ref{Assumption: Relative Exogeneity}
with the following assumption. 
\begin{assumption}
\label{Assumption: Relative Exogeneity alt} For all $t\in\mathbf{P}$,\\
 (i) $\sigma_{u,t}^{2}=\mathrm{Var}(u_{t})\rightarrow0$ and $\sigma_{Z,t}^{2}=\mathrm{Var}(Z_{t})\rightarrow0$,\\
 (ii) $\mathrm{Var}(g_{D,e}\left(e_{t},\,t\right))\neq0$ is finite,
\\
 (iii) $\mathbb{E}(\varphi_{Y,u}\left(Z_{t},\,u_{t},\,t\right)^{2})$
and $\mathbb{E}(g_{D,u}\left(Z_{t},\,u_{t},\,t\right)^{2})$ are decreasing
to zero as $\sigma_{Z,t}^{2},\sigma_{u,t}^{2}\rightarrow0$. 
\end{assumption}
Under this alternative framing of relative exogeneity, Lemma \ref{Lemma: Independence Y(d) and D}
can be modified to the following.
\begin{lem}
\label{Lemma: Independence Y(d) and D alt}Let Assumptions \ref{Assumption: Structural Form}\textendash \ref{Assumption: Structural shocks}
and \ref{Assumption: Relative Exogeneity alt} hold. Then, for all
$t\in\mathbf{P}$ and all $d\in\mathbf{D}$, $\mathbb{E}({Y}_{t}\left(d\right)({D}_{t}-\mathbb{E}({D}_{t})))$
is monotonically decreasing to zero as $\sigma_{u,t}^{2},\sigma_{Z,t}^{2}\rightarrow0$. 
\end{lem}
\noindent{\textit{Proof}.} The proof is essentially identical to
that of Lemma \ref{Lemma: Independence Y(d) and D} since the proof
of Lemma \ref{Lemma: Independence Y(d) and D} up until \eqref{Eq. (2) Proof Lemma}
does not rely upon the normalization of $Y_{t}(d)$ or $D_{t}$ by
$\sigma_{D,t}$ or $\sigma_{D}$ and 
\[
\mathbb{E}\left(\varphi_{Y,u}\left(Z_{t},\,u_{t},\,t\right)\left(g_{D,u}\left(Z_{t},\,u_{t},\,t\right)-\mathbb{E}\left(g_{D,u}\left(Z_{t},\,u_{t},\,t\right)\right)\right)\right)
\]
decreases to zero as $\sigma_{Z,t}^{2},\sigma_{u,t}^{2}\rightarrow0$
under Assumption \ref{Assumption: Relative Exogeneity alt}(iii).
$\square$

With Lemma \ref{Lemma: Independence Y(d) and D alt} in hand, the
analogs of Theorems \ref{Theorem: WATE Event Study}-\ref{Theorem: WATE Event Study with Relative Exogeneity}
that do not normalize $Y_{t}(d)$ or $D_{t}$ and replace Assumption
\ref{Assumption: Relative Exogeneity} with Assumption \ref{Assumption: Relative Exogeneity alt}
(for Theorem \ref{Theorem: WATE Event Study with Relative Exogeneity})
immediately follow by identical arguments.

\end{document}